\documentclass[a4paper,12pt]{article}

	\usepackage{jcappub}
	
	\usepackage[utf8]{inputenc}
	\usepackage{setspace}
	\usepackage{pdfpages}
	\usepackage{lmodern}
	\usepackage{xcolor,color,soul,colortbl}
	\usepackage{graphicx,graphics}
	\usepackage{indentfirst}
	\usepackage{amsmath,amssymb}
	\usepackage{multirow,bigstrut}
	\usepackage{fullpage}
	\usepackage{epstopdf}
	\usepackage{psfrag}
	\usepackage{pstool}
	\usepackage{slashed}
	\usepackage{bbold}
	\usepackage[makeroom]{cancel}
	\usepackage{xspace}
	\usepackage{rotating}
	
	\graphicspath{{./Images/}}
	
\def\be{\begin{equation}}
\def\ee{\end{equation}}
\def\bea{\begin{eqnarray}}
\def\eea{\end{eqnarray}}
\def\beann{\begin{eqnarray*}}
\def\eeann{\end{eqnarray*}}

\newcommand{\T}{\rule{0pt}{2.8ex}} 
\newcommand{\B}{\rule[-1.2ex]{0pt}{0pt}} 

\newcommand{\five}{MCHM$_5$ }
\newcommand{\fourt}{MCHM$_{14}$ }

\newcommand{\PH}{\ensuremath{h}\xspace}
\newcommand{\ttbar}{\ensuremath{t\overline{t}}\xspace}
\newcommand{\ttH}{\ensuremath{\ttbar\PH}\xspace}
\newcommand{\ttHH}{\ensuremath{\ttbar\PH\PH}\xspace}

\definecolor{tablered}{rgb}{1,0.2,0.2}
\definecolor{tableblue}{rgb}{0.5,0.5,1}
\definecolor{tableorange}{rgb}{1,0.6,0.2}
\definecolor{tablecyan}{rgb}{0.5,1,1}

\makeindex

\begin{document}

\title{Probing the Top-Higgs Sector with Composite Higgs Models at Present and Future Hadron Colliders}

\author{Carlos Bautista,$^{1,2}$}
\emailAdd{carlos.bautista@unesp.br}

\author{Leonardo de Lima,$^3$}
\emailAdd{leonardolima@utfpr.edu.br}

\author{Ricardo D'Elia Matheus,$^1$}
\emailAdd{matheus@ift.unesp.br}

\author{Eduardo Pont\'on,$^{1,2,}$\footnote{Eduardo greatly contributed to the work here presented but sadly passed away in the early stages of the writing of this paper. He will be sorely missed.} }

\author{Le\^onidas A. Fernandes do Prado,$^{1,4}$ and}
\emailAdd{leonidas.prado@cern.ch}

\author{Aurore Savoy-Navarro$^4$}
\emailAdd{aurore.savoy.navarro@cern.ch}

\affiliation{
  $^1$Instituto de F\'isica Te\'orica -- UNESP, S\~ao Paulo, Brazil}

\affiliation{
  $^2$ICTP South American Institute for Fundamental
  Research, S\~ao Paulo, Brazil}

\affiliation{
  $^3$Universidade Tecnol\'ogica Federal do Paran\'a, Toledo, Brazil}

\affiliation{
  $^4$IRFU-CEA, Universit\'e Paris-Saclay and CNRS-IN2P3, France}

\abstract{We study the production of $t{\overline t}h$ and $t{\overline t}hh$ at hadron colliders, in the minimal Composite Higgs Models, based on the coset $SO(5)/SO(4)$. We explore the fermionic representations $\bf 5$ and ${\bf 14}$. A detailed phenomenological analysis is performed, covering the energy range of the LHC and its High Luminosity upgrade, as well as that of a future 100 TeV hadron collider. Both resonant and non-resonant production are considered, stressing the interplay and complementary interest of these channels with each other and double Higgs production. We provide sets of representative points with detailed experimental outcomes in terms of modification of the cross sections as well as resonance masses and branching ratios. For non-resonant production, we gauge the relative importance of Yukawa, Higgs trilinear, and contact $t\bar{t}hh$ vertices to these processes, and consider the prospect for distinguishing the fermion representations from each other and from the Standard Model. In the production of top partners, we find that the three-body decay channel 
$W^+ W^- t$ becomes significant in certain regions of parameter space having a degenerate spectrum, and is further enhanced with energy. This motivates both higher energy machines as well as the need to go beyond the current analysis performed for the searches for these resonances.}

\newpage
\setcounter{footnote}{4}
\maketitle
\setcounter{footnote}{0}

\section{Introduction}

The discovery of the Higgs boson~\cite{Aad:2012tfa,
Chatrchyan:2012xdj} , has aroused the interest in measuring with high
precision the Higgs properties. This has led to greater emphasis on a strong
joint effort between theory and experiments to explore the Higgs sector.
The Large Hadron collider, LHC, at CERN, which is also a Top factory,  offers
an unique playground to perform the searches in the Top-Higgs sector both
now and with its upgrade into the High-Luminosity LHC.

Besides, the
importance of knowing in depth the Higgs sector, led to studying a Higgs factory
as the next machine in Particle Physics as well as the importance of a very
high energy hadron collider.
While all current measurements within their present precision are
consistent with a Standard Model (SM) Higgs boson, there is
room for deviations from the SM that could hold the key to a deeper
understanding of the phenomenon of electroweak symmetry breaking
(EWSB).
Furthermore, as the luminosity increases and experiments are progressively
upgraded, a number of Higgs-related processes
are becoming more and more accessible, in the next years to come.
An important open question is whether the
observed $125$~GeV scalar is a composite bound state of more
fundamental constituents, or whether it is elementary down to
distances much shorter than currently explored.  The experimental
program will be essential in illuminating this fundamental question.

The idea of Higgs compositeness received an important boost with the
construction of rather complete models that can be consistent with all
current measurements~\cite{Agashe:2004rs}.  Such constructions
incorporated a geometric solution to the hierarchy
problem~\cite{Randall:1999ee}, a dynamical mechanism for EWSB, and an
appealing understanding of the flavor structure observed in the SM. In
these scenarios, the Higgs boson is understood as a pseudo-Nambu
Goldstone boson (pNGB), somewhat analogous to the pions in QCD, but
also with important differences.  While much model building and
extensions have been proposed in the literature, in this work we focus
on the minimal setup, based on the symmetry breaking pattern $SO(5)
\to SO(4)$.  These are referred to as ``Minimal Composite Higgs
Models'' (MCHM).

It is known that there is considerable model-dependence associated
with the fermion sector of the MCHM, which is of relevance to our
work.  In particular, the top sector is expected to play a crucial
role in these models, given that the top quark couples most strongly
to the Higgs boson.  We focus on the production of one or two Higgs
bosons in association with a top/anti-top pair.  The $t{\overline t}h$
cross section is being actively measured~\cite{Aaboud:2018urx,
Sirunyan:2018hoz}, in many different channels, with values compatible with the SM prediction, within approximately 20\% uncertainty.  An experimental search for the $t{\overline t}hh$ process has been just performed, for the first time at the LHC ~\cite{LeonidasPhD}.
Such a process is of particular interest in the
present class of models, due to the generic prediction of charge 2/3
vectorlike ``top partners'' that can decay in the $th$ channel, thus
leading to the previous final state.  These resonances are already constrained by previous searches in this channel \cite{Aaboud:2018xuw} as well as in combination with $tZ$ and $bW$ decay channels \cite{Aaboud:2018pii, Sirunyan:2018omb}.

Since the top partner resonances are under active search with already defined bounds above 1 TeV,
we bring here attention to the fact that frequently a large fraction of the $t{\overline t}hh$ cross section is composed of non-resonant production, which then becomes an important discriminator for the MCHM models. This  component is also more directly related to the pNGB nature of the Higgs boson, unlike the vectorlike fermionic resonances. This work highlights
the complementary role between non-resonant $\ttHH$ production and the $\ttH$ and $hh$ channels, specially in terms of measuring the trilinear Higgs couplings.
We gauge the relative importance of Yukawa, trilinear Higgs and new contributions to the non-resonant $\ttHH$ channel, accessing quantitatively the correlation between $\ttH$ and $\ttHH$ cross sections.

To illustrate the interest and importance of these two processes for the Higgs sector and within the Composite Higgs context, this work presents a phenomenological analysis at the parton level, with two specific realizations of the MCHM, in the framework of the present and future hadron colliders.

The goal is to establish phenomenological differences between a model with minimal embedding of fermions (the MCHM$_{5}$) and the simplest one that can allow for an increased top Yukawa coupling (the MCHM$_{14}$). We also provide sets of representative points with a large coverage of the parameter space of those models. Beyond these two specific MCHM realizations used as a showcase to scan and thus explore experimentally the MCHM, we link these two cases to the effective field theory (EFT)appli cable in the limit of decoupled resonances. To do so and complete this study, we also present our results in terms of modifications to the SM couplings which are more directly comparable to experiment.

Therefore this overall study can be used to guide searches for new physics, bridging the BSM phenomenological and experimental goals.
Indeed, starting from the current results being achieved at the LHC (and promptly evolving), this study provides some defined directions and predictions to look for, over the next decades, at the 14~TeV High Luminosity phase of the LHC and at a future High energy Hadron collider in the range of 100~TeV to possibly 150~TeV, as in the proposed FCC-hh at CERN and SppC in China~\cite{Benedikt:2018csr,CEPC-SPPCStudyGroup:2015csa,JGao:2020}.

The features of the composite Higgs scenarios directly relevant for this work are reviewed in Section 2. Section 3 sets the overall phenomenological analysis strategy. It includes the definition of the parameter space for the two MCHM scales considered in this work, the simple physical observables relevant here, the strategy and tools to extract the representative points in the parameter space, the implementation of both models into the event generator and the operator analysis with its effects on the $\ttH$ and $\ttHH$ processes. The MCHM low scale scenario is described in Section 4 with the corresponding sets of representatives points in both models, their detailed experimental outcomes in terms of modification of the cross-sections, of the non-resonant components and new resonances. Similar outcomes are shown in Section 5 for the MCHM at high scale that will be only reachable with the future high energy hadron colliders. Section 6 describes the EFT perspectives, stressing for instance the correlation between some selected EFT parameters in the MCHM$_5$ and the  MCHM$_{14}$ both at low and high scales.

Section 7 concludes by stressing how this detailed phenomenological analysis of these two MCHM scenarios, serves as an useful showcase for the exploration of the BSM world by experiments, with well-defined but still broad scope guidelines. It highlights in this way, what can be already achieved at the presently running LHC and the forthcoming HL-LHC era. It emphasizes the unique importance of a future high energy hadron collider to explore in details the Top-Higgs sector.

\section{The Higgs as a Pseudo-Nambu Goldstone Boson}
\label{MCHM}

We will present only the features of composite Higgs scenarios that
are directly relevant to our study.  In particular, we will not
describe the spin-1 sector, which is fixed by the pattern of symmetry
breaking, in our case $SO(5) \to SO(4)$.  We refer the reader to the
complete review~\cite{Panico:2015jxa} where the general construction
and results for the MCHM bosonic sector are presented.  For our
purposes, it is sufficient to know that the breaking of the
electroweak (EW) symmetry can be parametrized in unitary gauge through the orthogonal
matrix
\bea
U &=&
\left(\begin{array}{ccc}
           \mathbb{1}_{3 \times 3} & \vec{0} & \vec{0} \\
           \vec{0}^T & \cos{\frac{h_0+h}{f}} & \sin{\frac{h_0+h}{f}} \\
           \vec{0}^T & -\sin{\frac{h_0+h}{f}} & \cos{\frac{h_0+h}{f}},
\end{array}\right)~,
\label{Umatrix}
\eea
where $h$ is the Higgs boson, $h_0 = \langle h \rangle$ is its
expectation value, and $f$ is the scale at which the breaking $SO(5)
\to SO(4)$ occurs.  In Eq.~(\ref{Umatrix}), $\mathbb{1}_{3 \times 3}$
is the $3 \times 3$ identity matrix and $\vec{0}$ is a 3-dimensional
null vector.  We also recall that in this model one finds
\bea
  M_W^2 &=& \frac{1}{4} \, g^2 f^2 \sin^2{\frac{h_0}{f}}~,
  \nonumber
\eea
which leads to the identification
\bea
f \sin{\frac{h_0}{f}} &\equiv& v ~=~ 246~{\rm GeV}~.
\eea
We will often use the variable
\bea
\xi &=& \frac{v^2}{f^2} ~=~ \sin^2{\frac{h_0}{f}}
\eea
to express our results.  It characterizes the deviations from a SM
Higgs due to compositeness.  The current bound consistent with Higgs data is $f \gtrsim 800~$GeV, or $\xi\lesssim 0.1$,~\cite{Falkowski:2013dza, Carena:2014ria, Buchalla:2014eca,
Sanz:2017tco, Liu:2017dsz, Banerjee:2017wmg, deBlas:2018tjm}.

Our main focus is on the fermion sector and its interplay with the
Higgs boson.  This depends on the composite resonances associated with
the top quark.  Up to EWSB effects, these resonances fall into
representations of the unbroken $SO(4)$.  One expects to find that
sets of these $SO(4)$ representations build up representations of
$SO(5)$ with the non-degeneracy arising from the spontaneous breaking
of $SO(5)$ (characterized by the scale $f$).  For concreteness, we
will focus on two possibilities:
\begin{itemize}

\item Resonances falling into a ${\bf 5}$ of $SO(5)$, which split into
$SO(4)$ multiplets as ${\bf 5} = {\bf 4} + {\bf 1}$.

\item Resonances falling into a ${\bf 14}$ of $SO(5)$, which split
into $SO(4)$ multiplets as ${\bf 14} = {\bf 9} + {\bf 4} + {\bf 1}$.

\end{itemize}
The first case allows for the minimal number of extra fermionic
degrees of freedom, while imposing the custodial protection of the
$Z{\overline b}_L b_L$ coupling~\cite{Agashe:2006at}, which is important
when considering EW precision measurements~\cite{Carena:2006bn,
Anastasiou:2009rv, Orgogozo:2012ct, Grojean:2013qca, Pich:2013fea,
Englert:2014uua, Croon:2015wba}.  The second possibility is
non-minimal and has been considered in~\cite{Pomarol:2012qf,
Panico:2012uw, Montull:2013mla, Carena:2014ria, Carmona:2014iwa,
Kanemura:2016tan, Gavela:2016vte, Liu:2017dsz}.  It has been pointed
out that it allows for an enhancement of certain Higgs couplings
w.r.t.~to the SM~\cite{Montull:2013mla, Liu:2017dsz}, in contrast to
the minimal case (and other possibilities) which always leads to
suppressions.  One of our goals in this work is to contrast these two
scenarios from the point of view of searches for the $t{\overline t}h$ and
$t{\overline t}hh$ processes at the 14~TeV High Luminosity phase of the
LHC, and the High Energy $pp$ collider projected to run at 100~TeV, and even up to 150~TeV.

We describe the relevant features of the previous fermion embeddings
in the following sections.

\subsection{The Fermion Sector of the MCHM$_5$}
\label{MCHM5}

Considering only the top and its partners, the model is comprised by the ``elementary''
fields $q_L = (t_L, b_L)$ and $t_R$ together with a set of
``composite'' fermionic resonances.  The elementary fields have the
$SU(3)_C \times SU(2)_L \times U(1)_Y$ quantum numbers of the SM
left-handed (LH) top-bottom $SU(2)_L$ doublet and of the SM
right-handed (RH) top $SU(2)_L$ singlet, respectively.  The composite
resonances fall into $SO(5)$ representations that split into $SO(4)$
representations due to the spontaneous breaking at the scale $f$.
They contain a number of vector-like $SU(2)_L \times U(1)_Y$
representations, that depend on the fermion embedding. The smallest representation compatible with custodial symmetry is the ${\bf 5}$ of $SO(5)$, whose decomposition under $SO(4)$ is given by a fourplet, $\Psi_4$, and a singlet, $\Psi_1$.  We
denote the corresponding states by
\bea
\Psi_4 &\sim& (X_{5/3}, X_{2/3}, T, B)~,
\nonumber \\ [0.5em]
\Psi_1 &\sim& \tilde{T}~,
\label{comp5content}
\eea
 The subindex denotes the electric charge.  While not explicitly
 indicated, the $T$ and $\tilde{T}$ states have charge $2/3$ and $B$
 has charge $-1/3$. The states $(T,
 B)$ transform as a $SU(2)_L$ doublet and have hypercharge $Y =
 1/6$, i.e. they have the same SM quantum numbers as the elementary
 field $q_L$, while an exotic $SU(2)_L$ doublet with $Y = 7/6$ is composed of the $(X_{5/3}, X_{2/3})$ states.  Finally, $\tilde{T}$ has the SM quantum numbers of
 $t_R$.

The elementary sector is simply described by
\bea
\mathcal{L}_{\rm elem} &= \overline{q}_L i \slashed{D} q_L +\overline{t}_R i \slashed{D} t_R~,
\label{elem}
\eea
where $D$ stands for the $SU(3)_C \times SU(2)_L \times U(1)_Y$
covariant derivative.

We write the composite sector directly in terms of the $SO(4)$
multiplets:
\bea
\mathcal{L}^{\mathbf{5}}_{\rm comp} &=&
 \overline{\Psi}_{\mathbf{4}} i (\slashed{D}-i\slashed{e}) \Psi_{\mathbf{4}}-M_4 \overline{\Psi}_{\mathbf{4}} \Psi_{\mathbf{4}}+ \overline{\Psi}_{\mathbf{1}} i \slashed{D} \Psi_{\mathbf{1}}-M_1 \overline{\Psi}_{\mathbf{1}} \Psi_{\mathbf{1}}~.
\label{comp5}
\eea
Here, the covariant derivative contains only the gluon and hypercharge fields, that is, $D_\mu = \partial_\mu +i g_s G_\mu + i \frac{2}{3} g^\prime B_\mu$. The remaining electroweak interactions are inside the $d_\mu$ and $e_\mu$ symbols, which are defined in terms of the Maurer-Cartan form
\bea
i U^{-1} (\partial_\mu +i g_a A_\mu^a T^a) U &=& d_{\mu, \hat{a}} T^{\hat{a}} + e_{\mu, a} T^a ~\equiv d_\mu + e_\mu~~,
\eea
where $T^a$ are the generators of the unbroken $SO(4)$ and $T^{\hat{a}}$ are the broken generators. The gauge fields $A_\mu^a$ belong to the algebra of $SO(4)\cong SU(2)_L \times SU(2)_R$, in which the electroweak fields are embedded as follows: the W fields gauge $SU(2)_L$ while $B_\mu$ gauges $T_R^3$ and the remaining generators are ungauged. The hypercharge is then given by $Y=2/3+T_R^3$.
\footnote{The factor of 2/3 arises because in order to reproduce the SM fermion hypercharges
one needs to introduce an extra $U(1)_X$ factor, under which
$\Psi_{\bf 5}$ has charge $X = 2/3$.  For further details,
see~\cite{Panico:2015jxa}}  This covariant derivative allows for
the non-linear realization of the full $SO(5)$ symmetry in the kinetic
terms, even though the Lagrangian (\ref{comp5}) exhibits explicitly
only the $SO(4)$ symmetry~\cite{Coleman:1969sm, Callan:1969sn}.  However,
note that the SM covariant derivatives break the $SO(5)$ global
symmetry explicitly.  In App.~\ref{app:generators} we give the $SO(5)$
generators, including those of the gauged $SU(2)_L \times U(1)_Y$
subgroup. The $e_\mu$ symbol term contains corrections to the electroweak interactions of the resonances, due to compositeness. These are detailed in App.~\ref{app:vertices}. Apart from the ``kinetic'' terms, we include
separate mass terms for $\Psi_{\bf 4}$ and $\Psi_{\bf 1}$.  The
difference $M_4 - M_1$ arises from the spontaneous breaking of
$SO(5)$, which we do not describe here.  For our purposes, $M_4$ and
$M_1$ can be treated as independent phenomenological parameters.

As mentioned above, some of the fermionic resonances have the same SM
quantum numbers as the elementary fields, leading to the possibility
of mixing between them.  In order to write the elementary-composite
mixing terms, it is convenient to embed the elementary states using
$SO(5)$ notation as follows
\bea
Q_L^{\mathbf{5}} &=& \frac{1}{\sqrt{2}} \left[\begin{array}{c}
-i b_L\\
- b_L\\
-i t_L\\
t_L  \\
0
\end{array}
\right]~,
\hspace{1em}
T_R^{\mathbf{5}}= \left[\begin{array}{c}
0\\
0\\
0\\
0  \\
t_R
\end{array}\right]~,
\label{e:embed5}
\eea
while $\Psi_{\bf 4}$ and $\Psi_{\bf 1}$ are similarly written in
5-plet notation as (see App.~\ref{app:embeddings} for further details)
\bea
\Psi_{\mathbf{4}}= \frac{1}{\sqrt{2}} \left[\begin{array}{c}
-i B+i X_{5/3}\\
- B- X_{5/3}\\
-i T -i X_{2/3}\\
T -X_{2/3}\\
0
\end{array}
\right]~,
\hspace{1em}
\Psi_{\mathbf{1}} = \left[\begin{array}{c}
0\\
0\\
0\\
0  \\
\tilde{T}
\end{array}\right]~.
\label{e:res5}
\eea
In terms of these definitions, the mass mixing Lagrangian takes the
form
\bea
\mathcal{L}^{\mathbf{5}}_{\rm mix} &=&
f \, \overline{Q}_L^{\mathbf{5}} U \left[ y_{L4} \Psi_{\mathbf{4}} + y_{L1} \Psi_{\mathbf{1}} \right] + \mathrm{h.c.}
\nonumber \\ [0.5em]
&+& f \, \overline{T}_R^{\mathbf{5}} U \left[ y_{R4} \Psi_{\mathbf{4}} + y_{R1} \Psi_{\mathbf{1}} \right] + \mathrm{h.c.}
\label{Lmix5}
\eea
thus implementing the idea of \textit{partial
compositeness}~\cite{KAPLAN1991259}.

Finally, we include in our Lagrangian additional Higgs interactions involving the $d_\mu$ symbol, which are allowed by the symmetries at the lowest order of derivatives, and are expected to arise from integrating out heavy resonances not included in our low energy theory (see \cite{Panico:2015jxa} for further details). These are given by

\bea
\mathcal{L}^{\mathbf{5}}_{\rm int} &=&
-i \, c_L \, \overline{\Psi}_{\mathbf{4}}P_L \, \slashed{d} \, \Psi_{\mathbf{1}}-i \, c_R \, \overline{\Psi}_{\mathbf{4}}P_R \, \slashed{d} \, \Psi_{\mathbf{1}} +\mathrm{h.c.},
\label{Lint5}
\eea
where $P_{L,R} = (1 \mp \gamma_5)/2$ are the left and right projectors and $c_L$ and $c_R$ are couplings, expected to be order one.\footnote{If the strong sector respects parity, we expect $c_L = c_R$. We will take this as a simplifying assumption in the analysis of the \textbf{14} representation below.} These terms contain extra Higgs interactions\footnote{It is possible to trade these derivative Higgs couplings to new Yukawa-like terms by a field redefinition (see for instance, \cite{De_Simone_2013}), however, we do not take this approach here.}, detailed in App.~\ref{app:vertices}. We should emphasize, however, that these additional couplings do not affect the top Yukawa at tree level, as long as $c_{L\, R}$ are taken to be real. This can be seen by noting that the operator in  Eq. (\ref{Lint5}) is antisymmetric in $\Psi_{\mathbf{4}}, ~\Psi_{\mathbf{1}}$, such that terms with the same mass eigenstate fermion will cancel out between the operator and its complex conjugate. This holds in fact for similar operators built out of any fermion representation, see \cite{Montull:2013mla}, which will be relevant for the ${\bf 14}$ representation described below.

For this reason, these modifications will only be important in characterizing the extended fermionic sector of the models. We find that the shape of distributions is largely unaffected by these terms, baring the possibility of a tuned cancellation between the vertices of Eq. (\ref{Lint5}) and those of Eq. (\ref{Lmix5}), as we will see in section \ref{analysis}.

The complete Lagrangian (in the top sector) is
\bea
\mathcal{L} &=& \mathcal{L}_{\rm elem} + \mathcal{L}^{\mathbf{5}}_{\rm comp} + \mathcal{L}^{\mathbf{5}}_{\rm mix}+\mathcal{L}^{\mathbf{5}}_{\rm int}~.
\eea
The charge 2/3 mass matrix in the $\{{\overline t}_L,~{\overline T}_L,~{\overline
X}_{2/3, L},~\overline{\tilde{T}}_L\}$ vs $\{t_R,~T_R,~X_{2/3,
R},~\tilde{T}_R\}$ basis is then given by
\begin{align}
\mathcal{M}_{2/3}^{\mathbf{5}} =
\left[\begin{array}{cccc}
        0 & \frac{1}{2} y_{L4} f (1+\sqrt{1-\xi}) & \frac{1}{2} y_{L4} f (1-\sqrt{1-\xi}) & \frac{1}{\sqrt{2}} y_{L1} f \sqrt{\xi} \\
        -\frac{1}{\sqrt{2}} y_{R4} f \sqrt{\xi} & -M_4 & 0 & 0 \\
        \frac{1}{\sqrt{2}} y_{R4} f \sqrt{\xi} & 0 & -M_4 & 0 \\
        y_{R1} f \sqrt{1-\xi} & 0 & 0 & -M_1
      \end{array}\right]~.
\label{e:mass5}
\end{align}
Diagonalization of this matrix leads to the physical fermion
eigenstates, which are in general admixtures of the original
elementary and composite states.  The lightest one is identified with
the observed top quark.  Our numerical analysis follows from
this mass matrix, as described in subsequent sections.  The remaining
resonances have masses
\bea
X_{5/3}: &\hspace{1em}& M_{X_{5/3}} = M_4~,
\\ [0.5em]
B: && M_B = \sqrt{M_4^2 + y_{L4}^2 f^2}~.
\eea
%

\subsection{The Fermion Sector of the MCHM$_{14}$}
\label{MCHM14}

In the second scenario, instead of assuming that the composite states
span a ${\bf 5}$ of $SO(5)$, we assume that they span a ${\bf 14}$ of
$SO(5)$.  This multiplet decomposes under $SO(4)$ as a fourplet and a singlet, as in Eq.~(\ref{comp5content}), plus a nonet $\Psi_{\bf 9}$.  We denote the corresponding states by
\bea
\Psi_{\bf 9} &\sim& (U_{8/3}, U_{5/3}, U_{2/3}, V_{5/3}, V_{2/3}, V_{-1/3}, F_{2/3}, F_{-1/3}, F_{-4/3})~.
\eea
Under $SU(2)_L$, this nonet breaks into three triplets, $U$ with $Y = 5/3$, $V$ with $Y=2/3$ and $F$ with $Y=-1/3$.

Adding to Eq.~(\ref{comp5}) the nonet $\Psi_9$, we obtain (the precise structure of $\Psi_9$ is given
in App.~\ref{app:embeddings}):
\bea
\mathcal{L}^{\mathbf{14}}_{\rm comp} &=& \mathcal{L}^{\mathbf{5}}_{\rm comp} +
\mathrm{Tr}\left[\overline{\Psi}_{\mathbf{9}} \left(i \slashed{D}\Psi_{\mathbf{9}}-i\left[\slashed{e}, \Psi_{\mathbf{9}}\right]\right)\right]-M_9 \mathrm{Tr}\left[\overline{\Psi}_{\mathbf{9}} \Psi_{\mathbf{9}}\right]~.
\eea
To write the mixing between the elementary and composite sectors, it
is convenient to formally embed all the elementary and composite
states into ``${\bf 14}$'' representations of $SO(5)$, in analogy to
what was done for the ${\bf 5}$ case.  We denote the elementary
embeddings by $Q_L^{\bf 14}$ and $T_R^{\bf 14}$ and continue using the
notation $\Psi_{\bf 9}$, $\Psi_{\bf 4}$ and $\Psi_{\bf 1}$ for the
composite embeddings.  All of these become $5 \times 5$ traceless
symmetric matrices, whose precise form is given in
App.~\ref{app:embeddings}.  The mixing Lagrangian is then written as
\bea
\mathcal{L}^{\mathbf{14}}_{\rm mix} &=&
f \, {\rm Tr} \left[ U^\mathsf{T} \overline{Q}_L^{\mathbf{14}} U
\left( y_{L9} \Psi_{\mathbf{9}} + y_{L4} \Psi_{\mathbf{4}} + y_{L1} \Psi_{\mathbf{1}} \right) \right] + \mathrm{h.c.}
\nonumber \\ [0.5em]
&+& f \, {\rm Tr} \left[ U^\mathsf{T} \overline{T}_R^{\mathbf{14}} U
\left( y_{R9} \Psi_{\mathbf{9}} + y_{R4} \Psi_{\mathbf{4}} + y_{R1} \Psi_{\mathbf{1}} \right) \right] + \mathrm{h.c.}
\label{Lmix14}
\eea
We also include extra $d_\mu$ symbol interactions allowed by the symmetries
\bea
\mathcal{L}^{\mathbf{14}}_{\rm int} &=&{-i \, c_4 \, \overline{\Psi}_4 \, \slashed{d} \, \Psi_1}
-i \, c_9 \, \overline{\Psi}_{\mathbf{9}}^{ij} \, \slashed{d}^i \, \Psi_{\mathbf{4}}^j-i \, \frac{c_{T9}}{4\pi f} \, \overline{\Psi}_{\mathbf{9}}^{ij} \, d^i_\mu d^{j\, \mu} \, \tilde{T} +\mathrm{h.c.},
\label{Lint14}
\eea
where $c_4$, $c_9$ and $c_{T9}$ are order one couplings and $i,~j$ are $SO(4)$ indices. Here, for simplicity we take the strong sector to be parity symmetric. We also expect the two derivatives term with $c_{T9}$ to be subdominant in most channels, since it is suppressed by an extra power of the cutoff $\Lambda \lesssim 4\pi f$. The explicit form of these vertices as well as those arising from the $e_\mu$ symbol in the kinetic term are reported in App.~\ref{app:vertices}.

The complete Lagrangian (in the top sector) is
\bea
\mathcal{L} &=& \mathcal{L}_{\rm elem} + \mathcal{L}^{\mathbf{14}}_{\rm comp} + \mathcal{L}^{\mathbf{14}}_{\rm mix} + \mathcal{L}^{\mathbf{14}}_{\rm int}~,
\eea
which leads to the charge 2/3 mass matrix in the $\{{\overline t}_L,~{\overline
T}_L,~{\overline X}_{2/3, L},~\overline{\tilde{T}}_L,~{\overline U}_{2/3,
L},~{\overline V}_{2/3, L},~{\overline F}_{2/3, L}\}$ vs $\{t_R,~T_R,~X_{2/3,
R},~\tilde{T}_R,~U_{2/3, R},~V_{2/3, R},~F_{2/3, R}\}$ basis:
{ \small
\begin{align}
&\mathcal{M}_{2/3}^{\mathbf{14}} =
&\left[\begin{array}{ccccccc}
        0 & \frac{1}{2} y_{L4} f a_+ & -\frac{1}{2} y_{L4} f a_- & -\frac{\sqrt{5}}{4} y_{L1} f s_{2h} & -\frac{1}{2} y_{L9} f b_- &-\frac{1}{2} y_{L9} f s_{2h} &\frac{1}{4} y_{L9}f b_+ \\
        \frac{\sqrt{5}}{4} y_{R4} f s_{2h} & -M_4 & 0 & 0&0&0& 0\\
        -\frac{\sqrt{5}}{4} y_{R4} f s_{2h} &0& -M_4  & 0&0&0& 0\\
        y_{R1}f\left(1-\frac{5}{4} s^2_h \right) & 0 & 0 & -M_1&0&0&0 \\
        \frac{\sqrt{5}}{4} y_{R9} f s^2_h & 0 & 0 &0 & -M_9&0&0 \\
        -\frac{\sqrt{5}}{4} y_{R9} f s^2_h &0 &0 &0 &0&-M_9&0 \\
        \frac{\sqrt{5}}{4} y_{R9} f s^2_h & 0 & 0 &0 & 0&0&-M_9
      \end{array}\right]~,
\label{e:mass14}
\end{align}
}
where we defined
\bea
s^2_{h} = \xi~,
\hspace{1em}
s_{2h} = 2\sqrt{\xi}\sqrt{1-\xi}~,
\hspace{1em}
a_\pm = 1 \pm \sqrt{1-\xi} - 2\xi~,
\hspace{1em}
b_\pm = \sqrt{\xi} \left(1 \pm \sqrt{1-\xi} \right)~.
\nonumber
\eea
The charge $-1/3$ mass matrix in the $\{{\overline b}_L,~{\overline B}_L,~{\overline
V}_{-1/3, L},~{\overline F}_{-1/3, L}\}$ vs $\{B_R,~V_{-1/3, R},~F_{-1/3,
R}\}$ basis takes the form
\bea
\mathcal{M}_{-1/3}^{\mathbf{14}} &=&
\left[\begin{array}{ccc}
       y_{L4} f \sqrt{1-\xi} &  - \frac{1}{\sqrt{2}} y_{L9} f \sqrt{\xi} &  \frac{1}{\sqrt{2}} y_{L9} f \sqrt{\xi} \\
       -M_4 & 0 & 0 \\
       0 & -M_9 & 0 \\
       0 & 0 & -M_9
\end{array}\right]~.
\label{e:mass14B}
\eea
The remaining states have masses
\bea
X_{5/3}: &\hspace{1em}& M_{X_{5/3}} = M_4~,
\\ [0.5em]
U_{8/3}, U_{5/3}, V_{5/3}, F_{-4/3}: && M_{U_{8/3}} = M_{U_{5/3}} = M_{V_{5/3}} = M_{F_{-4/3}} = M_9~.
\eea
As in the ${\bf 5}$-plet case, the previous mass matrices form the
fundamental input to our phenomenological analysis.

\subsection{Partial Compositeness and Higgs Couplings}
\label{PartialCompositeness}

The above models incorporate the partial compositeness paradigm
of~\cite{KAPLAN1991259}, via linear mixing of the elementary fields
$q_L$ and $t_R$ with composite operators transforming as singlets,
4-plets or nonets of the $SO(4)$ symmetry as described by Eqs.~(\ref{Lmix5}) and (\ref{Lmix14}).  In addition to giving rise
to the top mass, the same operators are responsible for the top-Higgs
Yukawa coupling, which is of central importance to this work.\footnote{As said previously, Eqs.~(\ref{Lint5}) and (\ref{Lint14}) do not affect the top Yukawa at tree level.}

\begin{figure}[h]
\centering
\includegraphics[width=0.3\textwidth]{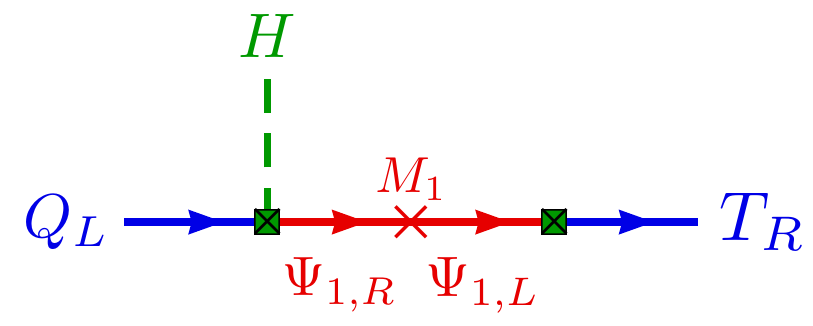}
\hspace{5mm}
\includegraphics[width=0.3\textwidth]{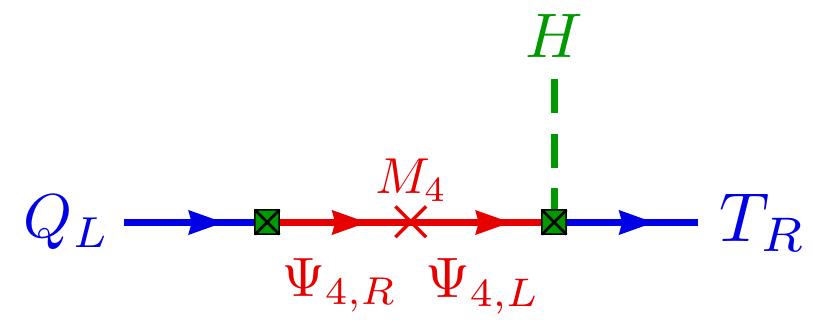}
\hspace{5mm}
\includegraphics[width=0.3\textwidth]{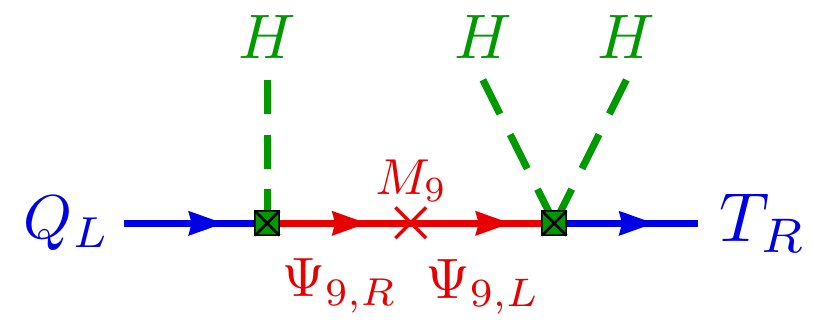}
\caption{Top-Higgs Yukawa coupling through mixing with singlet, 4-plet
and nonet resonances (the mixing is represented by the green squares).
The first two cases lead to a SM-like coupling to the Higgs (for $H
\ll f$), while the nonet exchange leads to a cubic, non-renormalizable
coupling $\overline{q}_L \tilde{H} t_R H^\dagger H$, at leading order in
$H/f$.}
\label{fig:partialcomp}
\end{figure}

The mechanism is illustrated diagrammatically in
Fig.~\ref{fig:partialcomp}, where $H$ is the Higgs doublet [see
Eq.~(\ref{HDoublet})].  Each green box represents an insertion of the
corresponding operator in Eqs.~(\ref{Lmix5}) or (\ref{Lmix14}), to
leading order in $H/f$.  For example, the mixing of the singlet
$\Psi_1$ with $T_R$ can happen at 0-th order in $H$, while the
$\Psi_1$-$Q_L$ mixing requires an insertion of the Higgs field, which
transforms as a ${\bf 4}$ of $SO(4)$: ${\bf 4}_{Q_L} \otimes {\bf 4}_H
\supset {\bf 1}$.  Similarly, the mixing of the 4-plet $\Psi_4$ with
$Q_L$ can happen at 0-th order in $H$, but requires an $H$-insertion
for the mixing with $T_R$: ${\bf 4}_{\Psi_4} \otimes {\bf 4}_H \supset
{\bf 1}$.  Both cases lead to a linear, SM-like coupling ${\overline q}_L
\tilde{H} t_R$, plus corrections non-linear in $H$.

The mixing with a nonet resonance is qualitatively different,
requiring one $H$ insertion for the $Q_L$-$\Psi_9$ mixing and
\textit{two} insertions for the $T_R$-$\Psi_9$ mixing: ${\bf 4}_{Q_L}
\otimes {\bf 4}_H \otimes {\bf 9}_{\Psi_9} \supset {\bf 1}$ and ${\bf
4}_{H} \otimes {\bf 4}_H \otimes {\bf 9}_{\Psi_9} \supset {\bf 1}$,
respectively.  As a result, the leading order coupling thus induced is
the non-SM like, non-renormalizable operator ${\overline q}_L \tilde{H} t_R
H^\dagger H$.

The top mass is obtained by replacing $H$ by its vev, leading to a factor of three in the ratio of the top mass to the top Yukawa (associated to the ${\overline t} t h$
operator) when it is induced by
the ``cubic'' interaction than for the ``linear'' interaction.  The
presence of the various channels simultaneously can then lead to an
enhancement of the top Yukawa coupling w.r.t.~the SM. One should
also notice that when $M_1 = M_4$, the linear
coupling cancels out,\footnote{This is the $SO(5)$ symmetric limit of
the MCHM$_5$.  See further comments after Eq.~(\ref{detmt}).} and the
leading order is cubic~\cite{Liu:2017dsz}.

Although we will not use it in our numerical analysis, useful approximate expressions
for the top mass in the MCHM$_{5}$ and in the MCHM$_{14}$ are given by
\bea
&m_t^{\bf 5} &\approx \frac{1}{\sqrt{2}} \sqrt{\xi} \sqrt{1-\xi} \, \frac{y_L y_R f^2}{\sqrt{Z_L Z_R} |M_1|} \left|1 - r_1\right| \nonumber \\
&m_t^{\bf 14} &\approx \frac{\sqrt{5}}{2} \sqrt{\xi} \sqrt{1-\xi} \, \frac{y_L y_R f^2}{\sqrt{Z_L Z_R} |M_1|} \left|1 - r_1 - \frac{3 r_1 + 5 r_9 - 8 r_1 r_9}{4 r_9} \, \xi \right|~,
\label{mtapprox}
\eea
where we defined the mass ratios $r_1 = M_1 / M_4$ and $r_9 = M_9 / M_4$
and took $y_{Li} \equiv y_L$ and $y_{Ri} \equiv y_R$ for $i = 1,4,9$ (see
Section~\ref{parameterspace}). The ``wavefunction renormalization'' factors
have the form $Z_{L,R} = 1 + y_{L,R}^2 f^2 / M^2_{4,1} + {\cal O}(\xi)$. The
expression (\ref{mtapprox}) displays explicitly the behavior described above.

The same effect has an impact on the coupling of the Higgs to two
gluons ($ggh$), normalized to the SM top contribution, which is given
by:
\bea
c_g &=& \frac{v}{2} \, \frac{d}{dh_0} \log {\rm det}\left( \mathcal{M}^\dagger \mathcal{M} \right)~,
\eea
where $\mathcal{M}$ is the fermion mass matrix, assuming all states
are much heavier than the Higgs boson (for our purposes, light state
contributions can be neglected).

For the MCHM$_5$, using $\mathcal{M} = \mathcal{M}_{2/3}^{\mathbf{5}}$
in Eq.~(\ref{e:mass5}), this gives
\bea
c_g^{\bf 5} &=& \frac{1-2\xi}{\sqrt{1-\xi}}~.
\label{cg5}
\eea
For the MCHM$_{14}$, one gets
\bea
c_g^{\bf 14} &=& c_g^t + c_g^b~,
\label{cg14}
\eea
where
\bea
c_g^t &=&
\frac{
4 \left(1 - r_1\right) r_9
- \left(9 r_1 + 23 r_9 - 32 r_1 r_9\right) \xi
+ 4 \left(3 r_1 + 5 r_9 - 8 r_1 r_9\right) \xi ^2
}
{\sqrt{1-\xi }
\left[4 \left(1 - r_1 \right) r_9 - \left(3 r_1 + 5 r_9 - 8 r_1 r_9\right) \xi \right]}~,
\label{cgt14}
\eea
arises from the charge 2/3 sector, $\mathcal{M}_{2/3}^{\mathbf{14}}$
in Eq.~(\ref{e:mass14}), and
\bea
c_g^b &=&
\xi \sqrt{1-\xi} \, \frac{y_L^2 f^2 \left(1 - r_9^2 \right)}
{r_9^2 \left(M_4^2 + y_L^2 f^2 \right) + y_L^2 f^2 \left(1 - r_9^2\right) \xi}~,
\label{cgb14}
\eea
arises from the charge $-1/3$ sector,
$\mathcal{M}_{-1/3}^{\mathbf{14}}$ in Eq.~(\ref{e:mass14B}).  For the
latter, we explicitly removed the zero-mode (the physical bottom
quark) and neglected its contribution to the $ggh$ coupling.

When $M_1, M_4 \gg M_9$ (or $M_1 = M_4$), one finds
\bea
c_g^t &=& \frac{3-4 \xi }{\sqrt{1-\xi }}
~\approx~
3 - \frac{5}{2} \, \xi~,
\eea
while
\bea
c_g^t &=& 1 - \frac{3 r_1 + 8 r_9 - 11 r_1 r_9 }{2 \left(1 - r_1\right) r_9} \, \xi + {\cal O}(\xi^2)
\eea
in all other cases, reflecting the underlying cubic versus linear
coupling of the top quark to the Higgs field.  In all cases, $c_g^b
\sim \xi \ll 1$.  Global constraints on the gluon fusion process from
Higgs measurements, which allow for about 20\% deviations from unity
in $c_g$ at the 95\% C.L.~\cite{Khachatryan:2016vau}, will then also
impose constraints on the allowed deviations in the top Yukawa
coupling from the SM limit.

\subsection{Higgs Decays}
\label{HiggsDecays}

While the amplitudes of the processes $t{\overline t} h$ and $t{\overline t}
hh$ are determined by the Lagrangians written in the previous
sections, it is necessary to specify how the light families of the SM
are treated in the context of the composite Higgs models in order to
take into account the possible modifications in Higgs decays.  Such
deviations are expected to be small, since the observed $125$~GeV
resonance is known to exhibit SM-like properties.  The dominant Higgs
decay channels are then as in the SM: $h \to b{\overline b}, W^+ W^-, gg,
\tau^+ \tau^-, c{\overline c}, ZZ$.  We neglect the decays into
$\gamma\gamma$, $\gamma Z$ and $\mu^+\mu^-$, which have branching
fractions ranging from 0.23\%, 0.15\% down to 0.01\%, as well as even
rarer decay channels.

Given the importance of the $b{\overline b}$ channel and the fact that
$b_L$ is embedded into $SO(5)$ together with $t_L$, we must also
specify how $b_R$ fits into the models.  There are several ways to
proceed.  Rather than exploring the various possibilities, we choose
to supplement the RH bottom with composite resonances that span a
${\bf 10}$ of $SO(5)$ for both the MCHM$_5$ and the MCHM$_{14}$.  Such
models, called MCHM$_{5,5,10}$ and MCHM$_{14,14,10}$, were introduced
in~\cite{Carena:2014ria}.  For the remaining fermions, we choose to
replicate the scheme employed for the third family.  We also assume
that the lepton sector follows the same scheme as the quark sector.
Furthermore, we assume that the mixing angles between the elementary
and composite states associated with the light families are small, as
in ``anarchy'' models of flavor (see e.g. \cite{Agashe_2005,Gherghetta_2000}).

Under the previous assumptions, it is found that the couplings of the
composite Higgs to the vector bosons and to light $f{\overline f}$ pairs
are controlled by two model-dependent functions that depend only on
$\xi$.  In the MCHM$_5$ the coupling of the Higgs to a pair of gluons
depends also only on $\xi$, but in the MCHM$_{14}$ it depends on
additional microscopic parameters, as shown in
Subsection~\ref{PartialCompositeness}.

The partial widths are then simply obtained by rescaling the SM ones. For the MCHM$_5$, one
finds~\cite{Carena:2014ria}
\bea
\Gamma(h \to b {\overline b}) &=& F_2(\xi)^2 \, \Gamma_{\rm SM}(h \to b {\overline b})~,
\nonumber \\ [0.5em]
\Gamma(h \to c {\overline c}) &=& F_1(\xi)^2 \, \Gamma_{\rm SM}(h \to c {\overline c})~,
\nonumber \\ [0.5em]
\Gamma(h \to \tau^+ \tau^-) &=& F_2(\xi)^2 \, \Gamma_{\rm SM}(h \to \tau^+ \tau^-)~,
\\ [0.5em]
\Gamma(h \to VV) &=& F_2(\xi)^2 \, \Gamma_{\rm SM}(h \to VV)~,
\nonumber \\ [0.5em]
\Gamma(h \to gg) &=& F_1(\xi)^2 \, \Gamma_{\rm SM}(h \to gg)~,
\nonumber
\eea
where
\bea
F_1(\xi) &=& \frac{1-2\xi}{\sqrt{1-\xi}}~,
\hspace{1cm}
F_2(\xi) ~=~ \sqrt{1-\xi}~.
\label{Ffunctions}
\eea
For the MCHM$_{14}$, the bottom channel is controlled by $F_1$ instead
of $F_2$, and the $ggh$ coupling is controlled by $c_g^{\bf 14}$ of
Eq.~(\ref{cg14}) instead of $F_1(\xi)$.

The total Higgs width in the MCHM models under consideration can then
be written as
\bea
\Gamma_5(h) &=& \left\{ F_2(\xi)^2 \left[ {\rm BR_{\rm SM}(b{\overline b})} + {\rm BR_{\rm SM}(VV)} + {\rm BR_{\rm SM}(\tau^+ \tau^-)} \right] \right.
\nonumber \\
&& \mbox{} + \left. F_1(\xi)^2 \left[ {\rm BR_{\rm SM}(gg)} + {\rm BR_{\rm SM}(c{\overline c})} \right] \right\} \Gamma_{\rm SM}(h)~,
\\ [0.5em]
\Gamma_{14}(h) &=& \left\{ F_2(\xi)^2 \left[{\rm BR_{\rm SM}(VV)} + {\rm BR_{\rm SM}(\tau^+ \tau^-)} \right] \right.
\nonumber \\
&& \mbox{} + \left. F_1(\xi)^2 \left[ {\rm BR_{\rm SM}(b{\overline b})} + {\rm BR_{\rm SM}(c{\overline c})} \right]
+ (c_g^{\bf 14})^2 \, {\rm BR_{\rm SM}(gg)} \right\} \Gamma_{\rm SM}(h)~,
\eea
and the branching fractions can also be expressed in terms of the
functions $F_1$, $F_2$, $c_g^{\bf 14}$, and SM quantities.  These
branching fractions are all that is needed to take into account the
effects of compositeness in Higgs decays.

\section{ Phenomenological Analysis Strategy}
\label{analysis}

\subsection{Parameter Space}
\label{parameterspace}

The top sector of the models described in Sections~\ref{MCHM5} and
\ref{MCHM14} is controlled by the (vector-like) mass parameters,
$M_i$, and by several dimensionless couplings, $y_{Li}$,  $y_{Ri}$, as well as $c_L$ and $c_R$ for the $\text{\bf 5}$ and $c_4$, $c_9$ and $c_{T9}$ for the $\text{\bf 14}$.
These parameters are a priori complex.  However, not all phases are
physical.  In order to identify the number of physical phases we can
proceed as follows.  We can start by absorbing the phase of each $M_i$
(thus making it real and positive) by redefining the phases of
$\Psi_{i,L}$ or $\Psi_{i,R}$.  This leaves one free phase in each
such pair, say in $\Psi_{i,R}$, that can be adjusted to absorb the
phase of the corresponding $y_{Li}$ (thus making all of the $y_{Li}$
real and positive).  Finally, we can absorb the phase of one of the
$y_{Ri}$ into $T_R^5$ or $T_R^{14}$.  We conclude that there  are three
(five) physical phase(s) in the MCHM$_5$ (MCHM$_{14}$).  Alternatively,
we can choose all the $y_{Li}$ and $y_{Ri}$ to be real and positive,
putting the physical phases in $M_1,~c_{L,R}$ for the MCHM$_5$, and in $M_1,~M_4,~c_4,~c_9,~c_{T9}$ for the MCHM$_{14}$.  In this work, for simplicity, we will
assume that all parameters are real, which amounts to imposing CP conservation in the strong sector\footnote{It is also worth noting that not imposing CP conservation leads to severe constraints. These (as well as the flavor structure of the models) are beyond the scope of our analysis, and we refer the reader to \cite{Redi_2011} for an example of a composite Higgs model addressing these issues.}.  This leaves three physical signs in the case of the MCHM$_5$ and five signs in the MCHM$_{14}$. We will choose these signs to be sign$(M_1)$, sign$(c_L)$ and sign$(c_R)$  in the first case and sign$(M_1)$, sign$(M_4)$, sign$(c_4)$, sign$(c_9)$ and sign$(c_{T9})$ in the second.
Finally, in order to simplify our analysis, we will disregard the derivative couplings in eqs.~\ref{Lint5} and \ref{Lint14} until the end of Sec.~\ref{analysis}, which leaves us with $M_i$, $y_{Li}$ and  $y_{Ri}$. The effect of the neglected operators will be considered separately in Sec. \ref{newops}.

While the above parameters respect the $SO(4)$ symmetry, in general
they violate the $SO(5)$ symmetry.  The $SO(5)$ symmetric limit
corresponds to $M_1 = M_4$ and $y_{L1} = y_{L4}$, $y_{R1} = y_{R4}$
for the MCHM$_5$, and $M_1 = M_4 = M_9$ and $y_{L1} = y_{L4} =
y_{L9}$, $y_{R1} = y_{R4} = y_{R9}$ for the MCHM$_{14}$.  It turns out
that deviations from the $SO(5)$ symmetric limit in the dimensionless
couplings corresponds to a ``hard'' breaking of the symmetry, in the
sense that the Higgs effective potential is finite when the $SO(5)$
symmetry relations are satisfied, but becomes UV sensitive when not\footnote{This can be seen from the trace of $\mathcal{M}_{2/3}^\dagger \mathcal{M}_{2/3}$, which becomes independent of the Higgs in this limit, signifying the cancellation of quadratic divergences to the potential.}.
Deviations from the $SO(5)$ symmetric limit in the $M_i$, on the
contrary, correspond to a ``soft'' breaking, and the Higgs potential
remains IR dominated in that case.  This motivates us to focus on the
case where
\bea
y_{Li} \equiv y_L~,
\hspace{1cm}
y_{Ri} \equiv y_R~,
\eea
for all $i$, as a way to reduce the number of independent parameters.  Small deviations from this limit mean
that one can expect additional UV dependent contributions to the Higgs
potential.  Such contributions can affect the region of parameter
space that leads to EWSB and to a Higgs mass $m_h = 125$~GeV. Thus, we
take the point of view that imposing that the Higgs mass be reproduced
by strictly adhering to the case of a calculable Higgs potential is
overly restrictive in the context of our collider study.  For this reason  we will simply fix the mass of the Higgs, and leave
the study of points that reproduce the Higgs mass in the strictly
calculable limit to future work.

In conclusion, we are left with the following set of parameters:
\begin{itemize}

\item MCHM$_5$: $f$, $|M_1|$, $|M_4|$, ${\rm sign}(M_1)$, $y_L$ and $y_R$.

\item MCHM$_{14}$: $f$, $|M_1|$, $|M_4|$, $|M_9|$, ${\rm sign}(M_1)$, ${\rm sign}(M_4)$, $y_L$ and $y_R$.

\end{itemize}
One of these parameters can be further fixed by requiring that the top
mass be reproduced.  We choose to fix $y_R$ in this way.  Our
procedure is to require that
\bea
{\rm Det} (\mathcal{M}_{2/3} \mathcal{M}^T_{2/3} - \overline{m}^2_t  \mathbb{1}) = 0~,
\label{detmt}
\eea
which we solve numerically for $y_R$ for each choice of the parameters
other than $y_R$.  When there are no real solutions to this equation,
we discard the parameter point.  For example, in the $SO(5)$ symmetric
limit ($M_1 = M_4$ for the MCHM$_5$ and $M_1 = M_4 = M_9$ for the
MCHM$_{14}$, and equal couplings as we are assuming), one can easily
check that the mass matrices given in Subsections~\ref{MCHM5} and
\ref{MCHM14} have vanishing determinant: the top mass is only induced
in the presence of $SO(5) \to SO(4)$ breaking in the composite sector.
Thus, for parameters close to this symmetry enhanced point it can
become difficult to accommodate the observed top mass.

For $\overline{m}_t$ in Eq.~(\ref{detmt}), we take $\overline{m}_t = 150$~GeV, the running top mass at
the scale of the resonances, which will be typically around $2-3$~TeV \footnote{Later, in section \ref{MCHMHighScale}, we will consider resonances with masses up to tens of TeV, such that strictly speaking, a different running top mass should be picked depending on the energy reached in each parameter space point. However, we find that the effect of the choice of top mass at this scale is negligible and would be masked in comparison with the spread in physical parameters obtained from the numerical scan, so we simply fix $\overline{m}_t = 150$~GeV.}.
On the other hand, in the $t\overline{t}h$ and $t\overline{t}hh$ production
the relevant
scales are of the order of a couple hundred GeV. We therefore
distinguish between the high-scale running top mass (relevant for the
diagonalization of the mass matrix), and a low scale running top mass,
relevant to the physical processes of interest.  We take for the
latter the pole top mass of $m_t = 173$~GeV, which also enters in
kinematical quantities.  To first approximation, this takes into
account the running between the two scales.

Our strategy to extract the physical quantities is straightforward:
given values for the $M_i$ and for $y_L$, we find $y_R$ from
Eq.~(\ref{detmt}).  We then diagonalize numerically the fermion mass
matrix to obtain the spectrum, and the unitary transformations $U_L$
and $U_R$ such that
\bea
U_L \mathcal{M}_{2/3} U_R^\dagger = {\rm diag}(m_t, M_{T^{(1)}}, , M_{T^{(2)}}, \ldots)~,
\eea
with all the physical masses real and positive.  This is done with
Mathematica~\cite{mathematica}.  We also treat the charge $-1/3$
sector in the MCHM$_{14}$ numerically.\footnote{The bottom mass in
Eq.~(\ref{e:mass14B}) vanishes.  Although one can easily incorporate a
finite bottom mass, its effect in the diagonalization is negligible.
The correct couplings between the Higgs boson and the bottom quark are
taken into account as described in Section~\ref{HiggsDecays}.} The
physical spectrum and the rotation matrices are the main input to the
rest of the numerical analysis.

For example, we can obtain other quantities of interest, such as the
Yukawa matrix in the mass eigenbasis
\bea
Y^{\rm mass}_{2/3} = U_L Y^{\rm gauge}_{2/3} U_R^\dagger~,
\eea
where
\bea
Y^{\rm gauge}_{2/3} &=& \frac{d}{d h_0} \mathcal{M}_{2/3}
\eea
is the Yukawa matrix in the gauge eigenbasis.  The most relevant
quantity will be the $(1,1)$ entry in $Y^{\rm mass}_{2/3}$, which
corresponds to the top Yukawa coupling.  It includes exactly all
tree-level effects arising from the Higgs compositeness (the
dependence through $\sqrt{\xi} =\sin{\frac{h_0}{f}}$), as well as the
mixing with the vector-like resonances.

The non-linear dependence on $h$ in composite Higgs models leads to
interactions between top pairs and a number of Higgs bosons.  The
$tthh$ vertex, whose Feynman rule is given by $i$ times the
$(1,1)$ entry of $\frac{1}{2}d^2 \mathcal{M}_{2/3}/dh_0^2$, after rotating to the
mass eigenbasis, also enters in our analysis.

\subsection{MCHM Scales, low versus high}
\label{sec.scans}

The parameter space of \five and \fourt is explored here in two steps. The first step considers the parameter space relevant for the reach of the LHC machine. This step includes two running operation stages of the LHC, i.e from now until 2023-2024 with about 400 $fb^{-1}$ total integrated luminosity at 13 and 14 TeV, and from 2026 to about 2038, with 10 times more luminosity and 14 TeV CM energy (may be slightly more) with the HL-LHC.
The region of the parameter space corresponding to the overall LHC machine operation (from now until the end of the HL-LHC) is labelled as the ``Low Scale MCHM'', as the dimensionful parameters will take values of a few TeV. This will be the focus for the remainder of this section and section~\ref{results}.

The second step of this analysis is extended to the ``High Scale MCHM''. This relates to the future hadron colliders in project, expected to run at CM energies around 100~TeV or even higher ~\cite{Benedikt:2018csr,CEPC-SPPCStudyGroup:2015csa,JGao:2020}.
For the MCHM high scale regime, the starting hypotheses are either:
\begin{enumerate}
    \item No new physics is discovered at the HL-LHC, i.e. within the possible reach in mass and/or precision of this collider, or
    \item Some evidence (3$\sigma$ effect) is found such as new high mass resonance(s) or a deviation from the SM for $\mu(t\overline{t}h)$, i.e., the top Yukawa, or
    \item A deviation on the $\mu(t\overline{t}h)$ is present at 5$\sigma$ and $\mu(\ttHH)$ is observed, but with sizeable uncertainty.
    Thus a higher energy $pp$ collider would allow higher precision measurements and looking for further effects.
\end{enumerate}
This is the subject of section~\ref{MCHMHighScale}.
But it is worth stressing already here, and when looking to the results in section~\ref{results}, that the points generated in the MCHM low scale parameter space, are generated both at 14 and at 100 TeV. This is simply because of the reasons listed above.

Here below are specified the ranges considered in each case and the reasoning behind them.

\subsubsection{Definition of the ranges for the Low Scale parameters}
\label{sec.scans.low}

For the MCHM$_5$, we consider the following ranges for the parameters:
\begin{align*}
|M_1|& \in [0.8, 3.0]~{\rm TeV}, &  M_4 &\in [1.2, 3.0]~{\rm TeV},  \\
f&\in [0.8, 2.0]~{\rm TeV},          &  y_L&\in [0.5, 3.0].
\end{align*}

For the MCHM$_{14}$, we use:
\begin{align*}
|M_1|& \in [0.8, 3.0]~{\rm TeV}, &  |M_4| &\in [1.2, 3.0]~{\rm TeV},  &  M_9 &\in [1.3, 4.0]~{\rm TeV},  \\
f&\in [0.8, 2.0]~{\rm TeV},          &  y_L&\in [0.5, 3.0].
\end{align*}
In order to remain in a perturbative regime, justifying the present tree-level analysis, we will take $y_L < 3$.
For the same reason, we also check that $y_R$, as determined by the
top mass, is below 4. The distribution of points within those ranges was not uniform, due to computing constraints, but we strive to cover most of the parameter space.

\subsubsection{Definition of the ranges for the High Scale parameters}
\label{sec.scans.high}

The range to be covered by the parameters for the High scale case is defined, for each considered MCHM scenario, by the possible reach of a high energy hadron collider order 100 TeV in CM and at least 20 $ab^{-1}$ total luminosity, taking also into account previous studies~\cite{LTWang2019}. Moreover this parameter space range must be linked continuously to the one defined for the Low Scale, which will already be mainly tackled by the HL-LHC; indeed, some showcase scenarios in the Low scale will remain of interest in the High scale as pointed out in the next two sections.

Given the above, for the MCHM$_5$ we consider:
\begin{align*}
|M_1|& \in [2, 30]~{\rm TeV}, &  M_4 &\in [2, 30]~{\rm TeV},  \\
f&\in [0.8, 8.0]~{\rm TeV},          &  y_L&\in [0.5, 3.0],
\end{align*}
and for the MCHM$_{14}$, we use:
\begin{align*}
|M_1|& \in [2, 30]~{\rm TeV}, &  |M_4| &\in [2, 30]~{\rm TeV},  &  M_9 &\in [2, 30]~{\rm TeV},  \\
f&\in [0.8, 8.0]~{\rm TeV},          &  y_L&\in [0.5, 3.0].
\end{align*}
%

\subsection{Simple Physical Observables}
\label{SimpleObservables}

%
\begin{figure}[h]
\centering
\begin{tabular}{cc}
\includegraphics[width=0.475\textwidth]{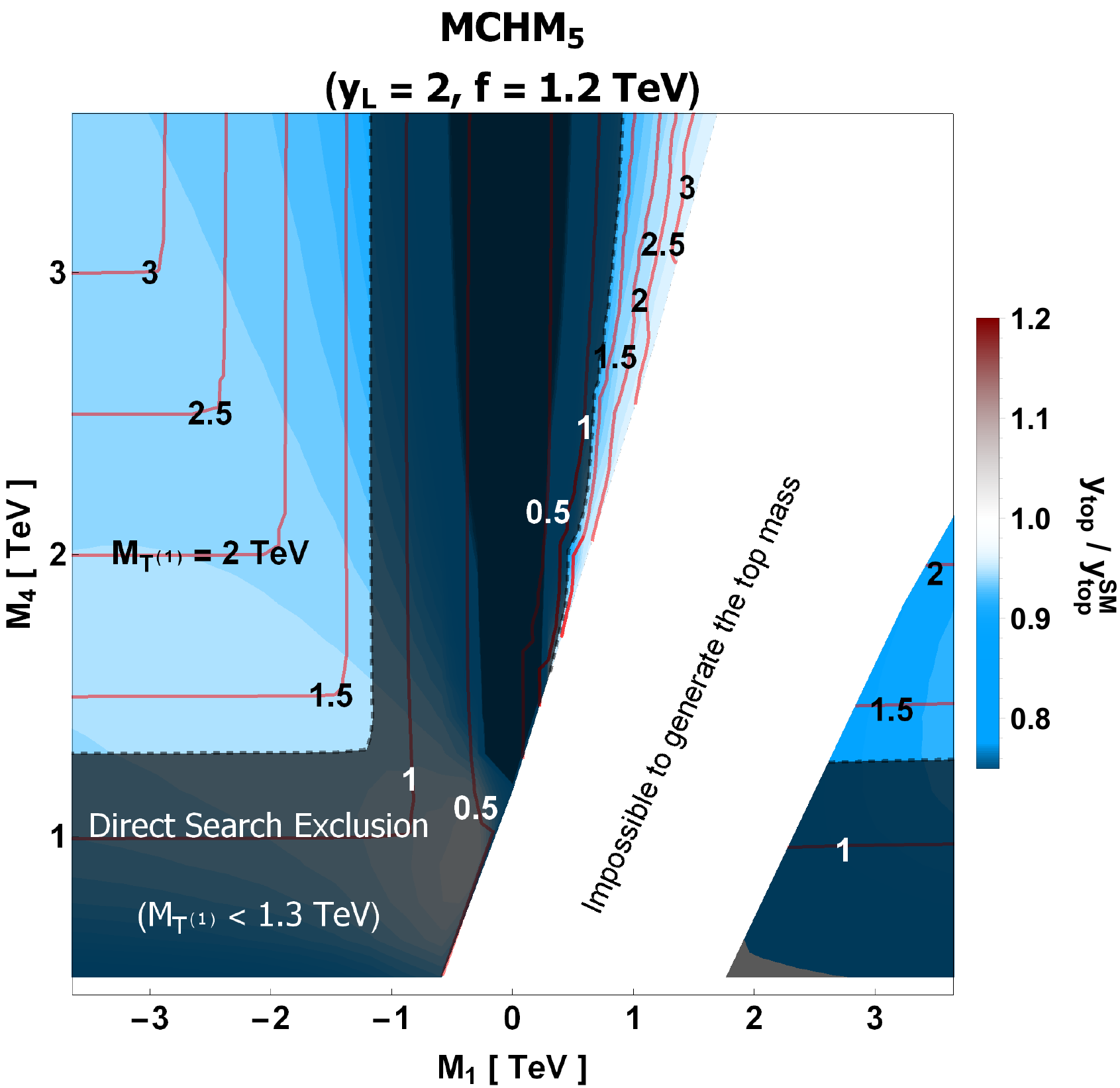} &
\includegraphics[width=0.485\textwidth]{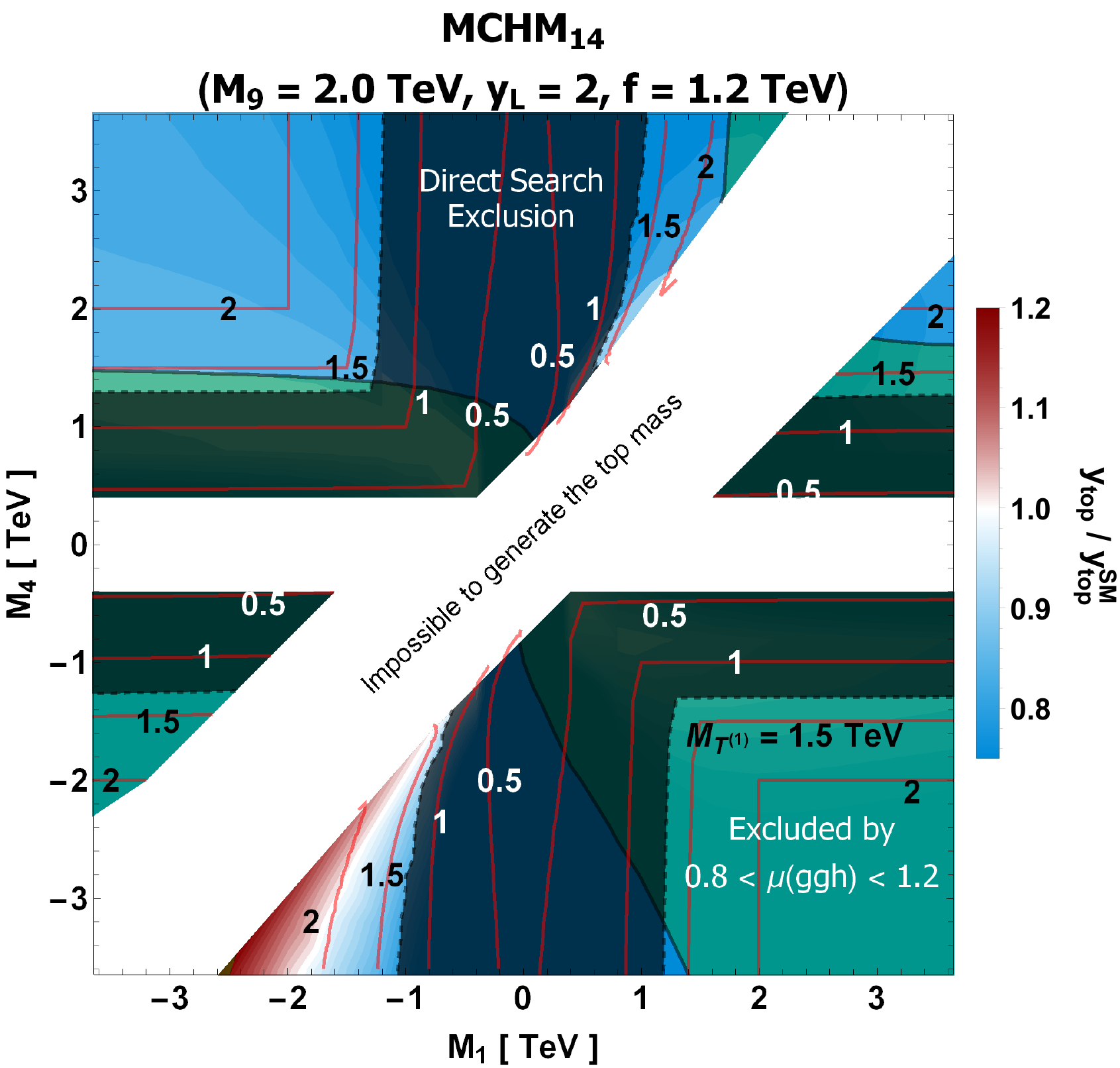} \\
\rule{0pt}{42ex}
\includegraphics[width=0.495\textwidth]{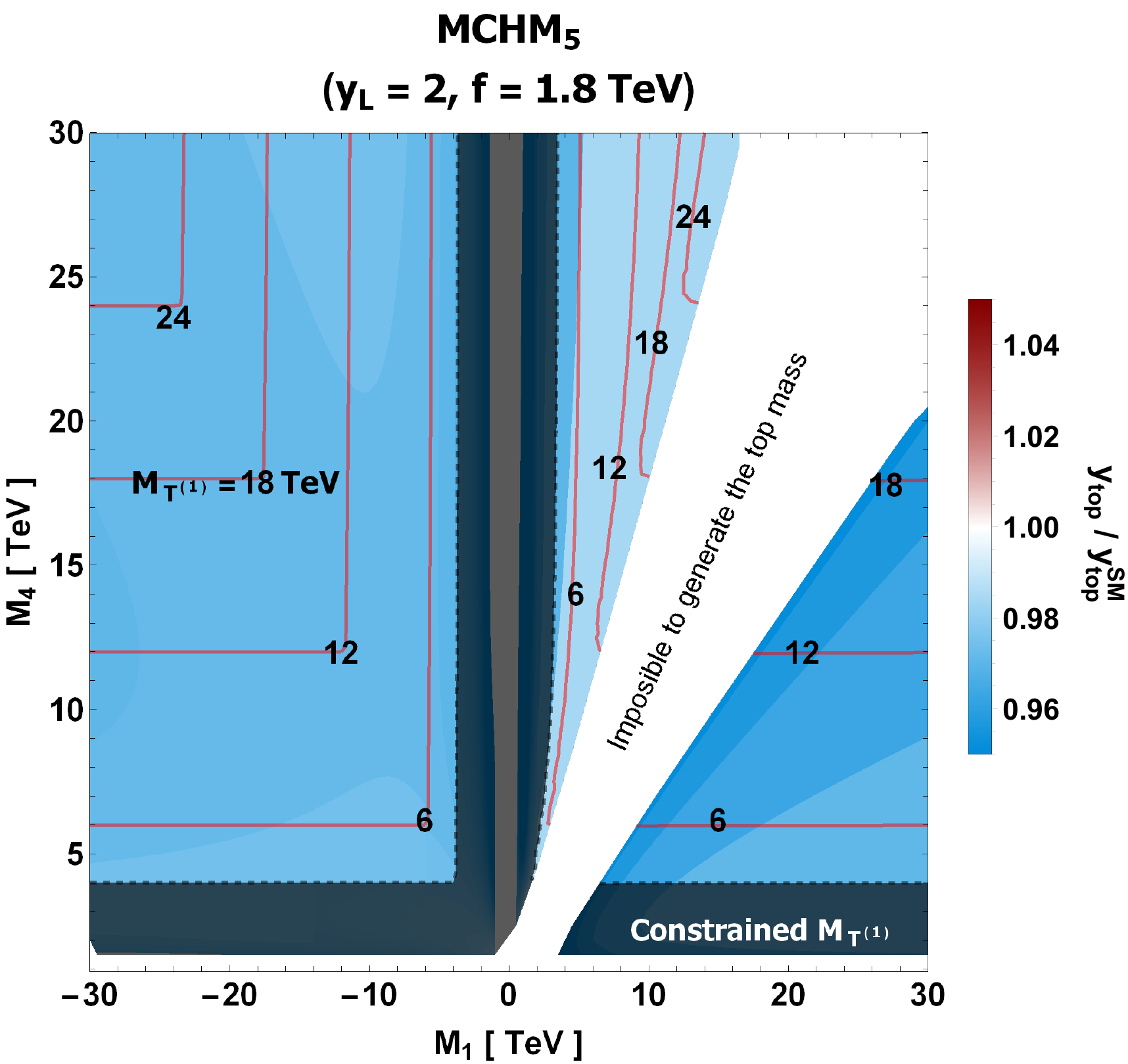} &
\includegraphics[width=0.495\textwidth]{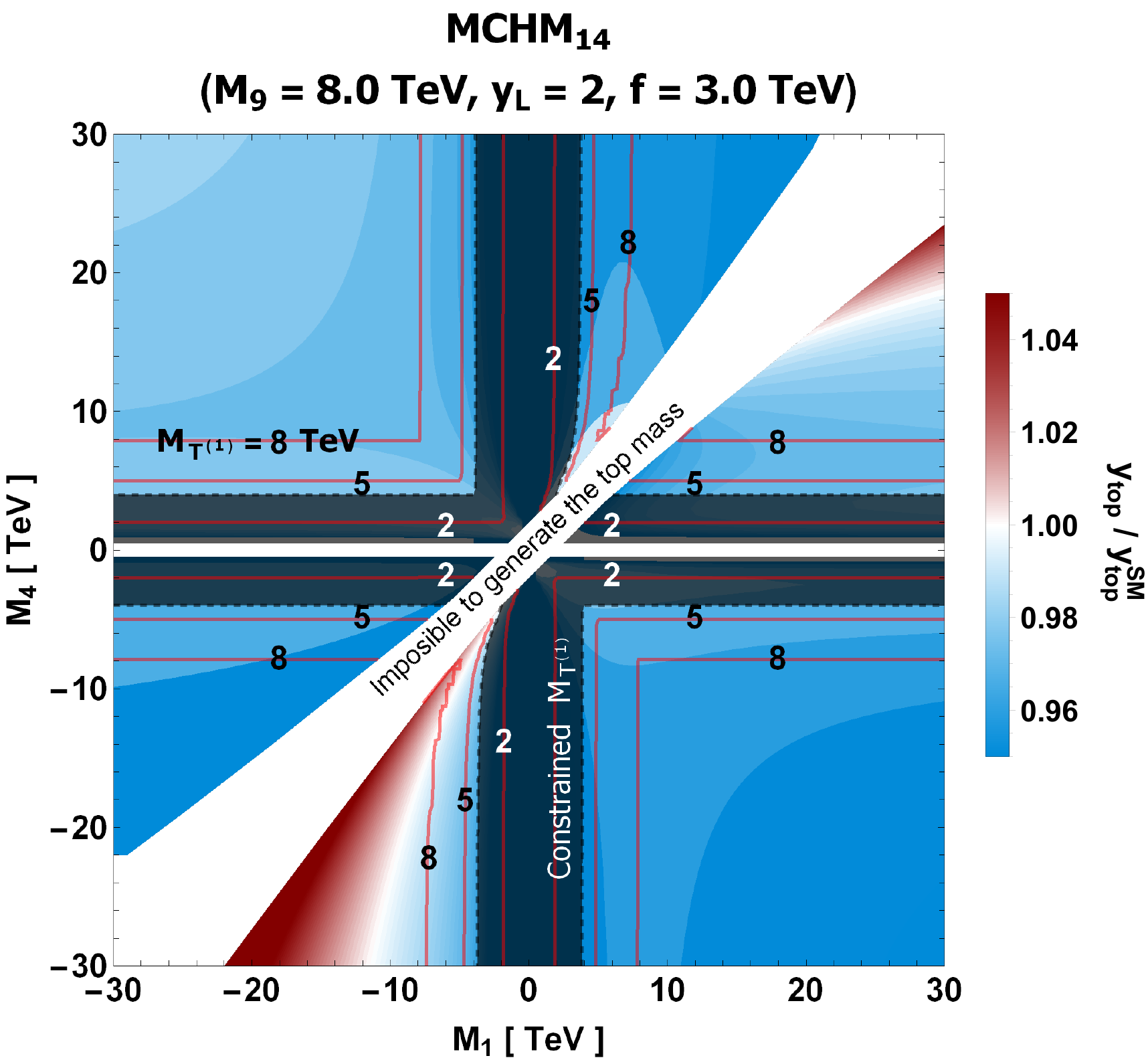}\\
\end{tabular}
\caption{
Normalized value of the Yukawa coupling of the top, $y_{\rm top} / y_{\rm top}^{\rm SM}$ in the  $M_1$-$M_4$ plane for the MCHM$_5$ (left) and MCHM$_{14}$ (right).
Top figures focus on ${\cal O}(1)$ TeV scale, bottom ones expand it to higher mass scales.
Also shown are contours of constant mass of the lightest top partner, $M_{T^{(1)}}$. Overlayed regions indicate constraints: the dark one is given by direct exclusion of top partners in the top plots, and by expected constraints in the HL-LHC in the bottom ones ($M_{T^{(1)}} < 4$ TeV); the green region is constrained by $\mu(ggh)$ measurements. In the white region, the top mass cannot be reached without violating perturbativity.
}
\label{fig:ytvsM1M4}
\end{figure}
The diagonalization of the $Q=2/3$ mass matrix leads to several
physical quantities of interest.  There is a rich spectrum of
vectorlike states.  Up to small EW symmetry breaking effects, these
vector-like masses are approximately given by:
\begin{itemize}

\item MCHM$_5$:
\bea
M_4~,
\hspace{1cm}
\sqrt{M_4^2 + y_L^2 f^2}~,
\hspace{1cm}
\sqrt{M_1^2 + y_R^2 f^2}~,
\nonumber
\eea

\item MCHM$_{14}$:
\bea
M_4~,
\hspace{1cm}
\sqrt{M_4^2 + y_L^2 f^2}~,
\hspace{1cm}
\sqrt{M_1^2 + y_R^2 f^2}~,
\hspace{1cm}
M_9~~~(\textrm{degeneracy} = 3)~.
\nonumber
\eea
\end{itemize}
We use the mass of the lightest state, which we call $T^{(1)}$, as a
proxy for the scale of the new physics.  There are important lower
bounds from direct searches for such resonances by both the
ATLAS~\cite{Aaboud:2018pii} and CMS~\cite{Sirunyan:2018omb}
collaborations.  These searches consider vectorlike top partner resonances
decaying exclusively in the $bW$, $tZ$ and $th$ channels.  These bounds
depend mildly on the decay branching fractions of the heavy state and
are roughly around $1.3$~TeV. However in the models under consideration here, one
finds sometimes additional decay channels, as we will show in section~\ref{results}, so these bounds must be taken with care.

We also consider the deviations from the SM of the top quark decay
width.  However, the previous direct bounds imply that such deviations
are well within the experimental uncertainties, and therefore do not
impose additional bounds on the models.

As mentioned earlier, a quantity of direct interest in our study is
the top Yukawa coupling.  Figure \ref{fig:ytvsM1M4} displays in the $M_1$-$M_4$ plane the top Yukawa normalized to its SM value, $y_{\rm top} / y_{\rm top}^{\rm SM}$, for MCHM$_5$ (left) and MCHM$_{14}$ (right). We consider two different slices in parameter space. For the figures at the top, we take $f = 1200$~GeV, $y_L = 2$ and $M_9 = 2$~TeV, and scan over $M_{1,~4} \leq 3.5~$TeV, while for the bottom figures, we fix $y_L=2$ and $f=1.8$ TeV for the MCHM$_5$ and  $y_L=2$, $f=3$ and $M_9=8$ TeV for the MCHM$_{14}$, and scan over the larger range $M_{1,~4} \leq 30~$TeV. The more restricted scan is relevant for the low energy survey as defined in Sec. \ref{sec.scans.low}, while the zoomed out scan refers to the high energy parameter ranges defined in Sec. \ref{sec.scans.high}. While in the MCHM$_5$  the top Yukawa is approximately determined by the $F_1(\xi)$ in
Eq.~(\ref{Ffunctions}), and is always suppressed compared to the SM, in the  MCHM$_{14}$ it displays richer behavior, with the possibility of an enhancement in certain parameter space regions, as emphasized by \cite{Liu:2017dsz}. Specifically, for the smaller range, only Region III allows an enhancement, while, for the larger scan range, the top Yukawa can also be enhanced in Region I. We also show red contours of constant mass of the lightest top partner resonance $M_{T^{(1)}}$. In the dark overlayed regions, the lightest top partner is approximately excluded by direct searches, assuming it decays only in the
$bW$, $tZ$ and $th$ channels~\cite{Aaboud:2018pii, Sirunyan:2018omb}, while the green overlay shows the region in which $\mu(ggh)$ differs from unity by 20\% or more, which is in tension with current Higgs coupling measurements \cite{Khachatryan:2016vau}. In the white area, the top mass cannot be reproduced by values of $y_R$ within our considered perturbative range.

\subsection{Strategy to select example points and benchmark points}
\label{selectedpts}

We select two classes of points to study the physics, in each of the two MCHM scenarios, for both the MCHM at low and at high scale. These are respectively labeled the ``example points'' and the ``benchmark points''.

The selected example points are chosen based on striking features or on how accessible they are in the near future (at the start of the HL-LHC) or towards the end of the HL-LHC or in the much longer term with a 100 TeV pp collider. The striking features are defined with the present results from the LHC experiments on the Top-Higgs sector or the prospect studies achieved for the HL-HE/LHC scenarios or the FCC-hh project. However they do not necessarily represent what is the typical behaviour of the parameter space we explored. They are picked according to our criteria of what is interesting and thus carry a bias. While that is very useful to see what kind of phenomenology can be produced by the model, and give insight on what is happening or could happen in the future, it is not the most comprehensive and extensive way of looking into the possibilities of the models.

On the other side, looking individually on all kinematic distributions for hundreds of points in parameter space is just unfeasible. An interesting compromise can be reached using the approach proposed in \cite{Carvalho:2015ttv}: one can use a statistical test to group points of the parameter space into ``clusters'' based on the similarity between the kinematic distributions of the final or intermediary states produced by them. This is called the {\it{``clustering strategy''}}.

Following this strategy, one can then use the same test statistic to choose one point from each cluster as a typical representative of that behaviour, a benchmark point. Analysis designed to search for those benchmark points will be guaranteed to cover all possible phenomena in the region of parameter space considered. We outline the main steps of this algorithm below and refer the reader to \cite{Carvalho:2015ttv} for further details and discussion.

The first task is, given two different points in parameter space and one or more kinematic distributions  generated by these points (the $p_T$ of the Higgs or top quark, invariant masses, etc.), to decide how similar the distributions are. We organize each of these sets of distributions in samples $S_{a}$, where $a$ identifies the point in parameter space, so that running over the bins of the sample is the same as running over all bins of all distributions included in the analysis. In order to decide on similarity between the samples we will use the following log-likelihood ratio:

\bea
TS_{ab} &=&-2 \sum^{N_{bins}}_{i=1} \left[
  log(n_{(i,a)}!)+log(n_{(i,b)}!)-2 log
  \Bigg( \frac{n_{(i,a)}+n_{(i,b)}}{2}!\Bigg)
\right],
\label{clus.ts}
\eea
where $n_{(i,a)}$ is the number of event counts in the i-th bin of sample $S_a$, $n_{(i,b)}$ is the same for sample $S_b$ and $N_{bins}$ is the number of bins in the sample. $TS_{ab}$ is zero for identical samples and increasingly more negative for increasingly different distributions, so if $TS_{ab} > TS_{cd}$ it means $S_a$ and $S_b$ are more mutually similar than $S_c$ and $S_d$ are.

Now, starting from a number of samples $N_{sample}$, we follow the steps below to organize them into clusters:

\begin{enumerate}
    \item We start with a number of clusters that is equal to the number of samples, thus with $N_{cluster} = N_{sample}$, each cluster containing exactly one sample.
    \item We obtain the cluster-to-cluster similarity between two clusters, defined as the minimum $TS_{ab}$ between members of those clusters: $TS^{min} = min_{ab}(\{TS_{ab}\})$ ($a$ runs over all samples in the first cluster and $b$ does the same for the second cluster).
    \item We calculate $TS^{min}$ between all possible pair of clusters, and find the two clusters with the highest $TS^{min}$. Merge these two clusters into one. The number of clusters $N_{cluster}$ diminishes by one.
    \item We repeat step 3 until the desired $N_{cluster}$ is obtained.
\end{enumerate}

For each step in the clustering algorithm we can also choose one special sample within each cluster that is the best representative of its behaviour. We do that by calculating $TS^{min}_a = min_{b}(\{TS_{ab}\})$
for each sample $S_a$ in the cluster, where $b$ runs over every element in the cluster except $a$. The sample with the highest $TS^{min}_a$ is the benchmark sample and the equivalent point in parameter space will be called a benchmark point.

The appropriate final number of clusters is a compromise between a very fine grained view, with a unwieldy number of very homogeneous clusters (in the limit we go back to $N_{cluster} = N_{sample}$), and the opposite extreme, with just a few clusters that contain a huge number of samples that are very heterogeneous in behaviour (thus with a benchmark sample that will not be a very good representative of the whole group). What can be done is to run this algorithm all the way down to $N_{cluster} = 2$  clusters, keeping record of each step. That way one can examine each of the different realizations and decide on the ideal number. The same can be said about which kinematic distributions to include in the samples used for the clustering: we find the most interesting ones by experimentation.

\subsection{Implementation of both MCHM models into the Event Generator}
\label{mg}

We have implemented both models in FeynRules
(v2.3)~\cite{Alloul:2013bka} and produced an associated UFO file for
each model, that can be interfaced with MadGraph 5
(v2.6.2)~\cite{Alwall:2014hca}.  The numerical input from the
diagonalization of the mass matrices is then fed via a custom-written
Python script into the param\_card.dat for processing within MG5.  We
simulate the $t{\overline t} h$ and $t{\overline t} hh$ processes in MG5.
We have also checked that the deviations from the SM in
the top quark properties are negligible, since the new physics is
rather heavy.

It is important to stress here that the generation framework we fully developed at the parton and LO level can be connected directly to detector simulations such as DELPHES (fast simulation) or detailed experiment full simulations such as the ones of ATLAS or CMS.
\subsection{The $t\overline{t}h$ Process}
\label{tth}

We start by considering the $t\overline{t}h$ process in the MCHM scenarios.
This is related in a very simple way to the same process in the SM.
The Feynman diagrams (at tree level) are identical in all the models,
involving top/anti-top pair production, with a Higgs boson radiated
from the top lines.  Radiation of the Higgs boson from initial state
$q{\overline q}$ lines can be neglected due to the small Yukawa couplings
(and PDF suppressions for the heavier flavors).  As a result the
amplitude is simply proportional to the top Yukawa coupling.  The
cross section in the MCHM scenarios can then be simply expressed in
terms of the SM cross section as
\bea
\sigma_{\rm MCHM}(t\overline{t}h) &=& \left( \frac{y_t}{y_t^{\rm SM}} \right)^2 \, \sigma_{\rm SM}(t\overline{t}h)~.
\label{sigmatth}
\eea
All the modifications due to Higgs compositeness, or mixing with
vector-like fermions, enter only through the top Yukawa coupling.
Therefore,
as in other BSM cases, a modification in the total rate w.r.t. the SM is expected, but not in kinematic distributions. Besides as we will see in some example cases this deviation can be very small, and still compatible with MCHM. This means that a high precision (order $1\%$ or less) measurement of $\mu(t\overline{t}h)$ might be required (see sections \ref{results} and \ref{MCHMHighScale})

\subsection{The $t\overline{t}hh$ Process}
\label{tthh}

\begin{figure}[h]
\centering
\includegraphics[width=0.28\textwidth]{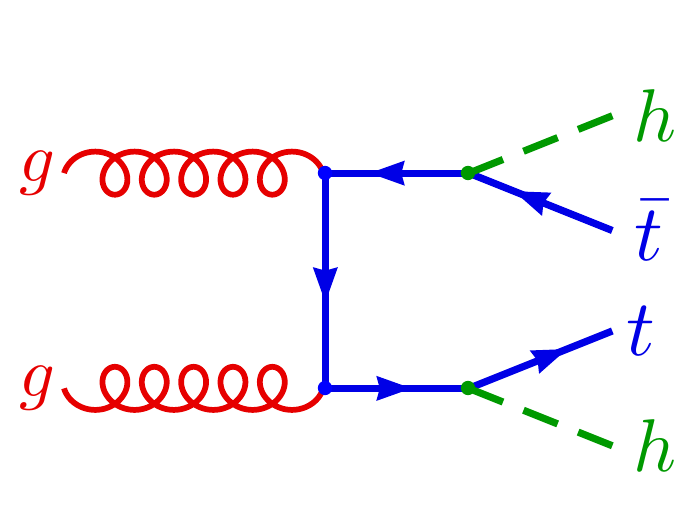}
\hspace{1cm}
\includegraphics[width=0.28\textwidth]{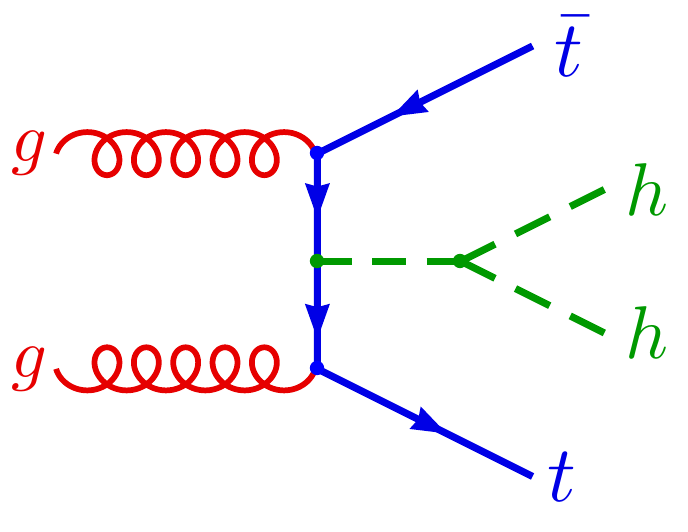}
\hspace{1cm}
\includegraphics[width=0.28\textwidth]{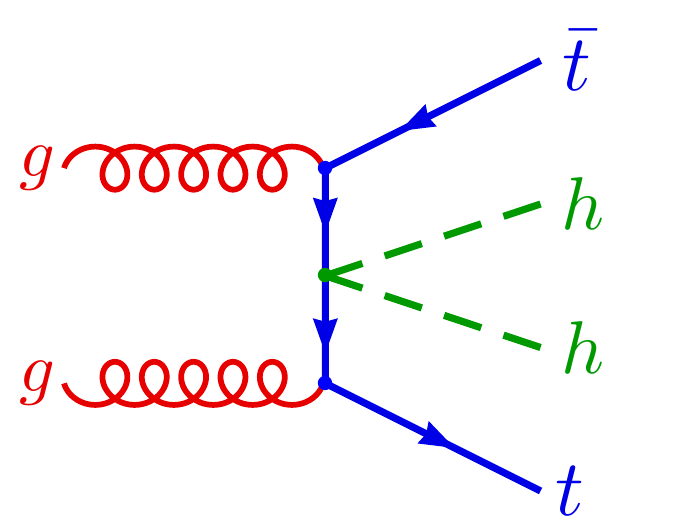}
\caption{Representative diagrams for the non-resonant $\overline{t}thh$
process, illustrating the three distinct physical subprocesses: the
Yukawa vertex, the Higgs trilinear self-coupling and the ``double
Higgs'' Yukawa vertex arising in composite Higgs scenarios.}
\label{fig:nrtthh}
\end{figure}

Next, we consider the $t\overline{t}hh$ process.  The additional radiated
Higgs boson allows for a richer dependence on the new physics than in
$t\overline{t}h$.  There are two qualitatively different contributions:
\begin{enumerate}

\item Resonant processes, in which vectorlike charge 2/3 resonances are produced, and subsequently decay in the $th$ channel.
The resonances can appear either in pairs (QCD pair production) or
singly (as an intermediate state in a fermion line, involving a
flavor-changing Yukawa interaction).  The process is, however, largely
dominated by QCD pair production.

\item Non-resonant processes, in which only the diagrams without intermediate top partners are included, see Fig. \ref{fig:nrtthh}.

\end{enumerate}
\begin{figure}[h]
\centering
\includegraphics[width=0.9\textwidth]{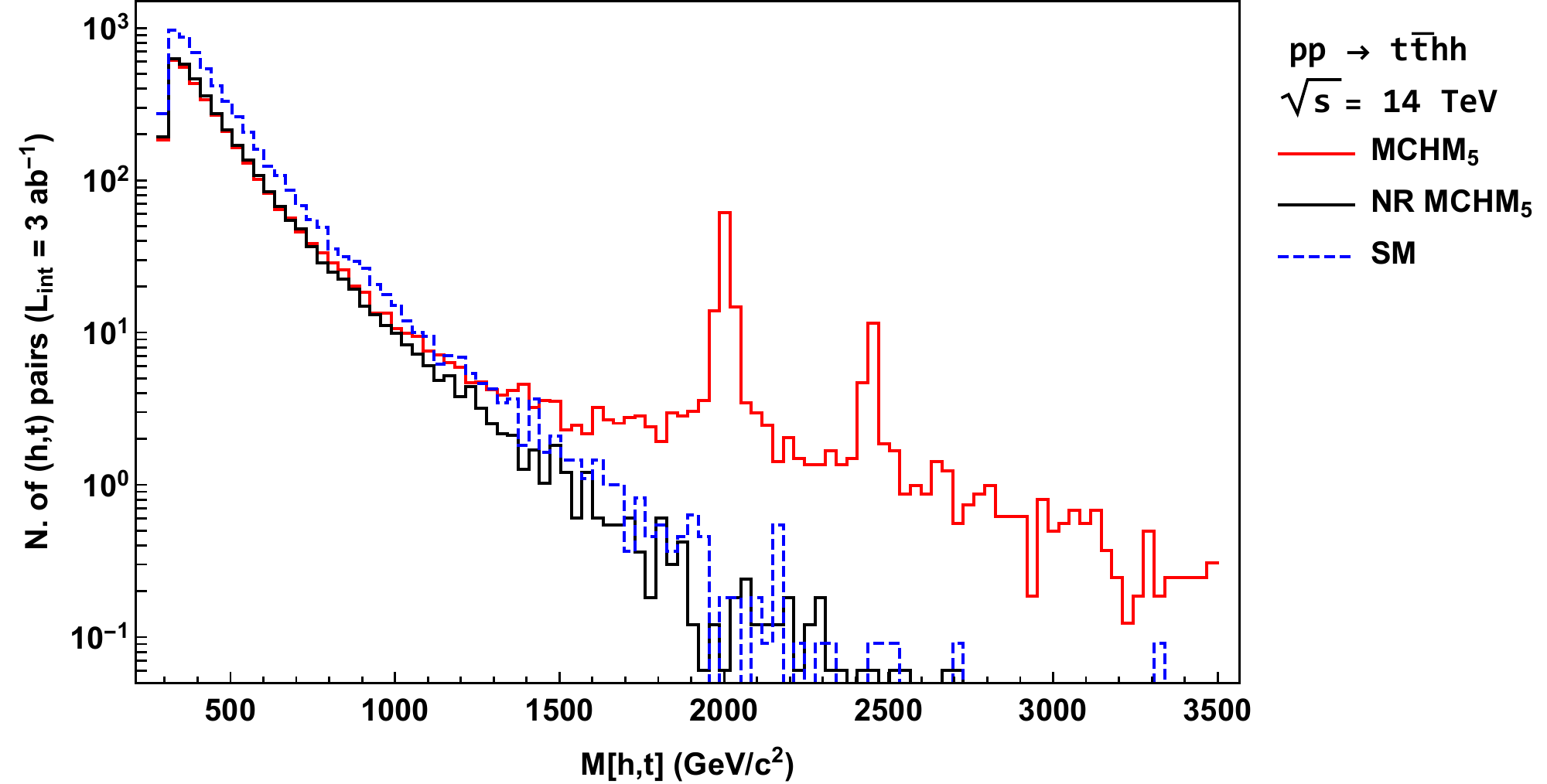}
\\ [0.5em]
\includegraphics[width=0.9\textwidth]{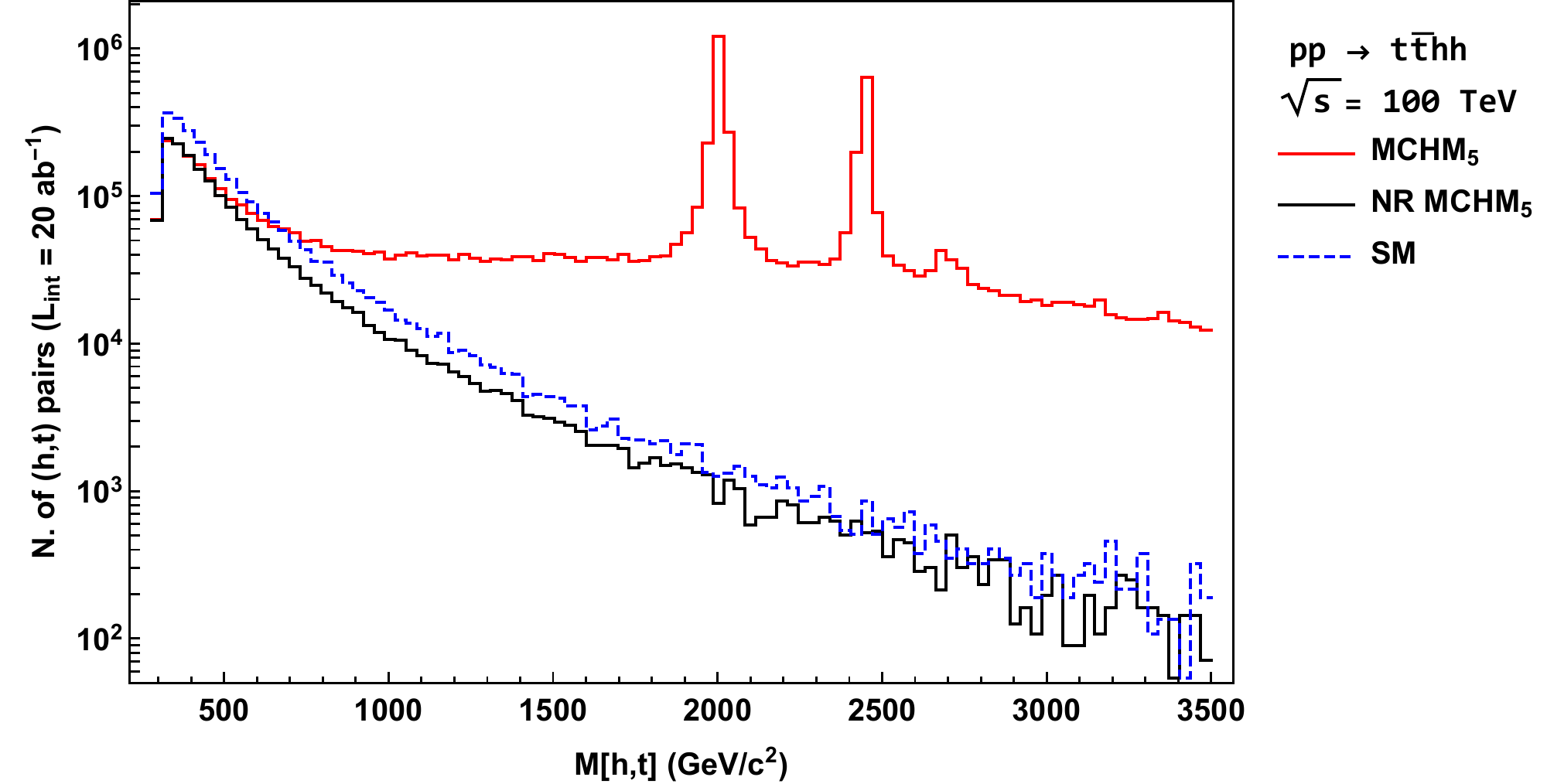}
\caption{Distribution of the invariant mass of the top quark and a Higgs boson in the MCHM$_5$ ($M_{1} = -2.5$~TeV, $M_4 =
2.0$~TeV, $f=1.0$~TeV, $y_L=1.5$).  The red continuous line shows the
distribution of the full $t\overline{t}hh$ process in the MCHM$_5$, while
the NR-$t{\overline t}hh$ cross section is shown in black.  For comparison, we also
show in dashed blue the SM $t\overline{t}hh$ distribution.  The upper (lower)
plot corresponds to $14$~TeV ($100$~TeV) CM energy.  Histograms generated
with MadAnalysis~5~\cite{Conte:2012fm}.}
\label{fig:htdist}
\end{figure}
The presence of resonant processes can lead to important enhancements
in the $t\overline{t}hh$ cross section w.r.t.~the SM, depending on their
mass.  The non-resonant process carries information that is distinct
from the resonant part, as discussed below.  It is therefore useful to
define a ``non-resonant cross section'' as obtained from this subset of
diagrams, which we label as ``NR-$t{\overline t}hh$''.  One can similarly define a
resonant cross section in terms of the diagrams involving QCD
vector-like pair production.  We find that, to an excellent
approximation, the total $t\overline{t}hh$ cross section is given by the
sum of these two cross sections.  The dependence of these two
subprocesses with kinematic variables (such as the $th$ invariant mass)
is different, so in principle they can be separated experimentally.

We show in Fig.~\ref{fig:htdist} the invariant mass distribution of $th$ for a point in our scan of the MCHM$_5$, where we display both the resonant and non-resonant contributions to the $t\overline{t}hh$
cross section, as well as the corresponding SM process for comparison.
We notice that
the NR-$t{\overline t}hh$ follows the SM cross section but displays a suppression and the relative importance of the resonant process w.r.t.~the non-resonant one increases with larger CM energies.
The cross section for both processes also increases significantly with the CM energy increase from 14 to 100~TeV. Likewise ($\mu(\ttHH)$ increases by a factor of 4 for the same increase in CM energy, while $\mu($NR-$\ttHH)$ does not change with energy.

\subsubsection{The Non-Resonant $t\overline{t}hh$ Process}
\label{NRtth}

The $p_T$ distributions for the NR-$t{\overline t}hh$ process show that the typical
scales involved are around 100~GeV. As explained before, the top
Yukawa coupling should be evaluated at around that scale for the
Yukawa vertices appearing in the NR-$t{\overline t}hh$ cross section.  For
simplicity, we use the top quark pole mass, which also appears in
kinematic quantities.

Depending on the vertices they contain, we may divide the diagrams in the MCHM scenarios into three categories, as
illustrated in Fig.~\ref{fig:nrtthh}:
\begin{enumerate}

\item Diagrams involving only the $t{\overline t}h$ vertex.

\item Diagrams involving both the $t{\overline t}h$ vertex and the trilinear Higgs self-interaction:\newline
$\lambda = \left[ (1 - 2 \xi)/\sqrt{1 - \xi} \right] \lambda_{\rm
SM}$.

\item Diagrams involving the $t{\overline t}hh$ vertex (``double Higgs'' Yukawa vertex).

\end{enumerate}
While the first two categories are composed of the same diagrams which appear in the corresponding SM process, the third category is particularly interesting, since the contact $t{\overline t}hh$ vertex has no counterpart in the SM, and is a direct consequence of the non-linear realization of the Higgs sector in composite models~\cite{Contino:2012xk}. For this reason, it would be extremely interesting if one
could get experimental evidence on this coupling.

In order to get a sense for the relative importance of the different
physical subprocesses, we have simulated the NR-$t{\overline t}hh$ cross section
turning off, in turn, the double Yukawa coupling and the trilinear
coupling.  Results of this study are shown in subsection \ref{NRtthhMCHM}.

\subsection{Operators Analysis}
\label{newops}
We move on to studying the effects of the high dimension operators in eqs.~\ref{Lint5} (controlled by the couplings $c_L$ and $c_R$) and \ref{Lint14} (controlled by $c_4$, $c_9$ and $c_{T9}$) on the two processes under consideration.
\subsubsection{Effect on the $t\overline{t}h$ Process}

As mentioned in section \ref{MCHM5}
the derivative operators of eqs.~\ref{Lint5} and \ref{Lint14} do not modify the $tth$ vertex at tree level, appearing only in vertexes involving the Higgs and one or more of the new fermionic resonances. Since the leading diagrams in this process do not involve these resonances the changes in their decay widths do not affect it.
Therefore, the cross section of the $t\overline{t}h$ process is not modified by the new operators at tree level.

\subsubsection{Effect on the $t\overline{t}hh$ Process}
The $t\overline{t}hh$ process cross section is affected in two ways.  First, the leading diagrams in this process contain resonances as intermediate states.  The change in their decay widths will therefore have an effect on the total cross section. The other source of modification arises from the Yukawa type vertex between the Higgs and two different flavour states which enters directly into the main diagrams.\\

\begin{figure}[h]
\centering
\includegraphics[width=0.47\textwidth]{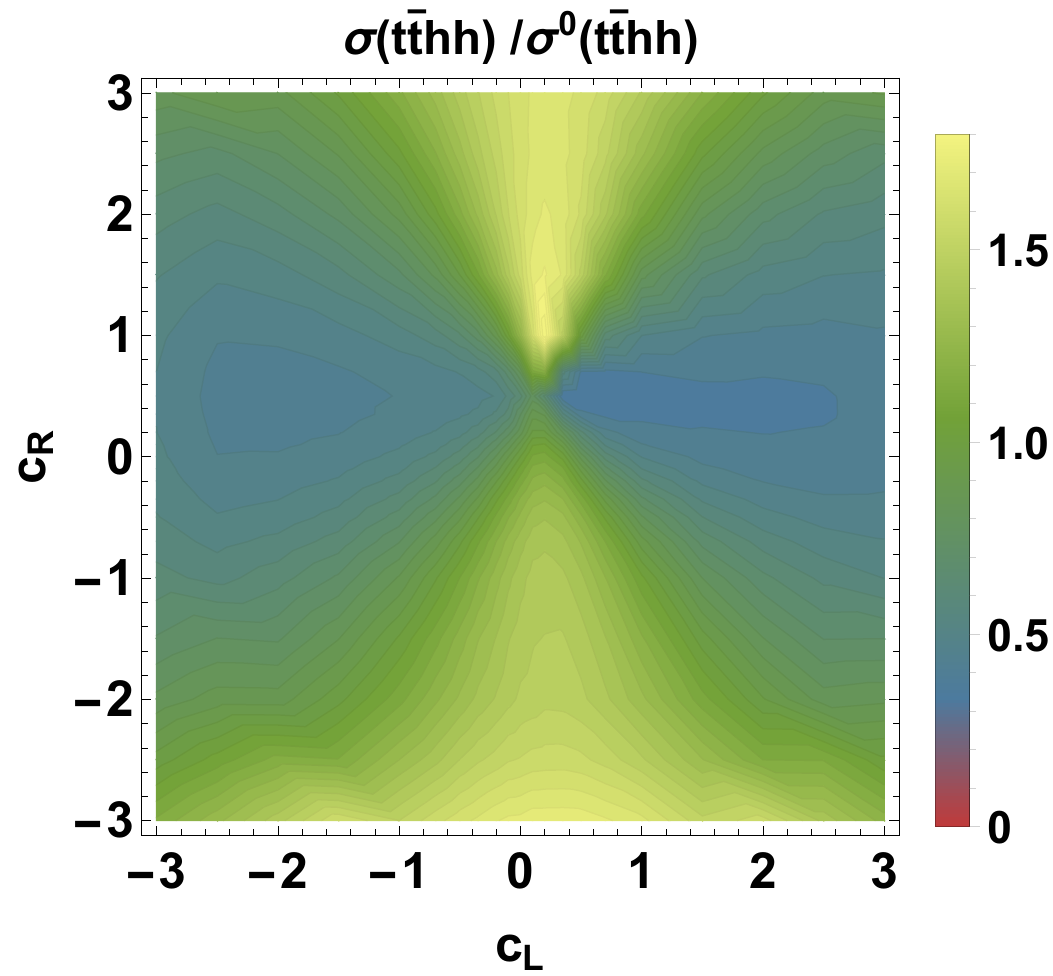}
\hspace{0.1cm}
\includegraphics[width=0.47\textwidth]{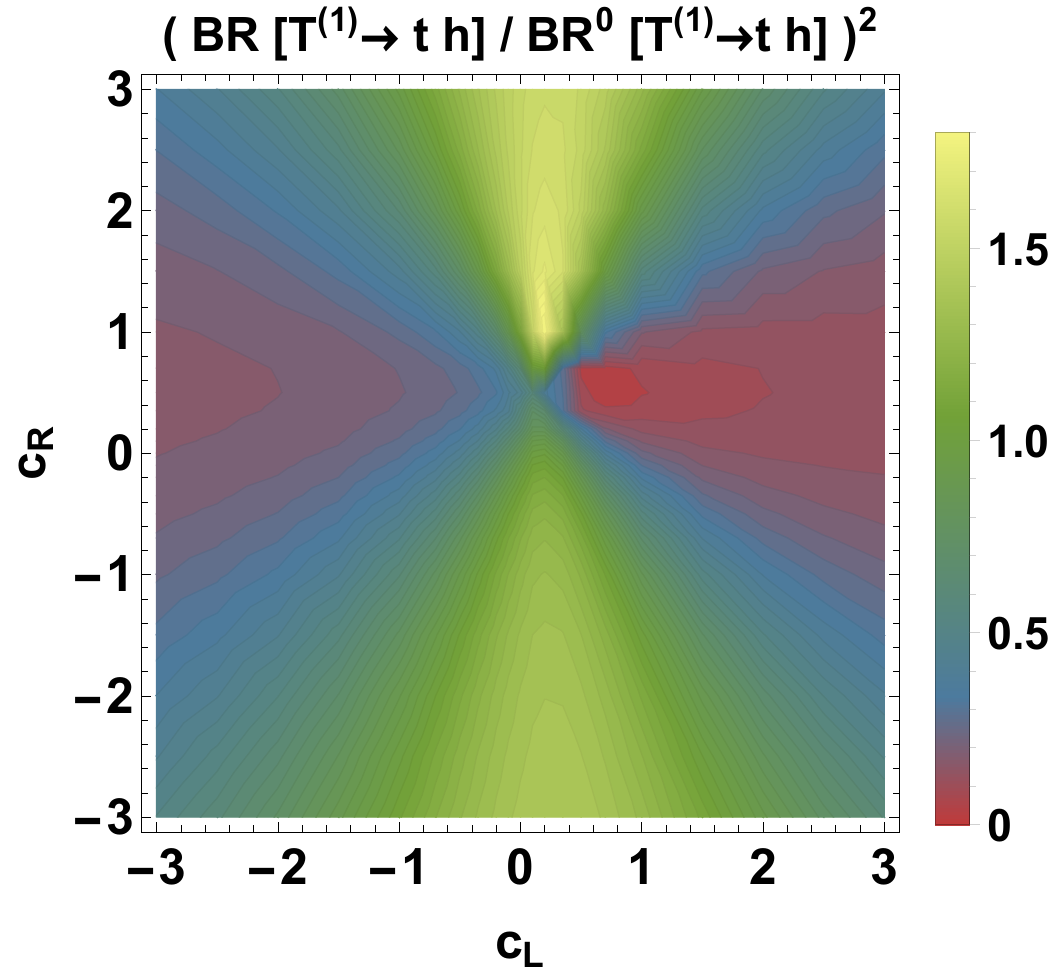}
\caption{Variation with $c_L$ and $c_R$ of the $t\overline{t}hh$ cross section, normalized to the cross section $\sigma^0 (c_R=0,c_L=0)$, in the MCHM$_5$ ($M_{1} = -0.96$~TeV, $M_4 =
1.4$~TeV, $f=1.2$~TeV, $y_L=0.88$).}
\label{fig:p3clcr}
\end{figure}

In order to have a better comprehension of the implications of the operators in the $t\overline{t}hh$ total cross section, we took some example points in the parameter space and did a scan in the region of the ($c_L$, $c_R$) plane compatible with perturbativity ($c_L$, $c_R$ $\in [-3,3]$). In Fig.~\ref{fig:p3clcr} we show the scan for one of the points. There, we see that there is a modification of the $t\overline{t}hh$ cross section by a factor that lies in the range [$ 0.3$, $ 2$]. This modification is dominated by the change in the branching ratio of the decay channel $T^{(1)}\rightarrow t h$, which is compatible with the fact that in most of the region $T^{(1)}$ is narrow. We can also see that the derivative operators can be tuned to strongly suppress the branching ratio of $T^{(1)}$ into $t h$, setting it to essentially zero in a small region (in red on Fig.~\ref{fig:p3clcr}).
It is also worth noting that the $t\overline{t}hh$ cross section does not go to zero in that tuned region, because even if we suppress the resonant production, there is still the NR-$t\overline{t}hh$ contribution, thus the absence of a red region in the left plot of Fig.~\ref{fig:p3clcr}.

\begin{figure}[h]
\centering
\includegraphics[width=\textwidth]{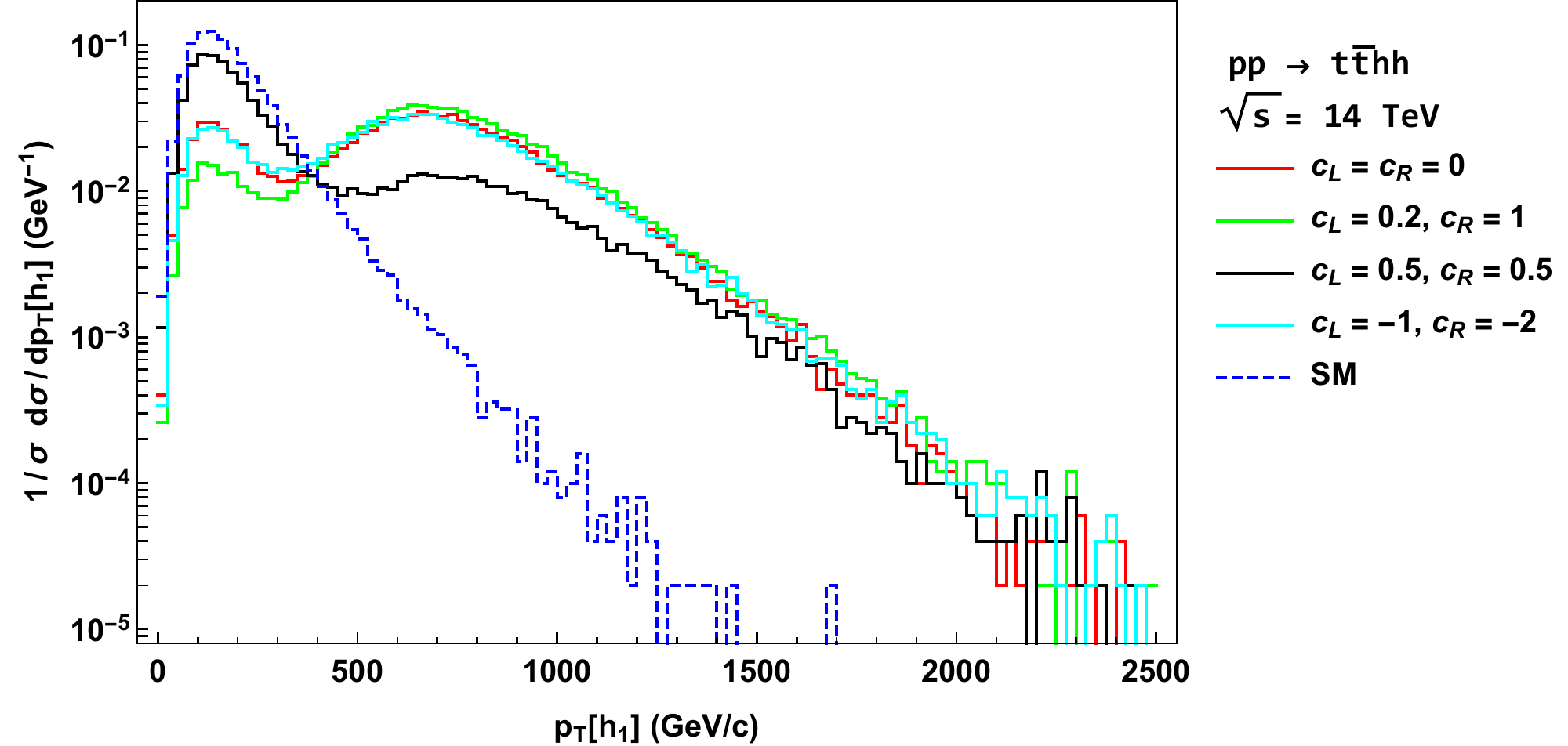}
\\ \vspace{0.5cm}
\includegraphics[width=\textwidth]{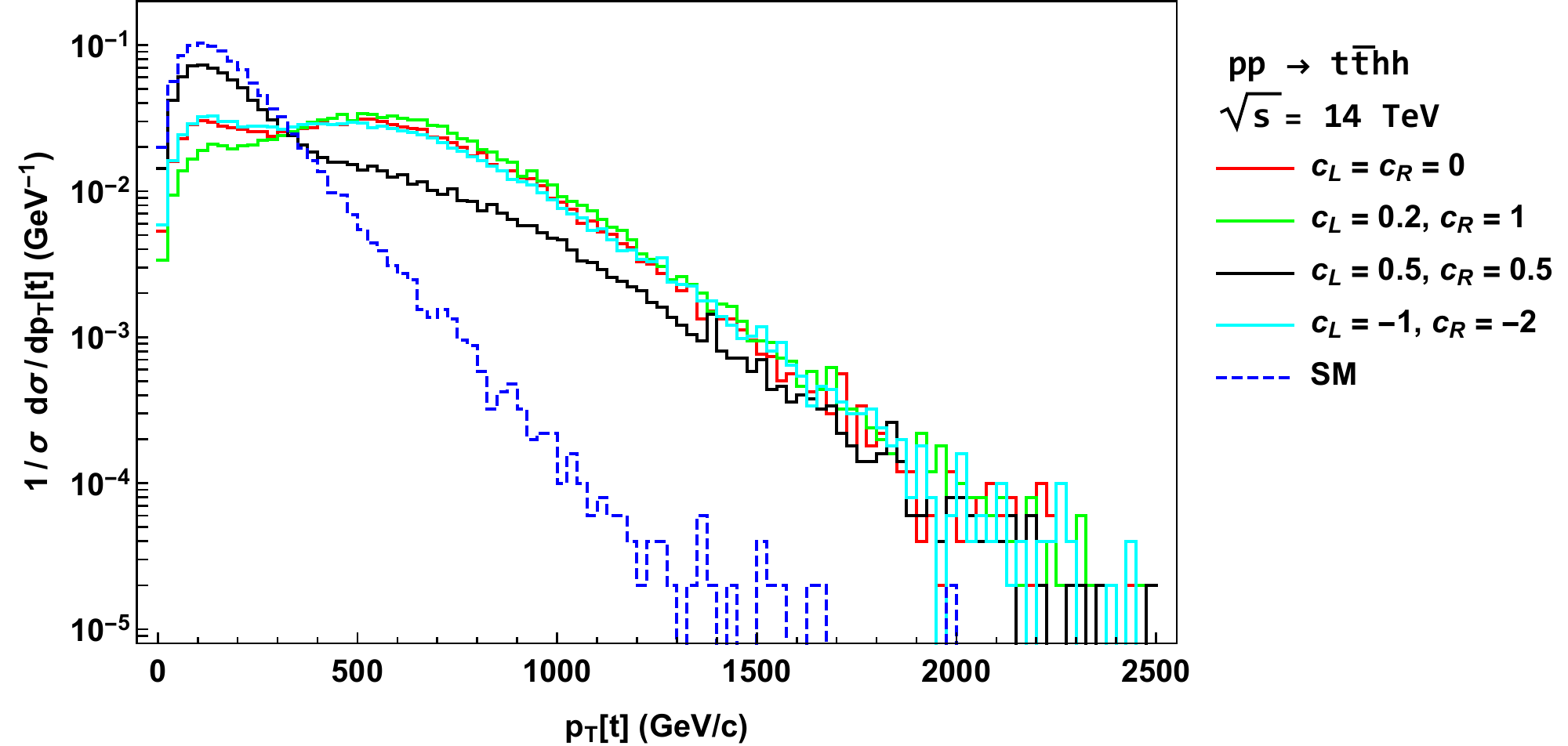}
\caption{$p_t$ distribution of the most energetic Higgs ($h_1$) and the top particle for $4$ combinations of $(c_L,\, c_R)$ in the MCHM$_5$ ($M_{1} = -0.96$~TeV, $M_4 =
1.4$~TeV, $f=1.2$~TeV, $y_L=0.88$). The SM is shown for comparison and all curves are normalized to unity area. Histograms generated
with MadAnalysis~5~\cite{Conte:2012fm}.}
\label{fig:ptdist}
\end{figure}

Figure \ref{fig:ptdist} shows the $p_t$ distribution of the most energetic Higgs and the top particle keeping the same \five parameters used in Fig.~\ref{fig:p3clcr} for $4$ combinations of $(c_L,\, c_R)$, as well as the SM distributions for comparison. One notices that the different combinations of $(c_L,\, c_R)$ produce similar kinematic distributions, except for the one with $(c_L=0.5,\, c_R=0.5)$ which is close to the fine tuned combination  which suppresses $T^{(1)}\rightarrow t h$. An identical situation is seen in other kinematic distributions and other points of the MHCM$_5$ parameter space, with the fine tuned combination of $(c_L,\, c_R)$ that suppresses the width being different for every \five point.

The conclusion is that we have two qualitatively different situations:

\begin{enumerate}
    \item For most values of $(c_L,\, c_R)$ the shapes of the distributions do not change and there is a change in the total cross section of a factor between $0.5$ to $6$ for 14 TeV and up to $8$ for 100 TeV.
    This is equivalent to a k-factor and  there is no advantage in including two extra parameters to that effect, since we are working at LO and a rescaling of cross sections is needed for comparison with experiment anyway.
    \item For a small region around a particular value of $(c_L,\, c_R)$, which is different for every point in the MHCM$_5$ parameter space, the branching ratio $BR[T^{(1)} \rightarrow t h]$ becomes very small (sometimes even zero), and the distributions rapidly change into those of the non-resonant $t\overline{t} hh$ production. It would be interesting to explore what happens to other decay channels near these points, but that is beyond the scope of this paper. We intend to return to this point in a future work focusing on the search for top-partners.
\end{enumerate}

For the reasons given above, we will take $c_L=c_R=0$, and ignore the effects of the derivative operators for the rest of our analysis. We have verified that the situation is qualitatively the same in the MCHM$_{14}$.

\section{MCHM at Low Scale}
\label{results}

We now proceed to the scan of the parameter space for the MCHM. As mentioned before, we start by looking at the ``Low Scale'' region, which was characterized in section~\ref{sec.scans.low}. We repeat the ranges here for convenience. For the MCHM$_5$, we scanned over:
\begin{align*}
|M_1|& \in [0.8, 3.0]~{\rm TeV}, &  M_4 &\in [1.2, 3.0]~{\rm TeV},  \\
f&\in [0.8, 2.0]~{\rm TeV},          &  y_L&\in [0.5, 3.0].
\end{align*}
For the MCHM$_{14}$, we used:
\begin{align*}
|M_1|& \in [0.8, 3.0]~{\rm TeV}, &  |M_4| &\in [1.2, 3.0]~{\rm TeV},  &  M_9 &\in [1.3, 4.0]~{\rm TeV},  \\
f&\in [0.8, 2.0]~{\rm TeV},          &  y_L&\in [0.5, 3.0].
\end{align*}

The results of the analysis of the scan over the parameters are presented in two ways, namely, the plots of a number of selected observables (Sec. \ref{scans}), completed by the selection of a number of example-points including some of their relevant physics characteristics (Sec. \ref{benchmarks}) and a broader survey of the parameter space and its benchmark points, made in Sec.~\ref{clusteringLScale}.

\subsection{Scanning Over Parameter Space}
\label{scans}

We show in this section a number of selected observables that highlight the behaviour of the MCHM in the scanned regions of the parameter space. For completing the results of this scanning over parameters analysis, these plots include the selected example-points, discussed in details in the next subsection, with their labeling as defined in Tables~\ref{fig:benchmarkTable1} and~\ref{fig:benchmarkTable2}  (Sec.~\ref{benchmarks}).
We will conveniently use the $M_1-M_4$ space displayed in Fig.~\ref{fig:ytvsM1M4} to define regions in the parameter space for both scenarios under consideration. For the MCHM$_5$ we will define the following two regions\footnote{We remind the reader that only the sign of $M_1$ is a free parameter in the MCHM$_5$, while the signs of both masses are free in the MCHM$_{14}$. See section~\ref{parameterspace}}:
\begin{itemize}
    \item[] Region I: $M_1,\,M_4>0$
    \item[] Region II: $M_1<0$, $M_4>0$
\end{itemize}
and for the MCHM$_{14}$ we will define the following four regions:
\begin{itemize}
    \item[] Region I: $M_1,\,M_4>0$
    \item[] Region II: $M_1<0$, $M_4>0$
    \item[] Region III: $M_1,\,M_4<0$
    \item[] Region IV: $M_1>0$, $M_4<0$
\end{itemize}
Each region is populated with about 200 points chosen at random, with points violating our conditions (see Sec.~\ref{parameterspace}) being disregarded. Each point in the MHCM$_5$ is given by a choice of $(f,|M_1|,|M_4|,\mbox{sign}(M_1),y_L)$ and a point in MHCM$_{14}$ by a choice of $(f,|M_1|,|M_4|, |M_9|,\mbox{sign}(M_1),\mbox{sign}(M_4),y_L)$. Each point is then passed to our implementation of the model in MadGraph, which gives us cross sections and distributions.

As shown in Eq.~(\ref{sigmatth}), the $t\overline{t}h$ cross section is
simply related to the SM one by a rescaling of the top Yukawa
coupling.  The deviations in the top Yukawa coupling from the SM limit
have two distinct origins:
\begin{itemize}

\item Deviations due to the compositeness nature of the Higgs boson,
which arise from the dependence on the Higgs through trigonometric
functions.  This depends only on $\xi$, but is model-dependent and can
in principle be used to distinguish the MCHM$_5$ from the MCHM$_{14}$.

\item Deviations arising from the mixing of the top quark with the new
$Q=2/3$ resonances.  This effect depends on all the microscopic
parameters of the model in a complicated manner through the
diagonalization of the mass matrix.  However, due to the fact that the
resonances must be much heavier than the top quark, the deviations
arising from the mixing are typically subdominant to the ones arising
from Higgs compositeness.
\end{itemize}

\begin{figure}[h]
\centering
\includegraphics[width=0.21\textwidth]{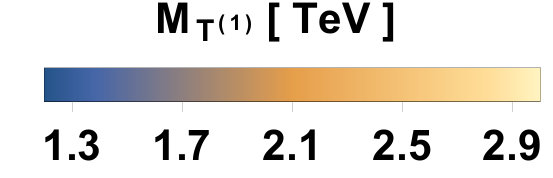}\\
\includegraphics[width=0.31\textwidth]{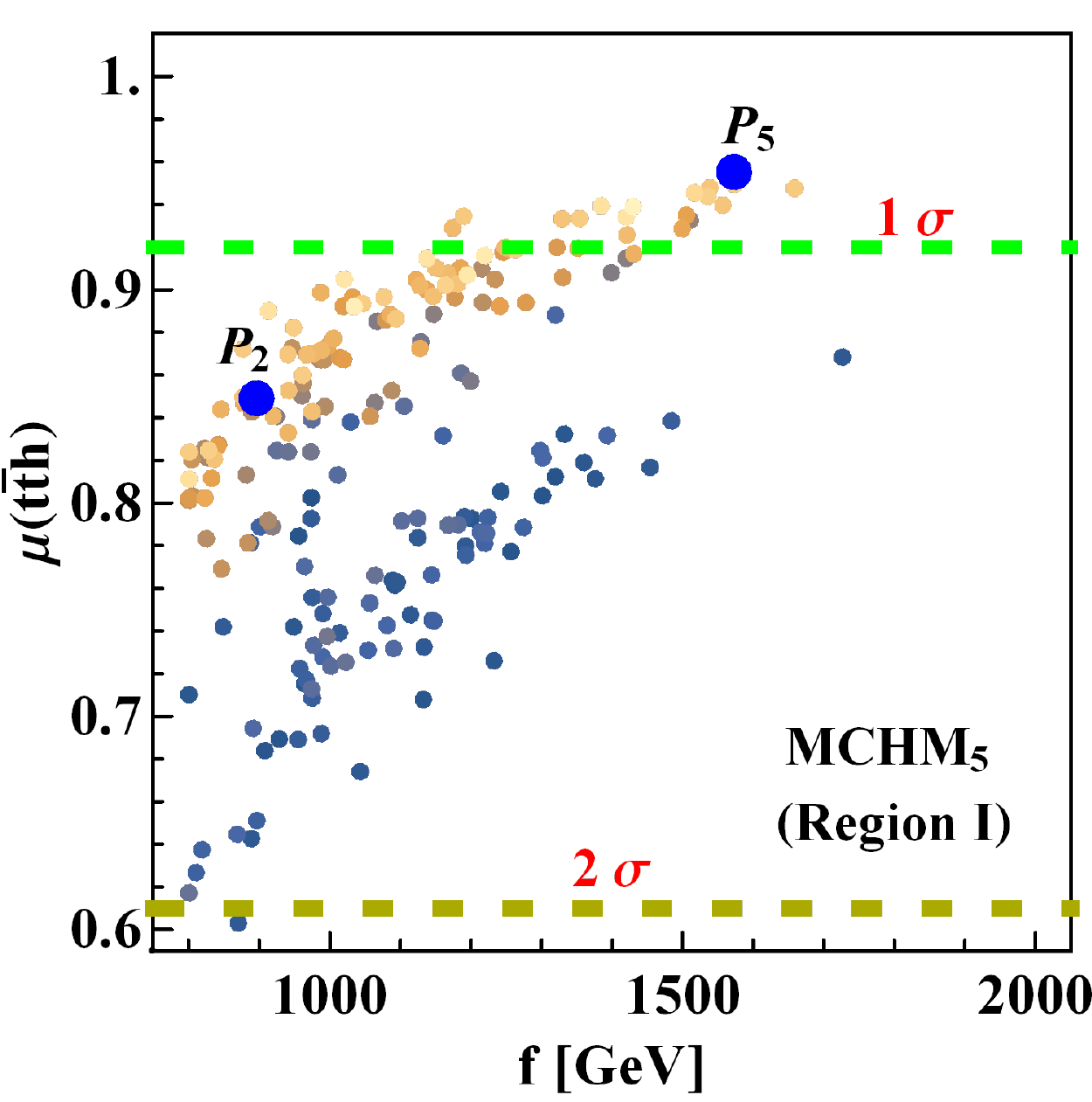}
\hspace{0.3cm}
\includegraphics[width=0.31\textwidth]{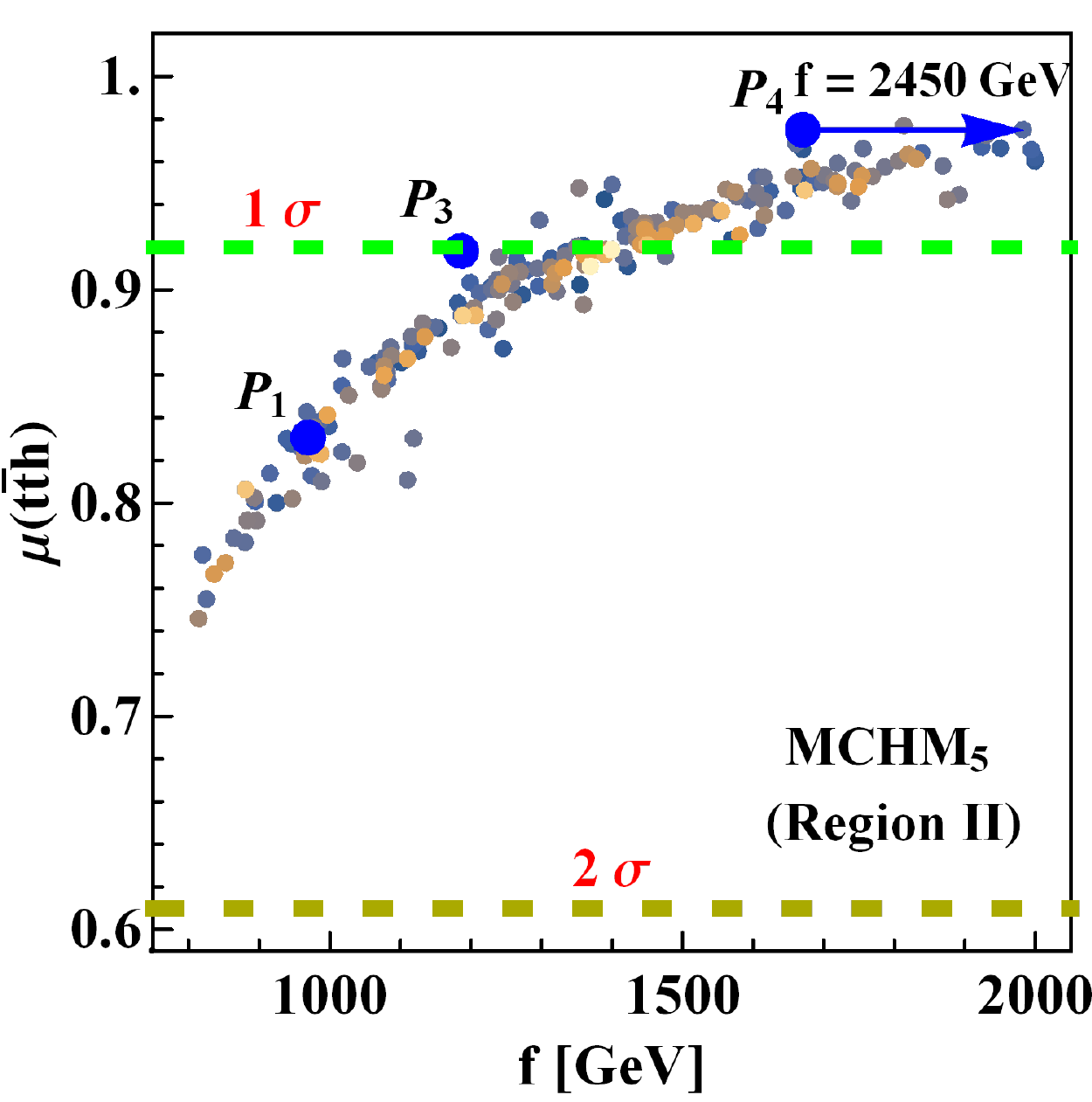}
\\ [0.3cm]
\includegraphics[width=0.31\textwidth]{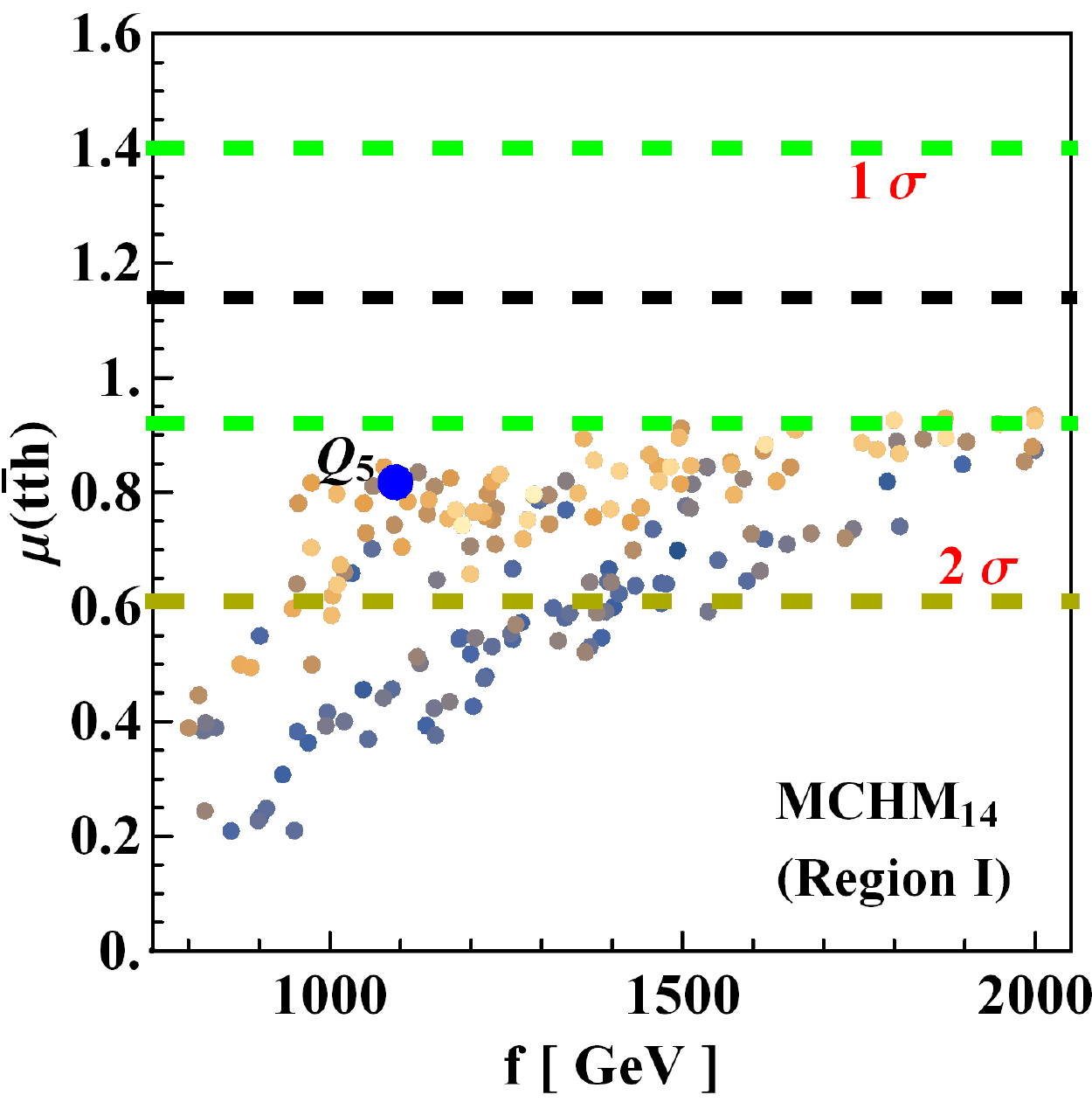}
\hspace{0.3cm}
\includegraphics[width=0.31\textwidth]{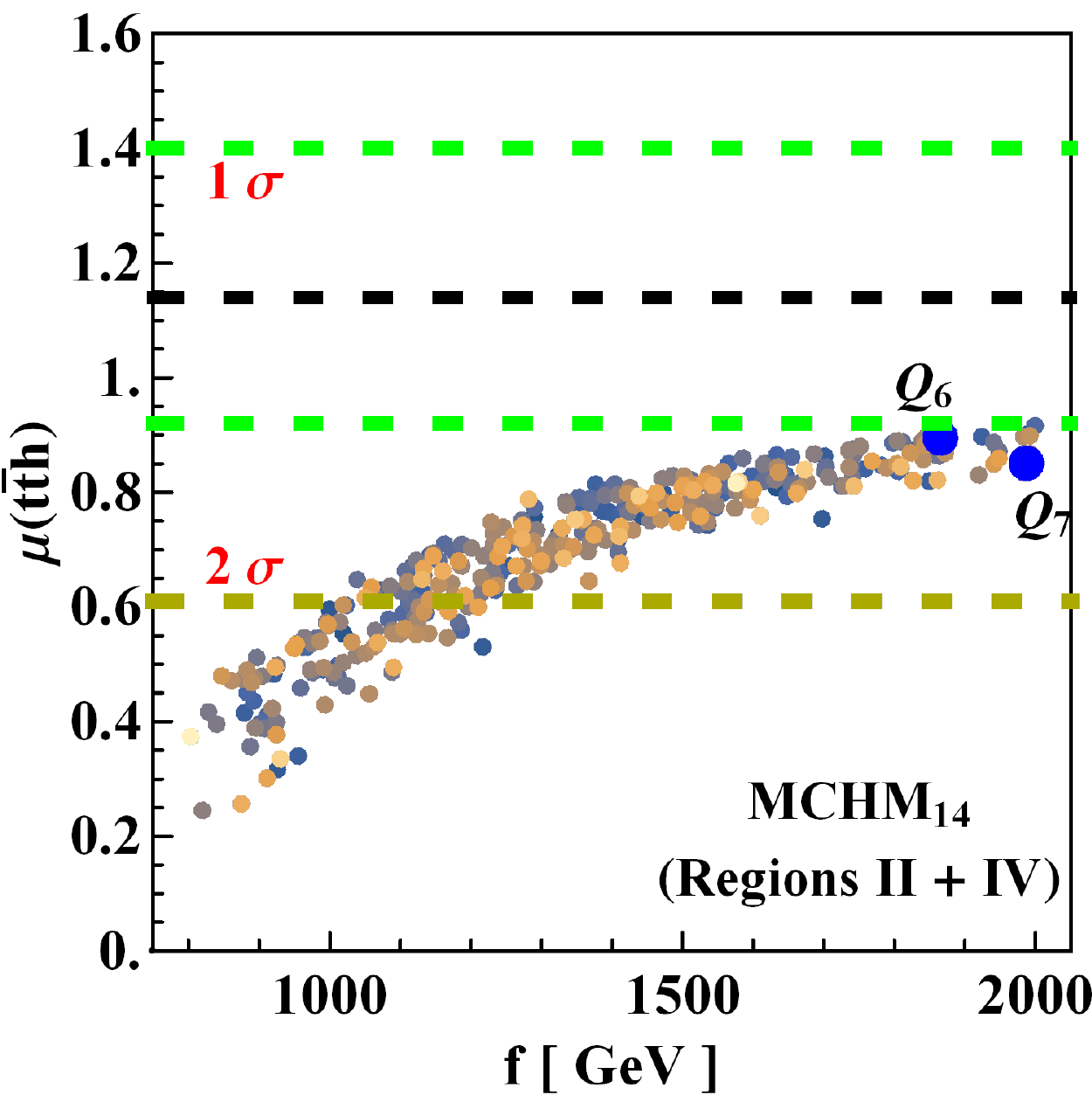}
\hspace{0.3cm}
\includegraphics[width=0.31\textwidth]{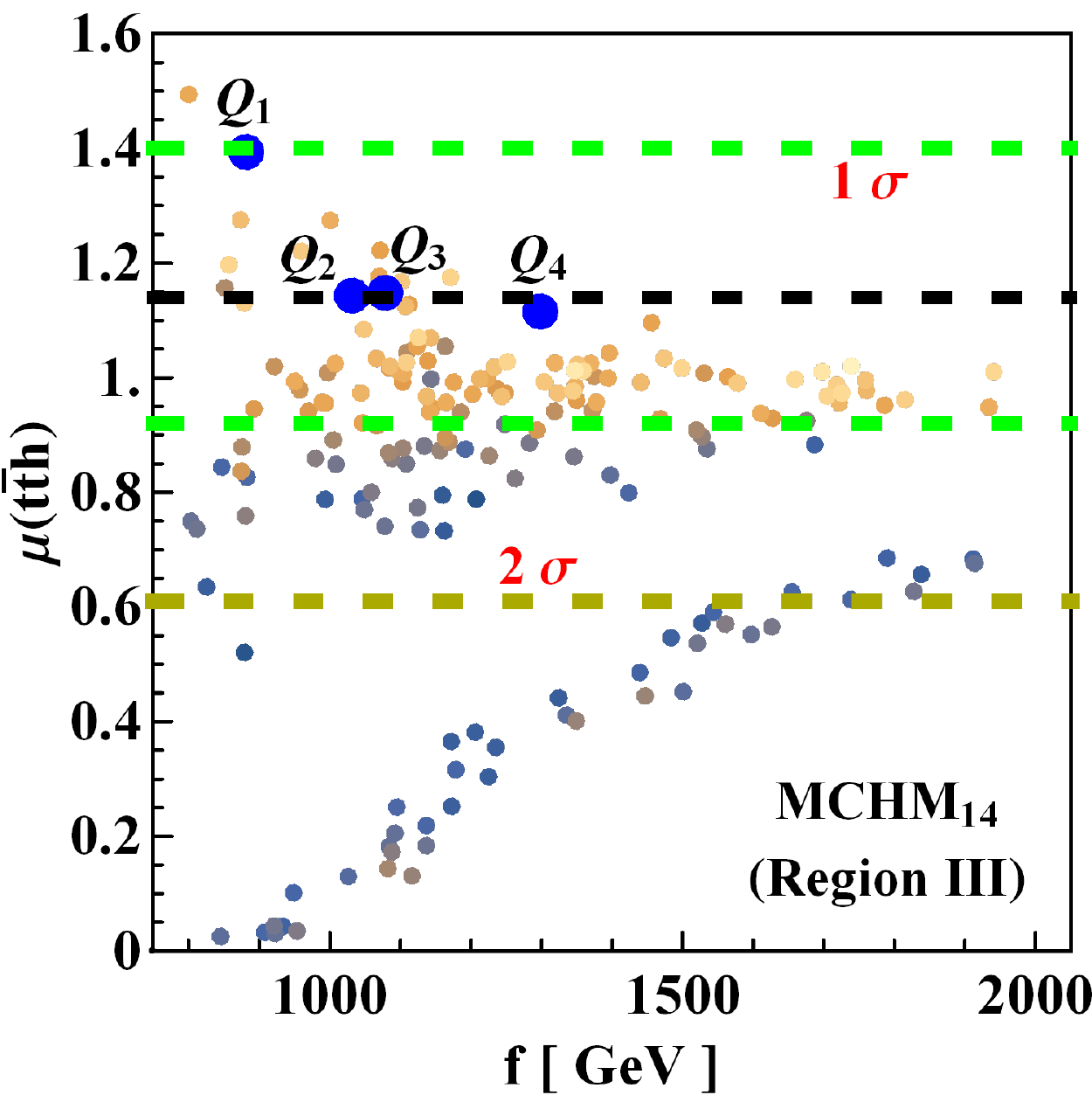}
\caption{Normalized $t\overline{t}h$ cross section as a function of $f$
for 14~TeV CM energies.  We also color code the lightest
vector-like mass. The upper left (right) plot corresponds to the Region I (II) of the MCHM$_5$. The lower left, central and right plots correspond to the Regions I, II + IV and III of the MCHM$_{14}$, respectively. The blue arrow in the upper right plot indicates that the example point $P_4$ is outside the horizontal range of the plot with $f = 2450$ GeV. The green (brown) dashed line shows the $1 \sigma$ ($2\sigma$) limits given by the CMS $\mu(t\overline{t}h)$ measurements while the black dashed line represents the central value~\cite{Sirunyan:2018hoz}.}
\label{fig:tthvsf}
\end{figure}

One concludes that the $t\overline{t}h$ cross section in the MCHM scenarios
is largely controlled by a single parameter, which we can take to be
the scale of global symmetry breaking, $f$.  This is illustrated in
Fig.~\ref{fig:tthvsf}. The signal strength $\mu(t\overline{t}h)$ of the production cross section is shown for $\sqrt{s}=14$ TeV but it does not depend on the CM energy at tree level, so the same results apply to $\sqrt{s}=100$ TeV. The CMS and ATLAS collaborations have presented results on the experimental measurements of $\mu(t\overline{t}h)$. Their reported best fits are: $\mu(t\overline{t}h)= 1.14_{-0.27}^{+0.31}$ for the combined $13$ TeV result at an integrated luminosity of $35.9 \text{ fb}^{-1}$ given by CMS~\cite{Sirunyan:2018hoz} and $\mu(t\overline{t}h)= 1.32_{-0.26}^{+0.28}$ for the combined $13$ TeV result at an integrated luminosity of up to $79.8 \text{ fb}^{-1}$ given by ATLAS~\cite{Aaboud:2018urx}. In Fig.~\ref{fig:tthvsf} we show the $1\sigma$ and $2\sigma$ limits from CMS (as ATLAS has not reported on their $2\sigma$ limits).

In Fig.~\ref{fig:tthvsf}, one can see that, in Region I of the MCHM$_5$, the points follow two distinct behaviours. The lower curve has most of its points (blue dots) corresponding to rather low M$_{T^{(1)}}$ masses (below 1.5 or 1.6 TeV) and $\mu (t\overline t h$) below 0.8, thus, in larger tension with the observed value while still within $2\sigma$ of it. Points with higher M$_{T^{(1)}}$ (brown dots) are spread around a second curve with $\mu$($t{\overline t}h$) larger than 0.8 and with M$_{T^{(1)}}$ greater than 2 TeV.

In contrast, in Region II of the MCHM$_5$, there is only one smooth curve with points with a relatively small dispersion and equally distributed over the whole scanned range in M$_{T^{(1)}}$. The selected example points in the MCHM$_5$ are indicated in both Regions ($P_1$, $P_3$ and $P_4$ in Region I and $P_2$ and $P_5$ in Region II). The important result is that in the MCHM$_5$ there is always a deficit in the $t{\overline t}h$ production cross section as compared to the SM. This is, indeed, a main feature of the MCHM$_5$.

For the MCHM$_{14}$, three different cases are identified concerning the evolution of this variable versus $f$ and M$_{T^{(1)}}$. Region I has some similarity with the corresponding Region I of the MCHM$_5$. The main difference is that, in the MCHM$_{14}$, $\mu$($t{\overline t}h$) can reach much smaller values (down to 0.2 if M$_{T^{(1)}}$ is smaller than $\sim$ 1.6 TeV (blue dots), and down to 0.4 even for higher M$_{T^{(1)}}$ masses). Thus, a fair fraction of all these scanned points have more than $2\sigma$ tension with the observed data. This case is represented by the example point $Q_5$ (see Table \ref{fig:benchmarkTable2} in Sec.~\ref{benchmarks}).

Regions II and IV of the \fourt are very similar to each other, and also to Region II of the MCHM$_5$, and are thus included in the same plot of Fig.~\ref{fig:tthvsf}. The main difference with the \five case lies in  a larger dispersion of the points and again the larger range in $\mu$($t{\overline t}h$) they cover (down to 0.2). Two example points, $Q_6$ and $Q_7$, have been selected and they are shown in this Figure (see Table \ref{fig:benchmarkTable2} in Sec.~\ref{benchmarks})

The last MCHM$_{14}$ scenario for this observable refers to Region III, which deserves special attention. Fig.~\ref{fig:ytvsM1M4} shows an increase of $t{\overline t}h$ production cross section as compared to the SM in a fraction of Region III. This is a main feature of the MCHM$_{14}$ as compared to the MCHM$_5$ or the MCHM$_{10}$. This feature is clearly visible in Fig.~\ref{fig:tthvsf}, where Region III of \fourt is the only one containing points with $\mu(t\overline{t}h) > 1$.

To further explore Region III, a special scan with 100 additional points was performed, extending M$_9$ down to 1.3 TeV. All of them are gathered in the lower right plot in Fig.~\ref{fig:tthvsf}. There are two curves. One curve gathers a major part of the low M$_{T^{(1)}}$ cases (blue dots). Some correspond to a dramatic deficit in $t{\overline t}h$ production cross section getting near to zero. Most of the points corresponding to $\mu$($t{\overline t}h$) larger or close to 1 correspond to larger M$_{T^{(1)}}$ masses i.e. masses larger than 1.8 - 2 TeV (brown points). However some rare blue points (lower M$_{T^{(1)}}$ masses) can also correspond to $\mu$($t{\overline t}h$) greater than 1. The highest $\mu$($t{\overline t}h$), above 1, are at relatively small $f$ value, as expected. The selected $Q_1$, $Q_2$, $Q_3$, $Q_4$ example points correspond to those cases but with a $\mu$($t{\overline t}h$) still within $1\sigma$ of the current LHC results.

The observable $\mu$($t{\overline t}h$) is thus a basic and key observable, not only to indicate that there is some BSM effect, but also to reject the MCHM$_5$ while keeping the MCHM$_{14}$ as still possible, if an enhancement w.r.t the SM is confirmed. If a deficit is instead observed, both MCHM scenarios will be possible, but the distinction between them is tricky and will depend on detailed phenomenology. More details are presented in Subsection~\ref{benchmarks} with the selected example points.

\begin{figure}[h]
\centering
\includegraphics[width=0.45\textwidth]{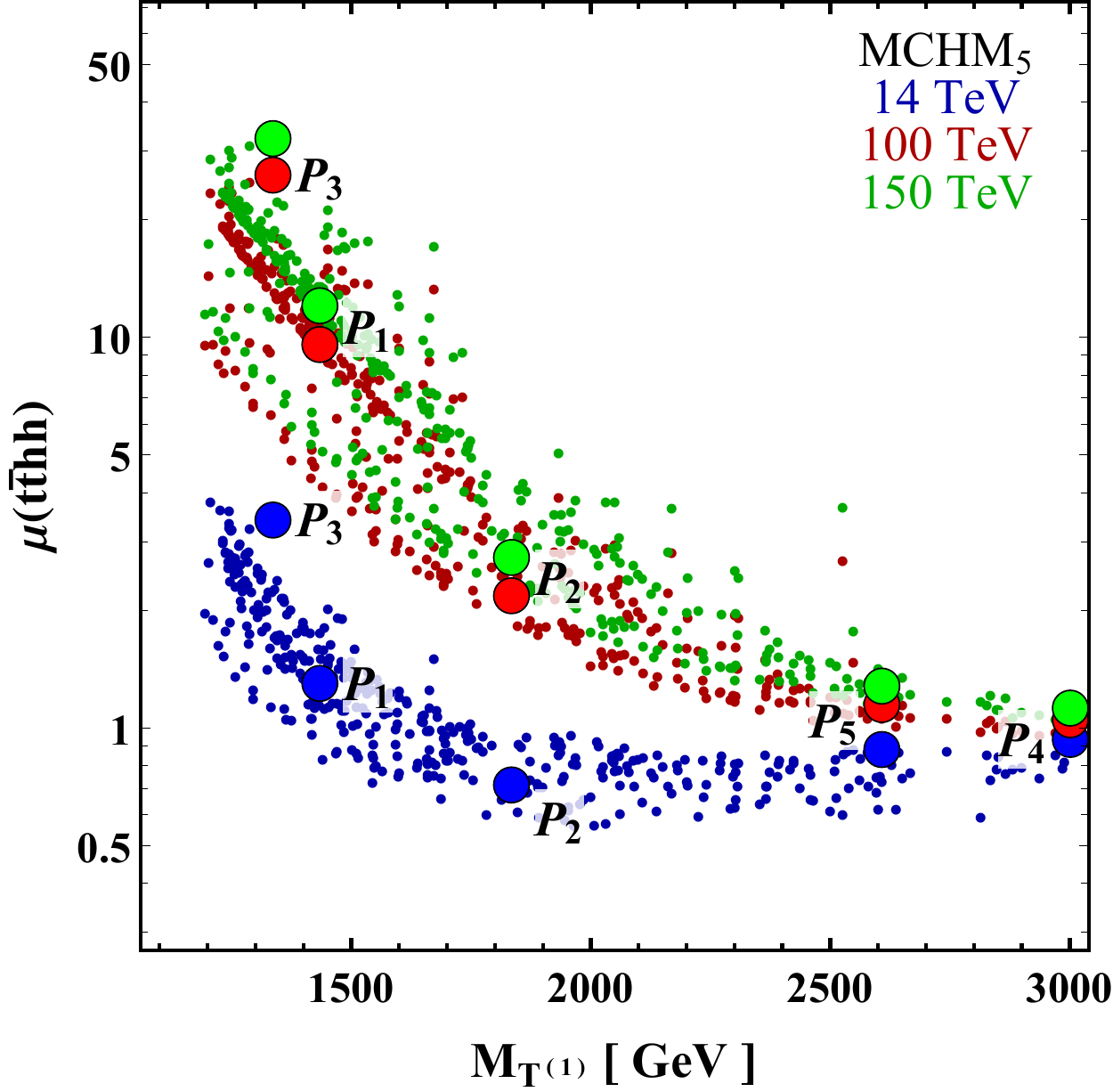}
\hspace{1cm}
\includegraphics[width=0.45\textwidth]{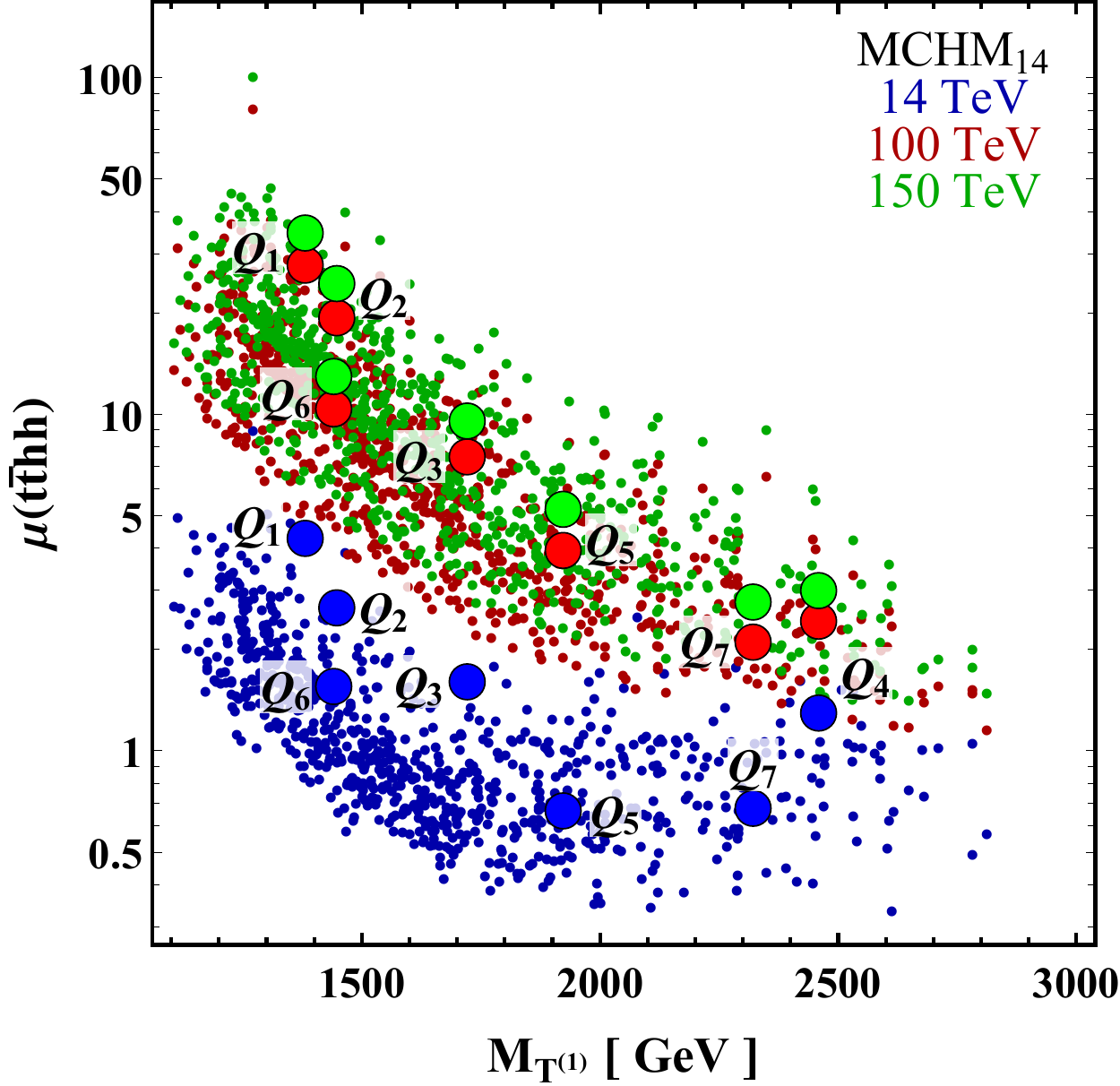}
\caption{Normalized $t\overline{t}hh$ cross section as a function of the lightest $Q = 2/3$ vector-like mass, for 14, 100 and 150~TeV CM energies.
The upper (lower) plots correspond to the MCHM$_5$ (MCHM$_{14}$).}
\label{fig:tthhvsMT4M4}
\end{figure}

Turning now to the $t\overline{t}hh$ process, we show in
Fig.~\ref{fig:tthhvsMT4M4} how the signal strength $\mu(t\overline{t}hh)$ depends on the mass of the lightest $Q=2/3$ resonance, for both MCHM scenarios and for different CM energies. There is a larger dispersion in the points of the MCHM$_{14}$. However it must be noted that all the points in the four MCHM$_{14}$ regions (about 1000 scanned points) are included in a single plot whereas only 400 (2 regions) are gathered in the MCHM$_5$ case. These plots show the expected result that for lighter resonances the resonant production can result in a significant enhancement of the total $t\overline{t}hh$ cross section.  This effect becomes more prominent for larger CM energies.

These effects are highlighted by the example points. For instance, $P_1$ and $P_3$ (in the MCHM$_5$ case) are showing a large enhancement in $\mu$($t{\overline t}hh$) when increasing the CM energy whereas $P_2$, $P_4$ and $P_5$ are not showing such an effect.

\begin{figure}[h]
\centering
\includegraphics[width=0.45\textwidth]{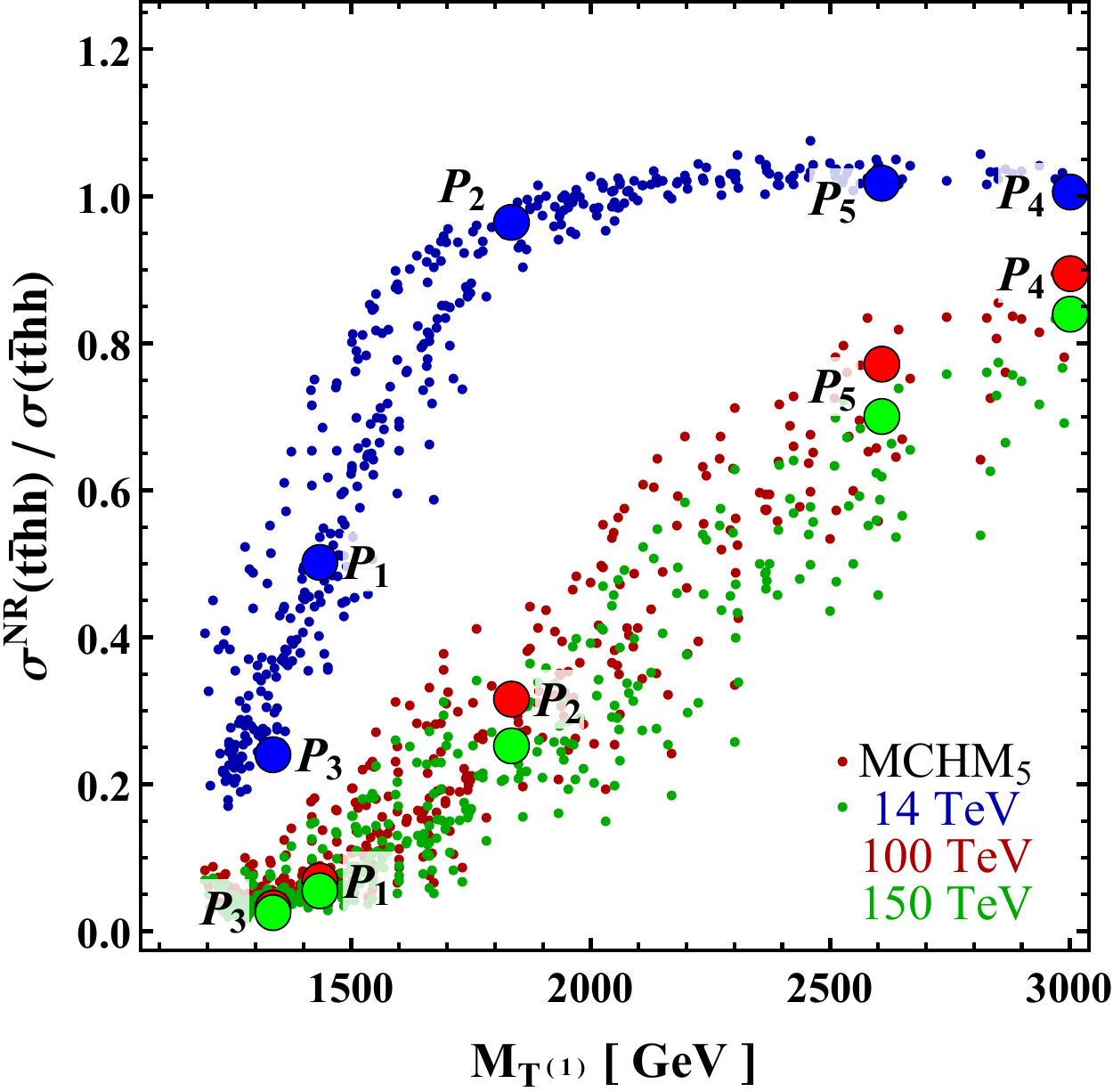}
\hspace{1cm}
\includegraphics[width=0.45\textwidth]{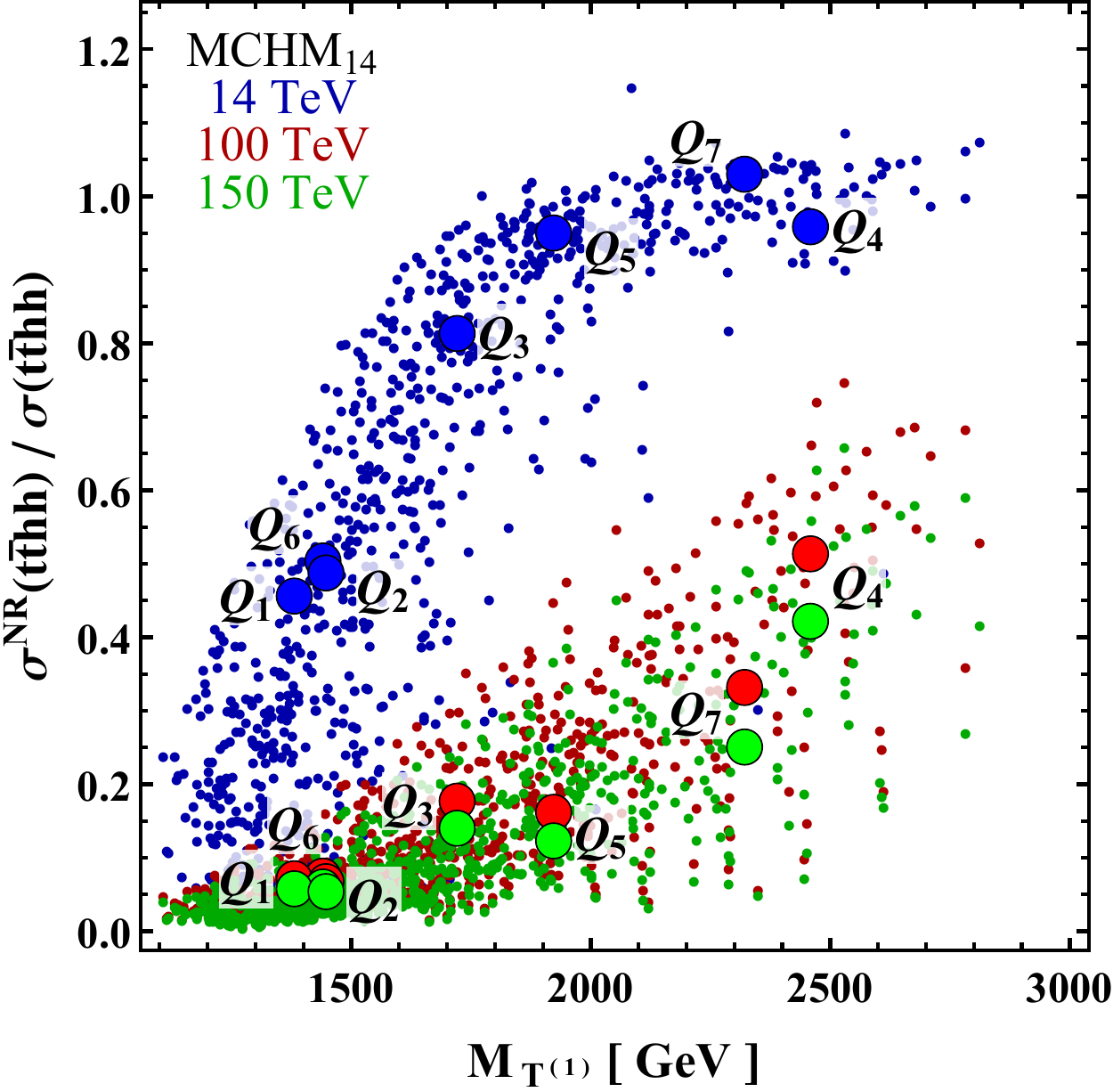}
\caption{Ratio between the non-resonant $t\overline{t}hh$ cross section and the total $t\overline{t}hh$ cross section as a function of the
lightest $Q = 2/3$ vector-like mass, for 14, 100 and 150~TeV CM energies.
The left (right) plots correspond to the MCHM$_5$ (MCHM$_{14}$).}
\label{fig:NRtthhvsMT4}
\end{figure}

To complete the results shown in Fig.~\ref{fig:tthhvsMT4M4}, Fig.~\ref{fig:NRtthhvsMT4} presents the ratio between the non-resonant contribution and the total cross section as a function of M$_{T^{(1)}}$ merging the points of all the corresponding regions for each of the MCHM cases. The trends are quite similar between each MCHM scenario with a larger dispersion of the points in the MCHM$_{14}$ (with again the caveat of 1000 scanned points for MCHM$_{14}$, versus 400 points for MCHM$_5$).

For heavier masses, the resonant production decreases and the total cross section can be dominated by the
non-resonant contribution.  Thus one expects deviations from the SM even when the resonances are rather heavy (say 3~TeV).

It is important to stress here that, except for cases with resonances close to the current direct search limit, we see that the non-resonant cross section
accounts for a significant fraction of the total cross section. It is
therefore of interest to search for deviations from the SM in this
quantity, in addition to the dedicated resonant searches.

\begin{figure}[h]
\centering
\includegraphics[width=0.45\textwidth]{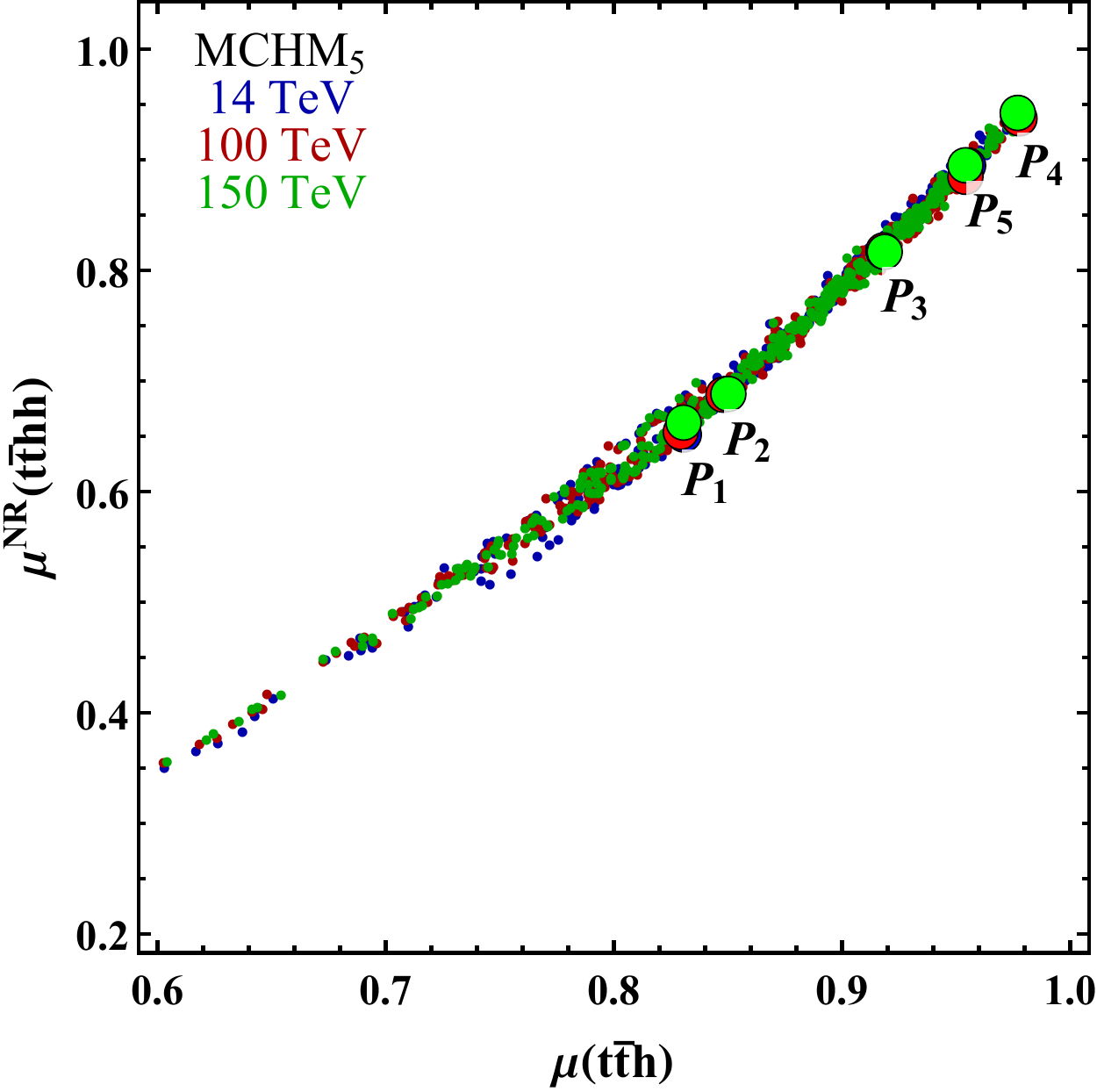}
\hspace{1cm}
\includegraphics[width=0.45\textwidth]{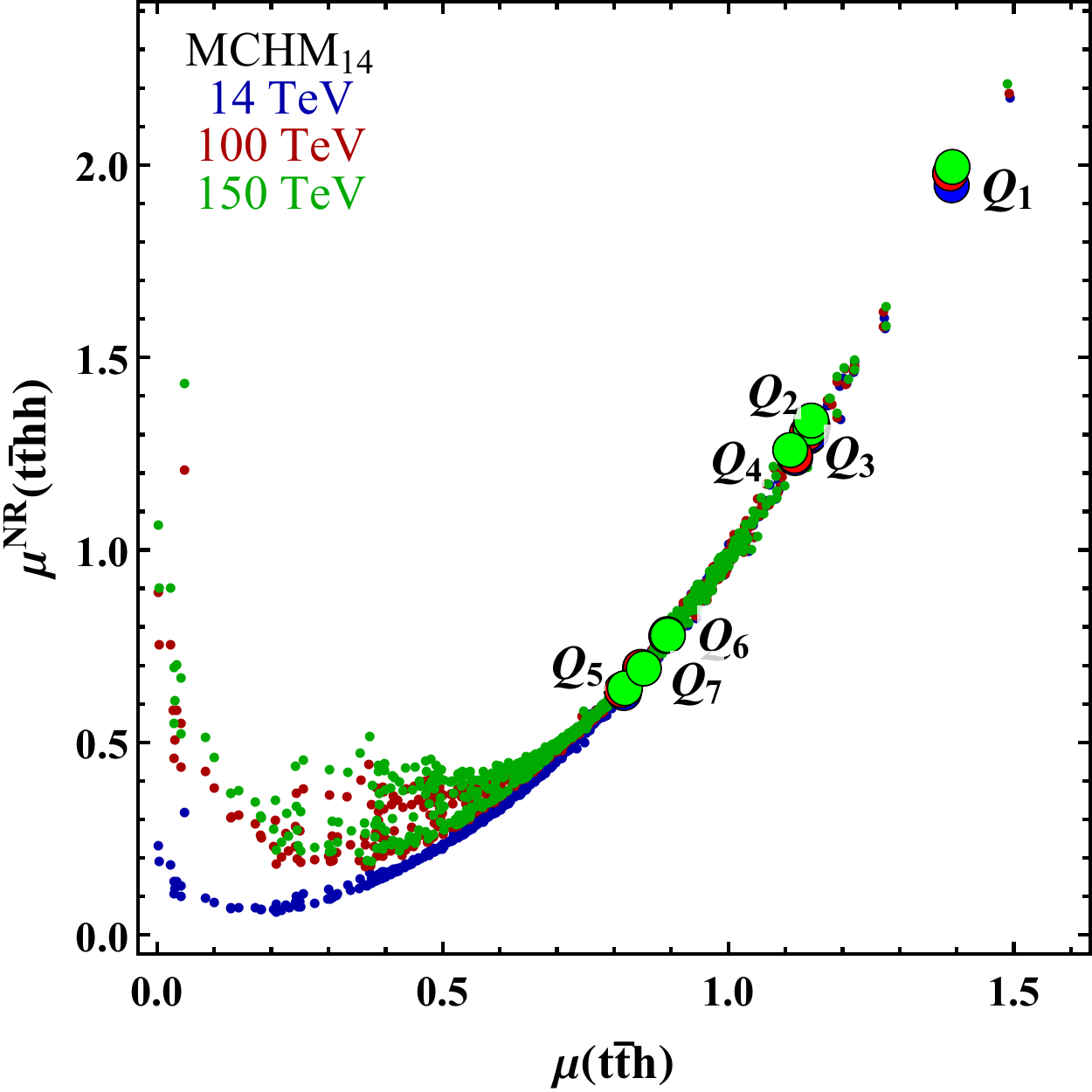}
\caption{Correlation between the normalized $t{\overline t}h$ and
non-resonant $t{\overline t}hh$ cross sections, for 14 and 100~TeV CM
energies. The left (right) plots correspond to the MCHM$_5$ (MCHM$_{14}$)}
\label{fig:nrtthhvstth}
\end{figure}

Finally, we show in Fig.~\ref{fig:nrtthhvstth} the NR-$t{\overline t}hh$
cross section (normalized to the SM $t\overline{t}hh$ cross section) as a
function of the normalized $t\overline{t}h$ cross section.  There is a
clear correlation, which reflects the fact that both are mainly
controlled by the top Yukawa coupling, as explained above. One can
note that the dispersion about the general trend is larger for MCHM$_{14}$ than for MCHM$_5$.

\subsection{Selection of some example points for each MCHM scenario}
\label{benchmarks}

In order to illustrate the physics of the MCHM scenarios we chose a number of
example-points.
The selection criteria in the $M_1$-$M_4$ plane (Fig.~\ref{fig:ytvsM1M4}), takes into account the
present experimental results including the LHC measurement of the $t{\overline t}h$ production process ~\cite{Aaboud:2018urx, Sirunyan:2018hoz}, its prospects at the start of the HL-LHC, after the end of Run 3 with 300 fb$^{-1}$ \cite{Cepeda:2019klc}, and the exclusion limits on the pair
production of heavy vector-like top partners currently obtained by ATLAS and CMS ~\cite{Aaboud:2018xuw,Aaboud:2018pii,Sirunyan:2018omb}. The plots resulting from the parameter
scans in section~\ref{scans} add an important input to this selection.

The MCHM parameters characterizing each of the example points and the main observable quantities are summarized in Table~\ref{fig:benchmarkTable1} and Table~\ref{fig:benchmarkTable2}, where we list the signal strengths for relevant energies, the spectra of vector-like fermionic resonances and the branching ratios of the lightest top partner.

\subsubsection{Selected example points and their main features for the MCHM$_5$}
\label{MCHM5ExamplePts}

Table~\ref{fig:benchmarkTable1} lists five selected points belonging either to Region I or Region II of the MCHM$_5$, with the $f$ scale ranging from about 900 GeV (``strong'' compositeness) up to 2.45 TeV
(``loose'' compositeness) and with different values of $y_L$. Note that similar scenarios
can be found in either region.
The chosen points are thus selected independently in one or the other case as reflected in Table~\ref{fig:benchmarkTable1}.

\begin{table}[!!th]
  \centering
  \scriptsize
  \begin{tabular}{|c|c|ccccc|}
    \cline{1-7}
    \multicolumn{2}{|c|}{} &  \cellcolor{tableblue} P$_1$ & \cellcolor{tablered} P$_2$ &  \cellcolor{tableblue} P$_3$ &  \cellcolor{tableblue} P$_4$ & \cellcolor{tablered} P$_5$ \\
    \hline
     \multirow{5}{*}{\rotatebox{90}{parameters}}
        & M$_1$(GeV) & -1317 & 800 & -960 & -3550 & 914 \\
        & M$_4$(GeV) & 1580 & 2311 & 1400 & 3000 & 2632 \\
        & f(GeV) & 969 & 896 & 1186 & 2450 & 1573 \\
        & y$_L$ & 1.66 & 1.80 & 0.88 & 1.00 & 2.36 \\
        & y$_R$ & 0.62 & 1.95 & 0.87 & 0.85 & 2.41 \\
    \hline
    \multicolumn{2}{|c|}{$\mu(\ttH)~\text{(All Energies)}$} & 0.83 & 0.85 & 0.92 & 0.98 & 0.96 \\
    \hline
    \multicolumn{2}{|c|}{$\mu(\ttHH)~\text{(14~TeV)}$} & 1.30 & 0.71 & 3.40 & 0.93 & 0.88 \\
    \multicolumn{2}{|c|}{$\mu(\ttHH)~\text{(100~TeV)}$} & 9.58 & 2.18 & 26.01 & 1.05 & 1.15 \\
    \hline
    \multicolumn{2}{|c|}{NR-\ttHH/\ttHH~(14~TeV)} & 0.50 & 0.97 & 0.24 & 1.01 & 1.02 \\
    \multicolumn{2}{|c|}{NR-\ttHH/\ttHH~(100~TeV)} & 0.07 & 0.32 & 0.03 & 0.90 & 0.77 \\
    \hline
    \multicolumn{2}{|c|}{$M_{T^{(1)}}~\text{(TeV)}$} & 1.44 & 1.83 & 1.34 & 3.00 & 2.61 \\
    \multicolumn{2}{|c|}{$M_{T^{(2)}}~\text{(TeV)}$} & 1.59 & 2.37 & 1.45 & 3.82 & 3.91 \\
    \multicolumn{2}{|c|}{$M_{T^{(3)}}~\text{(TeV)}$} & 2.25 & 2.83 & 1.76 & 3.99 & 4.56 \\
    \multicolumn{2}{|c|}{$M_{B^{(1)}}~\text{(TeV)}$} & 2.25 & 2.82 & 1.75 & 3.87 & 4.56 \\
    \multicolumn{2}{|c|}{$M_{X_{5/3}}~\text{(TeV)}$} & 1.58 & 2.31 & 1.40 & 3.06 & 2.63 \\
    \hline
    \multicolumn{2}{|c|}{$\Gamma_{T^{(1)}}$ (GeV)} & 24.7 & 95.2 & 4.1 & 26.7 & 16.5 \\
    \multicolumn{2}{|c|}{BR(T$^{\text{(1)}}$ $\to$th)} & 0.33 & 0.30 & 0.61 & 0.31 & 0.33 \\
    \multicolumn{2}{|c|}{BR(T$^{\text{(1)}}$ $\to$W$^+$b)} & 0.46 & 0.46 & 0.06 & 0 & 0.16 \\
    \multicolumn{2}{|c|}{BR(T$^{\text{(1)}}$ $\to$tZ)} & 0.22 & 0.20 & 0.26 & 0.30 & 0.26 \\
    \multicolumn{2}{|c|}{BR(T$^{\text{(1)}}$ $\to$W$^+$W$^-$t)} & 0 & 0.04 & 0.07 & 0.39 & 0.25 \\
    \hline
    \end{tabular}
    \caption{Properties of selected example points in the low scale MCHM$_5$. Red and blue column headings indicate points belonging to Region I and II respectively.}
    \label{fig:benchmarkTable1}
\end{table}

The first point in Table~\ref{fig:benchmarkTable1} ($P_1$) shows a non-resonant contribution accounting for almost half the
total cross section with a strong compositeness. The deficit in $\mu$($t{\overline t}h$) is relatively large
and a bit borderline in regards to the
estimated 1$\sigma$ uncertainty on this measurement by the end of Run 3 (with at least 300 fb$^{-1}$)~\cite{CMS:2018qgz, ATLAS:PhysPub}, and to the rather low masses of the two lightest heavy top partner and
the charge 5/3 resonance. This $P_1$-scenario will be fully scanned (also including its overall resonances spectrum) at the HL-LHC where a 1$\sigma$  uncertainty of $4.3\%$  is expected on $\mu$($t{\overline t}h$)~\cite{DiNardo:YR2019}.
On the contrary, the relatively slight increase of the $t{\overline t}hh$ production cross section
with respect to the SM at 14 TeV, might not be reachable even at high luminosity and may even be visible only when the discrepancy with the SM
further increases at higher CM energies. This makes this eventual scenario interesting to look at, even if
possibly quickly disregarded at a certain stage of the HL-LHC run.

Contrary to  point $P_1$,  point $P_2$ shows a high NR-$t{\overline t}hh$ and a clear deficit in $\mu$($t{\overline t}hh$); both effects are also visible for points $P_4$ and $P_5$. However, $P_2$, as it happens with $P_1$, presents a deficit in $\mu$($t{\overline t}h$) because both
have a rather low $f$ value (strong compositeness). Points $P_4$ and $P_5$, instead, have a high $f$ value which
translates into a $\mu$($t{\overline t}hh$) very close to 1.

Point $P_3$ has still a rather low $f$ value with, as striking features, the strong increase in $\mu$($t{\overline t}hh$) at the expense
of the low NR-$t{\overline t}hh$ contribution (see dominant decay of the lightest resonance into $th$) and $\mu$($t{\overline t}h$) getting
close to 1.  All the expected resonances, in this case, have relatively low mass well reachable at the HL-LHC (and even may be before, i.e. by end of the forthcoming Run 3). The HL-LHC increased luminosity will allow to measure the branching decay especially into $th$, predicted to be dominant with respect to $tZ$. Besides, the full HL-LHC dataset could indicate a possible excess in $\mu$($t{\overline t}hh$).

Points $P_4$ and $P_5$ are rather similar in terms of all the measurable quantities listed in this Table.
The NR-$t{\overline t}hh$ contribution is 100\% for both cases and will remain dominant even at higher energy accelerators. Moreover, while $\mu$($t{\overline t}h$) stays very close to 1 (due to a high $f$ value, especially for the point $P_4$), the deficit in
$\mu$($t{\overline t}hh$) could already be evidenced with the HL-LHC. Therefore, even if the points $P_4$ and $P_5$ look more like scenarios for the higher CM hadron colliders, a first breakthrough on such scenarios, especially for $P_5$ could be achieved by the end of
the HL-LHC. Finally, note that the lightest resonance in both cases has a low branching ratio into $Wb$
(especially $P_4$), whereas a more important 3-body decay. This enhanced $W^+\,W^-\,t$ channel occurs when $T^{(1)}$ comes from the fourplet and is thus almost degenerate with $X_{5/3}$ (both are controlled by $M_4$) as can be seen in the table. This effect will be specially important at higher energies, as we will discuss in detail in Sec. \ref{MCHMHighScale}.

It is interesting to note that, as expected in these models (see Subsections~\ref{MCHM14} and~\ref{SimpleObservables}), in all cases, the
resonances show a mass degeneracy between two or three of them (MCHM$_5$) or even more (MCHM$_{14}$),
in many cases due to EWSB. The separation in mass between those states can be of a few tens of GeV down to even a few hundreds of MeV.

The different scenarios described as example points for the  MCHM$_5$, present interesting features that
allow distinguishing them from each other. They represent a variety of cases, covering different
locations of the MCHM$_5$ parameter space; thus, they are  interesting for exploring this Minimal Composite Higgs Model.

\subsubsection{Selected example points and their main features for the MCHM$_{14}$}
\label{MCHM14ExamplePts}

The MCHM$_{14}$ parameter space involves four different cases in the $M_1$-$M_4$ plane (Fig.~\ref{fig:ytvsM1M4}).
A special attention is given to the Region III,  as it contains
the area with $\mu$($t{\overline t}h$) larger than 1. In this region, the
$M_9$ value ranges from 1.3 up to 4 TeV, whereas in the other ones the lowest $M_9$ value is 2 TeV. The first four points ($Q_1$ to $Q_4$) in Table~\ref{fig:benchmarkTable2} are the selected example points for this region.

\begin{table}[!!th]
  \centering
  \scriptsize
  \begin{tabular}{|c|c|cccc ccc|}
    \cline{1-9}
    \multicolumn{2}{|c|}{} & \cellcolor{tableorange} Q$_1$ & \cellcolor{tableorange} Q$_2$ & \cellcolor{tableorange} Q$_3$ & \cellcolor{tableorange} Q$_4$ &  \cellcolor{tablered} Q$_5$ & \cellcolor{tableblue} Q$_6$ & \cellcolor{tablecyan} Q$_7$\\
    \hline
     \multirow{6}{*}{\rotatebox{90}{parameters}}
        & M$_1$(GeV) & -1173 & -1054 & -1084 & -1579 & 976 & -1387 & 2998\\
        & M$_4$(GeV) & -1823 & -1826 & -1767 & -2512 & 1991 & 1443 & -2318\\
        & M$_9$(GeV) & 1382 & 1448 & 2036 & 2714 & 3096 & 3115 & 2875\\
        & f(GeV)     & 882 & 1032 & 1078 & 1298 & 1093 & 1865 & 1987\\
        & y$_L$      & 1.98 & 1.93 & 2.95 & 2.71 & 1.49 & 1.52 & 0.94\\
        & y$_R$      & 3.90 & 2.78 & 2.67 & 2.46 & 3.04 & 0.34 & 0.54\\

    \hline
    \multicolumn{2}{|c|}{$\mu(\ttH)~\text{(All Energies)}$} & 1.40 & 1.14 & 1.15 & 1.11 & 0.82 & 0.89 & 0.85\\
    \hline
    \multicolumn{2}{|c|}{$\mu(\ttHH)~\text{(14~TeV)}$} & 4.27 & 2.66 & 1.60 & 1.29 & 0.66 & 1.55 & 0.67\\
    \multicolumn{2}{|c|}{$\mu(\ttHH)~\text{(100~TeV)}$} & 27.70 & 19.32 & 7.46 & 2.42 & 3.93 & 10.36 & 2.09\\
    \hline
    \multicolumn{2}{|c|}{NR-\ttHH/\ttHH~(14~TeV)} & 0.46 & 0.49 & 0.81 & 0.96 & 0.95 & 0.50 & 1.03\\
    \multicolumn{2}{|c|}{NR-\ttHH/\ttHH~(100~TeV)} & 0.07 & 0.07 & 0.18 & 0.51 & 0.16 & 0.07 & 0.33\\
    \hline
    \multicolumn{2}{|c|}{$M_{T^{(1)}}~\text{(TeV)}$} & 1.38 & 1.45 & 1.72 & 2.46 & 1.92 & 1.44 & 2.32\\
    \multicolumn{2}{|c|}{$M_{T^{(2)}}~\text{(TeV)}$} & 1.38 & 1.45 & 2.01 & 2.70 & 2.47 & 1.52 & 2.82\\
    \multicolumn{2}{|c|}{$M_{T^{(3)}}~\text{(TeV)}$} & 1.41 & 1.46 & 2.04 & 2.71 & 3.09 & 2.96 & 2.87\\
    \multicolumn{2}{|c|}{$M_{B^{(1)}}~\text{(TeV)}$} & 1.38 & 1.45 & 2.02 & 2.70 & 2.53 & 2.98 & 2.84\\
    \multicolumn{2}{|c|}{$M_{X^{(1)}_{5/3}}~\text{(TeV)}$} & 1.38 & 1.45 & 1.77 & 2.51 & 1.99 & 1.44 & 2.32\\
    \hline
     \multicolumn{2}{|c|}{$\Gamma_{T^{(1)}}$ (GeV)} & 12.2 & 7.8 & 55.2 & 121.1 & 54.9 & 9.5 & 24.2 \\
    \multicolumn{2}{|c|}{BR(T$^{\text{(1)}}$ $\to$th)} & 0.39 & 0.28 & 0.44 & 0.38 & 0.50 & 0.42 & 0.37\\
    \multicolumn{2}{|c|}{BR(T$^{\text{(1)}}$ $\to$W$^+$b)} & 0.35 & 0.48 & 0.14 & 0.13 & 0.12 & 0.15 & 0.02\\
    \multicolumn{2}{|c|}{BR(T$^{\text{(1)}}$ $\to$tZ)} & 0.16 & 0.13 & 0.30 & 0.27 & 0.23 & 0.34 & 0.35\\
    \multicolumn{2}{|c|}{BR(T$^{\text{(1)}}$ $\to$W$^+$W$^-$t)} & 0.09 & 0.09 & 0.12 & 0.22 & 0.14 & 0.08 & 0.27\\
    \hline
    \end{tabular}
    \caption{Properties of selected example points in the low scale MCHM$_{14}$. Column headings indicate region, with red and orange meaning respectively Regions I and III (with same sign $M_1$ and $M_4$) and blue and cyan respectively for regions II and IV (with opposite sign $M_1$ and $M_4$).}
    \label{fig:benchmarkTable2}
\end{table}

The selection of the MCHM$_{14}$ points in the Region III requests, in addition to the criteria listed at the beginning of Subsection~\ref{benchmarks}, $\mu$($t{\overline t}h$) to be larger than 1,
this is the main difference between the two MCHM scenarios and also an important observable for the exclusion of the MCHM$_5$. Note that all these points correspond to a relatively low $f$ value, i.e., high compositeness.

The two first points are rather similar; they both correspond to low $f$ and $M_9$ values. $Q_1$ is close to the $1 \sigma$ experimental limits w.r.t the $\mu$($t{\overline t}h$) value. The three lightest associated resonances are already almost within the limits published by ATLAS and CMS. However, this point is a good example of a high increase in $\mu$($t{\overline t}hh$) (like also observed in some examples of the MCHM$_5$) and it includes 50\% of non-resonant contribution at 14 TeV. If one disregards the resonant contributions, it primarily differs from an equivalent MCHM$_5$ scenario by the increase in $\mu$($t{\overline t}h$), as compared to the SM.

The model parameters of point $Q_2$ have values very close to the ones of $Q_1$ but a slightly higher $f$ value (lower compositeness); it thus translates into a smaller $\mu$($t{\overline t}h$) value. Moreover, $\mu$($t{\overline t}hh$) decreases compared to $Q_1$, although it still stays relatively high.

Points $Q_3$ and $Q_4$, have both a relatively low $f$ value but higher $M_9$ values with about 2 TeV (for $Q_3$) and 2.7 TeV (for $Q_4$). Both have $\mu$($t{\overline t}h$) greater than 1 but well within the current experimental limits. Their selection is also based on the request for a high non-resonant contribution.

For completeness, the remaining three regions of the $M_1$-$M_4$ parameter space were considered. In each of them, a representative point is selected as summarized in Table~\ref{fig:benchmarkTable2}: The point $Q_5$ in Region I, $Q_6$ in Region II and $Q_7$ in Region IV.

In the overall covered space these 3 regions provide quite similar cases. Moreover, as shown in Fig.~\ref{fig:ytvsM1M4}, a subregion of Region IV is excluded because of the constraint on the $ggh$ coupling ($0.8<ggh/\text{SM}<1.2$). The three example points have different $f$ values (around 1 TeV for $Q_5$ and around 2 TeV for $Q_6$ and $Q_7$). They all have a relatively large $M_9$ (3 TeV).
The example points in these three regions show a $\mu$($t{\overline t}h$) value smaller than one and no strong increase in $\mu$($t{\overline t}hh$). The selected points $Q_5$ and $Q_7$ have large NR-$t{\overline t}hh$ contributions. Point $Q_6$ has only 50\% NR-$t{\overline t}hh$ contribution. All the NR-$t{\overline t}hh$ relative contributions decrease sharply at higher CM energies, as more phase space becomes available for the production of resonances.

The use of these preliminary observables shows that it will be difficult to disentangle between both MCHM scenarios if a deficit in $\mu$($t{\overline t}h$) is measured. A much more detailed analysis will be required. In some cases, the HL-LHC will perhaps provide a first indication, but a potential discovery will likely need
higher energy together with higher luminosity.

\subsubsection{NR-$t{\overline t}hh$ contributions in the MCHM$_5$ and the MCHM$_{14}$ }
\label{NRtthhMCHM}

In order to clarify the different contributions to the non resonant $t{\overline t}hh$ production, we simulated these contributions separately and summarized the results in Tables~\ref{fig:NRtthhMCHM5Table} and~\ref{fig:NRtthhMCHM14Table}. The ratios in those tables are obtained by turning off one or more couplings in the model in order to disregard particular classes of diagrams, which are indicated in the table.

\begin{table}[h]
\centering
    \resizebox{\linewidth}{!}{%
    \begin{tabular}{|cc|c|c|c|c|c|cc|}
    \hline
     &&P$_1$&P$_2$&P$_3$&P$_4$&P$_5$&\multicolumn{2}{c|}{Disregarded diagrams}\T\B\\
    \hline
    $\sigma_{\cancel{hh}}/\sigma_{\mbox{\tiny NR}}^{\mbox{\tiny \ttHH}}$ &(14~TeV) & 1.05 & 1.04 & 1.03 & 1.01 & 1.01  & \multirow{2}{*}[5mm]{\includegraphics[width=20mm, height=20mm]{Figures/tthh-Diagram-3.pdf}} & \rule{0pt}{10mm}\\
    $\sigma_{\cancel{hh}}/\sigma_{\mbox{\tiny NR}}^{\mbox{\tiny \ttHH}}$ &(100~TeV) & 1.05 & 1.03 & 1.03 & 1.01 & 1.01 & & \rule[-9mm]{0pt}{0pt}\\
    \hline
    $\sigma_{\text{Yuk}}/\sigma_{\mbox{\tiny NR}}^{\mbox{\tiny \ttHH}}$ &(14~TeV) & 0.86 & 0.85 & 0.84 & 0.82 & 0.82   &
    \multirow{2}{*}[5mm]{
    \includegraphics[width=20mm, height=20mm]{Figures/tthh-Diagram-3.pdf}}
    &
    \multirow{2}{*}[5mm]{
    \includegraphics[width=20mm, height=20mm]{Figures/tthh-Diagram-2.pdf}} \rule{0pt}{10mm}\\
    $\sigma_{\text{Yuk}}/\sigma_{\mbox{\tiny NR}}^{\mbox{\tiny \ttHH}}$ &(100~TeV) & 0.87 & 0.87 & 0.87 & 0.85 & 0.85   & & \rule[-9mm]{0pt}{0pt}\\
    \hline
    $\sigma^{\mbox{\tiny \ttHH}}_{\mbox{\tiny NR}}/\sigma^{\mbox{\tiny \ttHH}}_{\mbox{\tiny SM}} $& (14 TeV) & 0.65 & 0.69 & 0.82 & 0.94 & 0.90 &  &  \rule{0pt}{8mm}\\

    $\sigma^{\mbox{\tiny \ttHH}}_{\mbox{\tiny NR}}/\sigma^{\mbox{\tiny \ttHH}}_{\mbox{\tiny SM}} $& (100 TeV) & 0.65 & 0.69 & 0.82 & 0.93 & 0.89   & & \\

    $(y_t/y_t^{\mbox{\tiny SM}})^4$& & 0.69 & 0.72 & 0.85 & 0.95 & 0.91   & & \rule[-5mm]{0pt}{0pt}
    \\
    \hline
    \end{tabular}}
\caption{Study of NR-$t{\overline t}hh$ for the MCHM$_5$ points in Table~\ref{fig:benchmarkTable1}. The cross sections $\sigma_{\cancel{hh}}$ and $\sigma_{\mbox{\tiny Yuk}}$ are obtained by disregarding the classes of diagrams on the last column and $\sigma_{\mbox{\tiny NR}}$ is the total NR-$t{\overline t}hh$. The LO SM $t{\overline t}hh$ production is indicated by $\sigma^{\mbox{\tiny SM}}$ and  $\sigma^{\mbox{\tiny SM}}_{\mbox{\tiny Yuk}}$ means we disregarded the SM trilinear Higgs coupling. The top Yukawa couplings are indicated by $y_t$ and $y_t^{SM}$ in the MCHM and SM respectively.}
\label{fig:NRtthhMCHM5Table}
\end{table}
\begin{table}[h]
\centering
\begin{tabular}{|p{15mm} p{17mm}| c c c c c c c|}
\hline
    &&Q$_1$&Q$_2$&Q$_3$&Q$_4$&Q$_5$&Q$_6$&Q$_7$\T\B\\
    \hline
    $\sigma_{\cancel{hh}}/\sigma_{\mbox{\tiny NR}}^{\mbox{\tiny \ttHH}}$& (14~TeV) & 0.95 & 0.97 & 0.96 & 0.98 & 1.06 & 1.03 & 1.05 \T\\
    $\sigma_{\cancel{hh}}/\sigma_{\mbox{\tiny NR}}^{\mbox{\tiny \ttHH}}$& (100~TeV) & 0.93 & 0.96 & 0.95 & 0.96 & 1.05 & 1.03 & 1.05 \B\\
    \hline
    $\sigma_{\text{Yuk}}/\sigma_{\mbox{\tiny NR}}^{\mbox{\tiny \ttHH}}$& (14~TeV) & 0.81 & 0.82 & 0.81 & 0.81 & 0.86 & 0.84 & 0.85 \T\\
    $\sigma_{\text{Yuk}}/\sigma_{\mbox{\tiny NR}}^{\mbox{\tiny \ttHH}}$& (100~TeV) & 0.82 & 0.83 & 0.83 & 0.83 & 0.87 & 0.86 & 0.87 \B\\
    \hline
    $\sigma^{\mbox{\tiny \ttHH}}_{\mbox{\tiny NR}}/\sigma^{\mbox{\tiny \ttHH}}_{\mbox{\tiny SM}} $& (14~TeV) & 1.94 & 1.29 & 1.31 & 1.25 & 0.63 & 0.78 & 0.69 \T\\
    $\sigma^{\mbox{\tiny \ttHH}}_{\mbox{\tiny NR}}/\sigma^{\mbox{\tiny \ttHH}}_{\mbox{\tiny SM}} $& (100~TeV) & 1.98 & 1.30 & 1.32 & 1.25 & 0.64 & 0.78 & 0.69 \B\\
    $(y_t/y_t^{\mbox{\tiny SM}})^4$& & 1.94 & 1.30 & 1.31 & 1.24 & 0.67 & 0.80 & 0.72
    \\
    \hline
\end{tabular}
\caption{Study of NR-$t{\overline t}hh$ for the MCHM$_{14}$ points in Table~\ref{fig:benchmarkTable2}. The cross sections $\sigma_{\cancel{hh}}$ and $\sigma_{\mbox{\tiny Yuk}}$ are obtained by disregarding the classes of diagrams shown on Table~\ref{fig:NRtthhMCHM5Table} and $\sigma_{\mbox{\tiny NR}}$ is the total NR-$t{\overline t}hh$. The LO SM $t{\overline t}hh$ production is indicated by $\sigma^{\mbox{\tiny SM}}$ and  $\sigma^{\mbox{\tiny SM}}_{\mbox{\tiny Yuk}}$ means we disregarded the SM trilinear Higgs coupling. The top Yukawa couplings are indicated by $y_t$ and $y_t^{SM}$ in the MCHM and SM respectively.}
\label{fig:NRtthhMCHM14Table}
\end{table}

The $\sigma_{\cancel{hh}} / \sigma_{\mbox{\tiny NR}}$ ratios show that the effects of the double Higgs Yukawa coupling are typically at the couple to few percent level in the MCHM$_5$ and the MCHM$_{14}$ and hardly show any variation with CM energy increase.
We also find, by examining the $\sigma_{\mbox{\tiny Yuk}} /\sigma_{\mbox{\tiny NR}}$ ratios, that the effect of the trilinear Higgs self-interaction can be around $15\%$ in both MCHM$_5$ and MCHM$_{14}$.  For comparison, the effect of the trilinear Higgs self-interaction in the SM $t\overline{t}hh$ cross section
is about $20\%$, with a very mild CM energy dependence. Thus, it is largely the top Yukawa that governs the NR-$t\overline{t}hh$ (just as in the SM), which, to a first approximation, then scales as $(y_t / y_t^ {\rm SM})^4$. This correlation explains the behavior in Fig.~\ref{fig:htdist} and is evident in the last three lines of Tables~\ref{fig:NRtthhMCHM5Table} and~\ref{fig:NRtthhMCHM14Table}, where the enhancement or suppression in the cross section directly arises from the change in the top Yukawa.

The low contributions of the trilinear Higgs and the double Higgs Yukawa couplings present challenges. It makes the measurement of these two couplings harder, and both are important to characterize the compositeness of the Higgs (as opposed to the spectra of fermionic resonances) and are even harder to probe on the resonant production. Furthermore the shape of all kinematic distributions of the  NR-$t\overline{t}hh$ final states will be almost identical in shape to the SM one, changing only on the magnitude of the integrated cross section. The top Yukawa can be more directly accessed in the $t\overline{t}h$ channel, but it will be necessary to develop combined analyses between the $t\overline{t}h$ and the NR-$t\overline{t}hh$ to isolate the non-resonant contribution and extract information about these couplings.

Let's stress here a key-role of $t\overline{t}hh$ production process in the study of the self Higgs coupling and the important interplay between the measurement of the $\kappa_\lambda$, the tri-linear Higgs coupling normalized to the SM value, through the $hh$ process (both gluon fusion and Vector Boson Fusion) and of the  $t\overline{t}hh$ process. Even if the trilinear Higgs contribution to the $t\overline{t}hh$ process represents only about 15\% (slightly lower than in the SM case) of the total cross-section in the MCHM considered scenarios, it is not as difficult to access as the double Higgs Yukawa contribution at the percent level, and indeed it is worth it. The measurement of the $\kappa_\lambda$ parameter of the Higgs sector via several processes is becoming of increasing importance as it remains the experimentally least constrained Higgs parameter. This is due to the fact that the ``traditional'' way to access it at LHC, via the $hh$ production, is currently challenging, because of the still relatively low cross-sections and signature efficiency. In the years to come and over the whole HL-LHC era this will be indeed an essential experimental goal. In order to increase the experimental sensitivity reach, ATLAS and CMS experiments are already now searching, in addition to the $hh$ production through the gluon fusion process ~\cite{ATLAS:2020hh, CMS:2020hh}, for the production process through the Vector Boson Fusion, VBF($hh$) ~\cite{herrero:2019hhjj} and ~\cite{ATLAS:2020vbf}. Recently the need to look for the complementary contribution of $t\overline{t}hh$ on this specific topic is outlined both in view of the HL-LHC and even more of the FCC-hh at 100 TeV ~\cite{Englert:2015tthh, Durham:2019tthh}. It is worth noting that the SM cross-section of the VBF($hh$) and of $t\overline{t}hh$ processes are of the same order, i.e. at the fb level at 14 TeV. Besides, while both ATLAS and CMS are conducting searches for the VBF ($hh$), a search for the $t\overline{t}hh$ production has been carried on, for the first time, at CMS ~\cite{LeonidasPhD}, stressing the growing interest on the experimental side.

To conclude, from the phenomenological viewpoint the $t\overline{t}hh$  channel does not include destructive interference among diagrams unlike in the $hh$ case (see e.g. \cite{Gillioz_2012}), and if $\lambda > \lambda_{SM}$, it provides the leading channel where to observe an excess over the SM expectation. From the experimental viewpoint, the two tops in addition to the Higgs pair strengthen the signal efficiency with respect to the $hh$ and $hhjj$ signatures, making it accessible already now at LHC and furthermore at HL-LHC. Besides, the large increase with energy of its cross-section in the SM and even more in the MCHM case makes it an essential channel to study at FCC-hh for high precision and BSM measurements of the Higgs sector.

As a final remark, we wish to point out that the reader should be careful when looking for the separation between the resonant and the non-resonant production in Fig.~\ref{fig:htdist}. The separation looks clear for $\sqrt{s} = 14$ TeV and difficult for $\sqrt{s} = 100$ TeV, giving the wrong impression that there is little hope for exploring the NR-$t\overline{t}hh$ at future accelerators. That only happens because the same point in parameter space was used for both plots, and that is a point with low lying fermionic resonances which are produced abundantly at higher energies and dominate over everything else.
We explored also the hypothetical situation where no resonance was found at the end of the HL-LHC run, forcing us into regions of the parameter space where the top partners are heavier and the non-resonant production is more pronounced, as we shall see in Section~\ref{MCHMHighScale}.

\subsection{Cluster Analysis applied to the MCHM at low scale}
\label{clusteringLScale}

In the next two subsections we apply the clustering idea to the \five and the MCHM$_{14}$, using the parton level kinematic distributions of the $t\overline{t}hh$ process to do the cluster analysis described in Section~\ref{selectedpts}.

\subsubsection{Clustering of the MCHM$_5$}
\label{clusteringLS5}

In the case of the MCHM$_5$ we start with the points generated by the scan of Sec.~\ref{scans}: 400 points divided between Regions I and II. We then apply the following ``cuts'' to remove points that are already constrained at a $3\sigma$ level\footnote{
Neither CMS nor ATLAS report the 3$\sigma$ error directly, so the best we can do here is to assume a Gaussian error and estimate the  $3\sigma$ threshold simply by multiplying their $1\sigma$ intervals by 3, using the ATLAS/CMS combined value when available and the smallest one otherwise.}
\bea
0.33 \leq \mu(t\overline{t}h) \leq 2.07,
\label{const.tth}
\eea
\bea
M_{T^{(1)}} \geq 1.3 \text{ TeV}.
\label{const.m}
\eea

We also check if the points in the scan are excluded or not by the experimental measurements of the $t\overline{t}Z$ and $t\overline{t}W$ production cross sections. The latest measurement of the $t\overline{t}Z$ production, performed by the CMS collaboration, corresponding to an integrated luminosity of $77.5 \text{ fb}^{-1}$ and at $1\sigma$ level is $\sigma(t\overline{t}Z)=0.95\pm 0.08$ pb~\cite{Sirunyan_2020}. The $t\overline{t}Z$ signal strength was calculated dividing the measured cross section by the SM prediction $\sigma^{\text{SM}}=0.84~\pm~0.10$ pb, obtaining $\mu(t\overline{t}Z)=1.13\pm 0.16$. The latest measurement of the signal strength of the $t\overline{t}W$ production, performed by the ATLAS collaboration, corresponding to an integrated luminosity of $36.1~ \text{fb}^{-1}$ is $\mu(t\overline{t}W)=1.44\pm 0.32$ at $1\sigma$ level~\cite{Aaboud:2019njj}. All the scanned points survived these constraints at $3\sigma$ level\footnote{
Once again 3$\sigma$ is roughly estimated as three times the 1$\sigma$ intervals.
}.

That leaves us with 348 points at the start of the clustering algorithm.
Each one of these points was implemented in MadGraph 5 and events were generated for the production of $t\overline{t}hh$ at a fixed luminosity of 3000~fb$^{-1}$. At parton level we have only three different particles present: the top, the anti-top and two Higgs bosons (we focused on the most energetic one). Using MadAnalysis we obtained histograms for the following kinematic distributions:
\begin{itemize}
    \item invariant mass of the top/Higgs pair: $M[t,h_1]$;
    \item transverse momenta: $p_T[t]$ and $p_T[h_1]$;
    \item angular distances in the transverse plane: $\Delta_R[t,h_1]$ and $\Delta_R[t,\overline{t}]$
    \item angular distance to the beam axis: $\theta[t]$ and $\theta[h_1]$.
\end{itemize}

We then constructed samples from these distributions using single distributions or combinations containing two or three distributions. Since the main contribution to the total $t\overline{t}hh$ cross section is coming from the decays of top partners, all these kinematic variables are strongly correlated and that means that most combinations led to very similar clusters.

Initially one is led to consider only the invariant mass, as this is the one variable that makes the resonant structure evident, but we find that using only $M[t,h_1]$ for the clustering puts too much weight on the exact position of the peaks produced by the resonances, instead of more general behaviours like the two peak structure that shows up in the $p_T$ distribution of the Higgs. The result would be to have many benchmark points in the region with lighter and narrower resonances, and just one or two in the rest of the parameter space. Including at least one of the angular variables brings extra physical information and takes away that emphasis, resulting in benchmark samples that are more evenly spread. There is very little difference in regards to what angular distribution we choose, but the results with $\theta[t]$ resulted in samples being more evenly distributed among clusters.

We also experimented with the number of clusters and found out that with a small number of clusters ($N_{cluster} < 10$) we obtain one highly populated cluster that contains all the samples characterized by heavier ($M_{T^{(1)}} \gtrsim 1.5$ TeV) and broader resonances. This cluster results from the merger of two large clusters when we go from 11 to 10 clusters, but these two are already generated early on the clustering process, which means that stopping with $N_{cluster}$ much bigger than 11 only changes the low population clusters, that are already very homogeneous, so there is not much gain in increasing $N_{cluster}$.

\begin{figure}[h]
\centering
\includegraphics[width=0.95\textwidth]{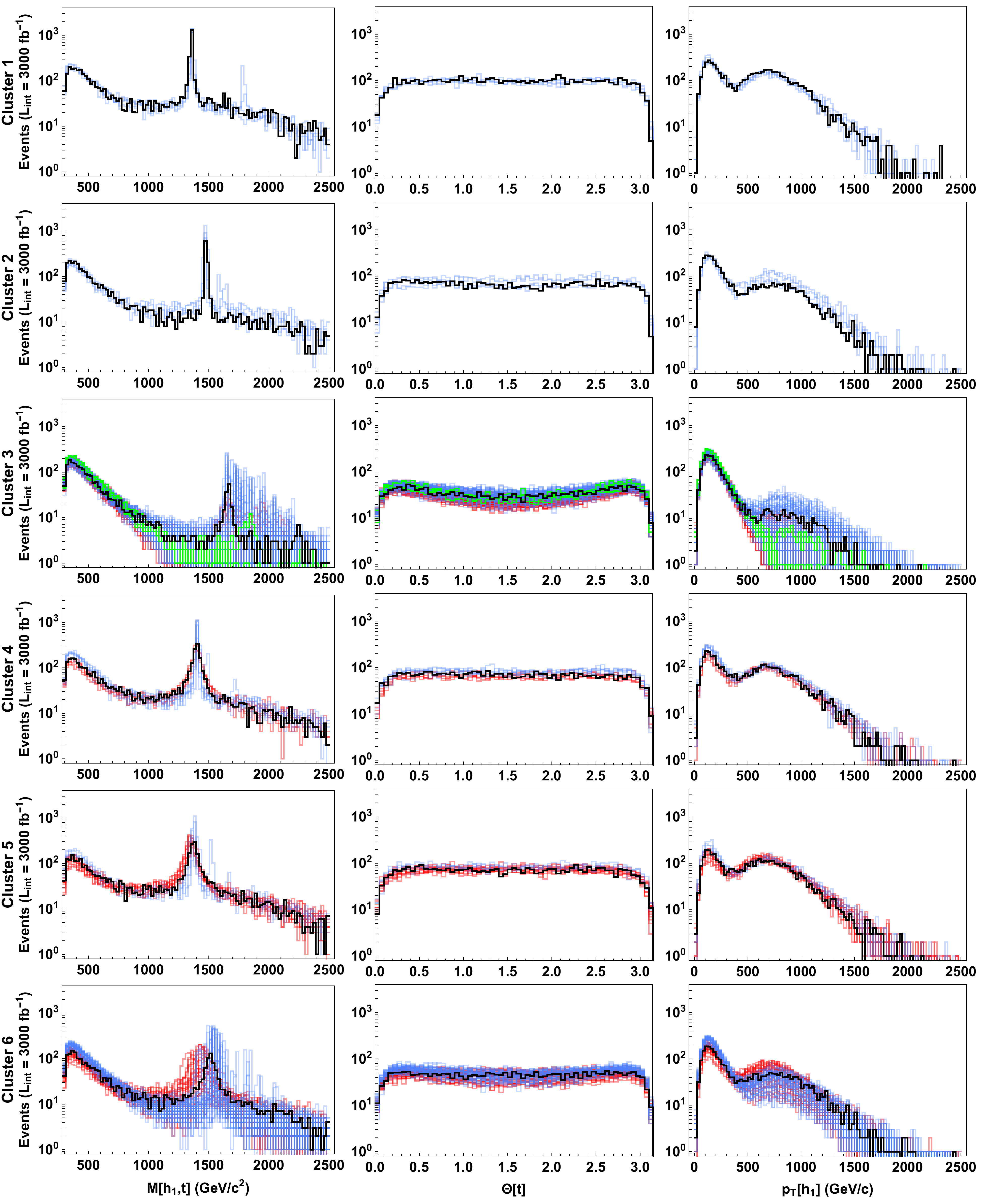}
\caption{Distributions at $\sqrt{s} = 14$ TeV for the invariant mass of the pair composed by top and most energetic Higgs (first column), angular distribution of the top (second column) and transverse momenta of the most energetic Higgs, organized in clusters by similarity (clusters 1 through 6). Red and blue curves indicate respectively points in Regions I and II of the MCHM$_5$, the benchmark point for each cluster is shown in black and the example points of table~\ref{fig:benchmarkTable1} are shown in green (cluster 3 contains $P_2$, $P_4$ and $P_5$). }
\label{bigclusterfig.mchm5.p1}
\end{figure}

\begin{figure}[h]
\centering
\includegraphics[width=0.95\textwidth]{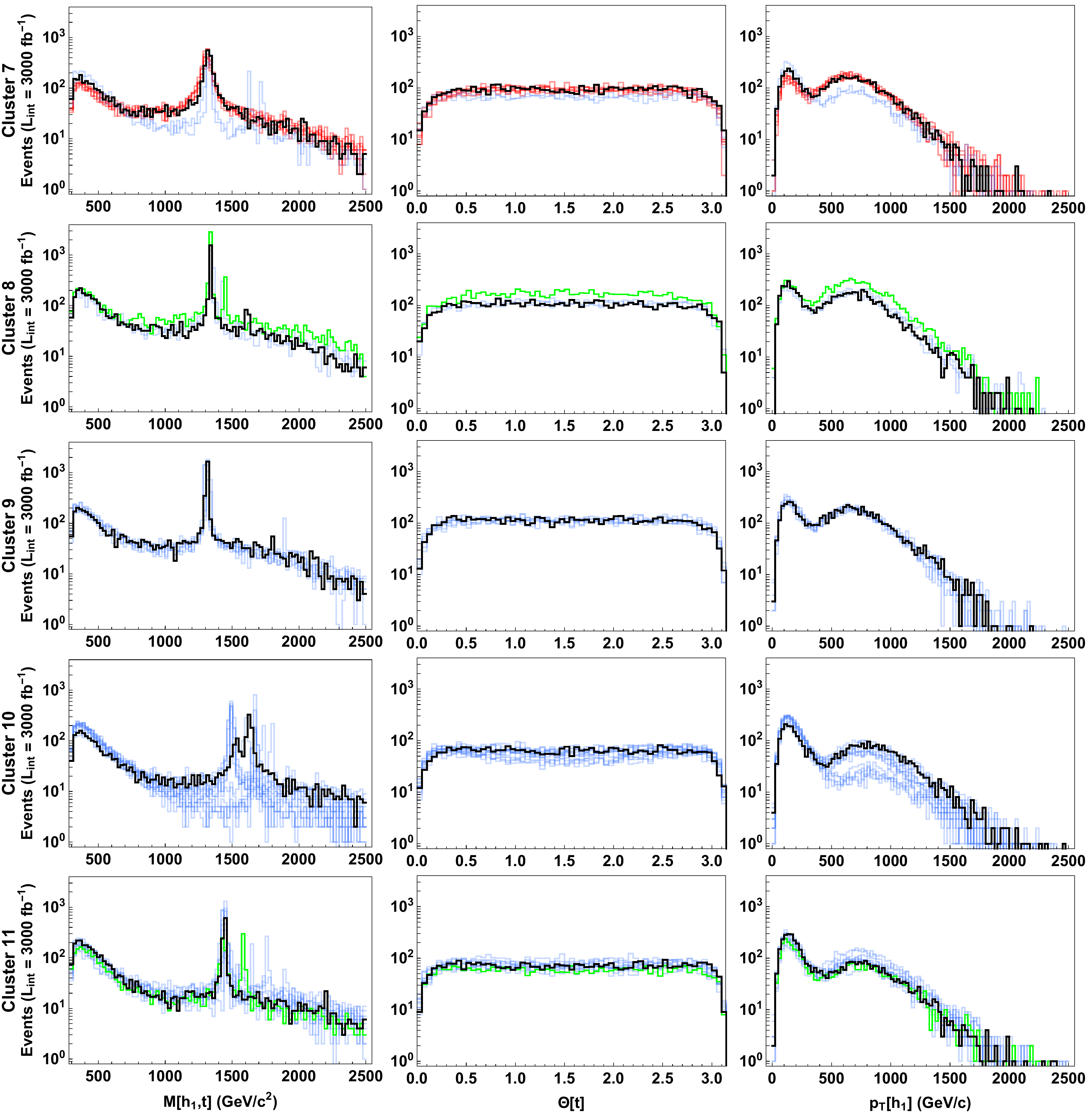}
\caption{Distributions at $\sqrt{s} = 14$ TeV for the invariant mass of the pair composed by top and most energetic Higgs (first column), angular distribution of the top (second column) and transverse momenta of the most energetic Higgs, organized in clusters by similarity (clusters 7 through 11). Red and blue curves indicate respectively points in Regions I and II of the MCHM$_5$, the benchmark point for each cluster is shown in black and the example points of Table~\ref{fig:benchmarkTable1} are shown in green ($P_3$ is in cluster 8 and $P_1 $ is in cluster 11). }
\label{bigclusterfig.mchm5.p2}
\end{figure}

We finally decided to stick with $N_{cluster} = 11$ and on using $M[t,h_1]$ and $\theta[t]$ for the cluster analysis of the \five, the resulting clusters are shown on Figs.~\ref{bigclusterfig.mchm5.p1} and~\ref{bigclusterfig.mchm5.p2},
where we also included $p_T[h_1]$ distributions to show their typical behaviour. We first note that most of the clusters are very homogeneous in distributions, in the sense that all curves in each plot are very similar. The benchmark points (black lines in the plots) will then be very good representatives of the behaviour of each cluster. This is true even for the $p_T[h_1]$ distributions,
which were not used as a criteria for clustering, and also for all the distributions we have checked (listed above). The only striking exception is cluster 3 (in Fig.~\ref{bigclusterfig.mchm5.p1}), which contains all samples with heavy fermionic resonances, with $M_{T^{(1)}} \gtrsim 1.65$ TeV, or no resonances at all decaying to $t\overline{t}hh$. That is a consequence of the comparatively low count of events in the resonant region, when compared to the non-resonant part of the distribution, which is very similar to them all. This could probably be fixed by doing a dedicated scan for points in that region and a separate clusterization, producing more clusters and benchmark points. We decided against it because we already covered that region with example points $P_2$, $P_4$ and $P_5$ of table~\ref{fig:benchmarkTable1}, which were in fact grouped in cluster 3 and are shown as green curves in Fig.~\ref{bigclusterfig.mchm5.p1}.


Figures~\ref{bigclusterfig.mchm5.p1} and~\ref{bigclusterfig.mchm5.p2} also allow us to survey the main features of the whole parameter space of the model. In particular one can see that points in Region I (colored in red) got separated from those of Region II (in blue). Region I is concentrated in clusters 3 to 7, while Region II dominates the rest of the clusters. One can also verify that the fermionic resonances produced in Region I are generally wider than resonances in Region II, this can be more clearly seen in clusters 4, 5 and 7 where resonances from both regions overlap.  In order to understand this, it is useful to look at an approximate expression for the top mass, given in Eq. (\ref{mtapprox}). This mass is generated in the MCHM$_5$ by mixing of the elementary fermion fields with the fourplet and singlet resonances. Looking at the first two diagrams in Fig. \ref{fig:partialcomp}, as well as the Lagrangian of Eq. (\ref{Lmix5}), we see this mixing leads to an insertion of $y_L$ times $y_R$, times a mass insertion of $M_1$ or $M_4$ for chirality. These diagrams must interfere destructively, since the pNGB Higgs vacuum misalignment is generated by $SO(5)$ breaking and hence must vanish in the $SO(5)$ symmetric $M_4 = M_1$ limit. This leads to a dependence $m_t \sim y_L y_R |M_4 - M_1|$, as shown explicitly in Eq. (\ref{mtapprox}). In Region I, there is a cancellation between same sign $M_1$ and $M_4$, such that larger values of $y_{L,R}$ are typically needed to generate the top mass\footnote{This is also responsible for the white region in Fig. \ref{fig:ytvsM1M4}, where the top mass cannot be reached without breaking our perturbativity bound on $y_{L,R}$.}. This enhanced mixing also leads, upon mass diagonalization, to a greater value for the $t\,  T^{(1)}\, h$ vertex, and hence, to wider resonances.

Another interesting general feature is the presence in many cases of more than one peak in the $M[h_1,t]$ distribution. While cluster 9 represents well the usual simplifying assumption used in top partner searches, namely the presence of only one resonance decaying to the $t h$, $t Z$ or $b W$ channels, many of the other clusters contain a sizeable presence of more complicated peak structures, coming specially from Region II. Cluster 10 is the perfect example of this, as its benchmark point has a double peak structure with the second resonance giving a stronger contribution than the lightest one. Most exclusion limits for top partners are obtained through analyses optimized for the situation in cluster 9, and it would be interesting to see how those limits change if more resonances are considered, specially if they overlap significantly.

\begin{figure}[h]
\centering
\includegraphics[width=0.45\textwidth]{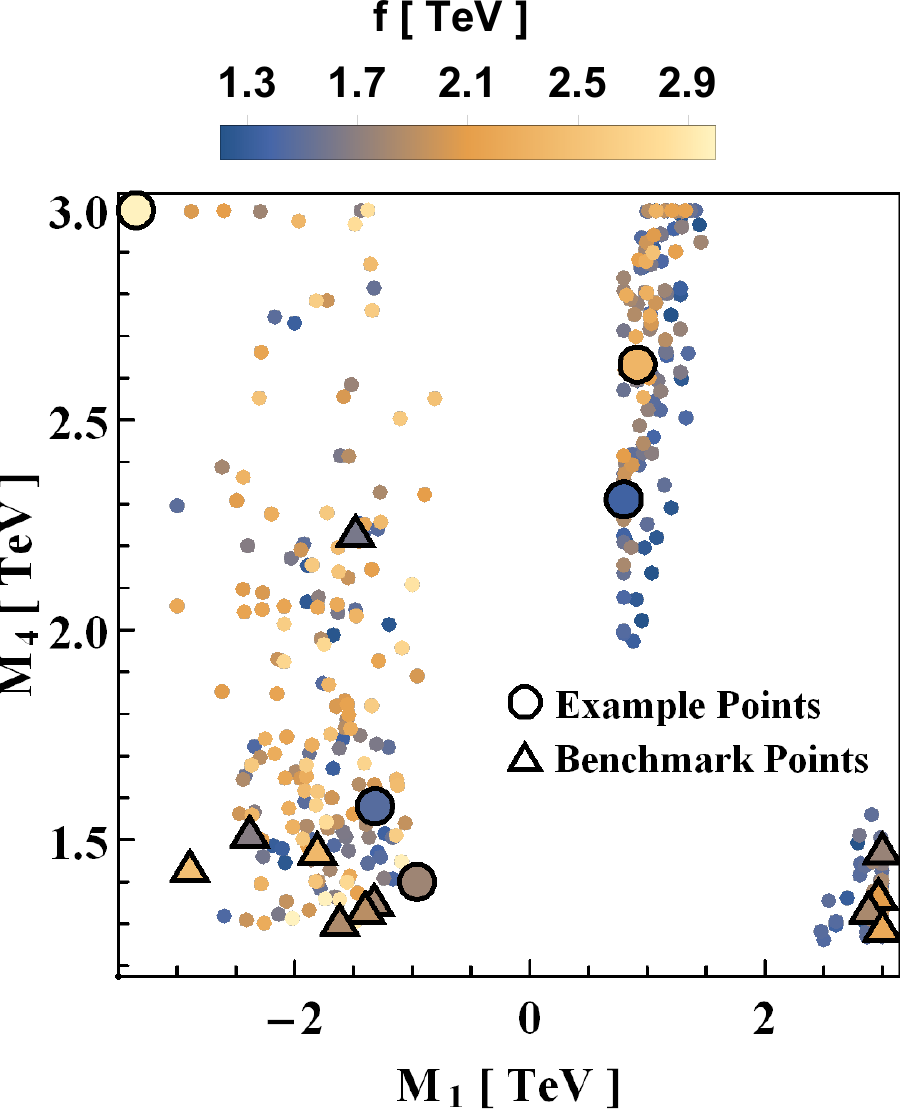}
\caption{Low scale scan  of the \five parameter space, including example points of Table~\ref{fig:benchmarkTable1}, benchmark points of Table~\ref{fig:benchmarkTableClusteringMCHM5} and the points consistent with constraints in Eqs.~\ref{const.tth} and~\ref{const.m}. }
\label{fig:mchm5.example.benchmark.points}
\end{figure}

In Fig.~\ref{fig:mchm5.example.benchmark.points} we show how the benchmark points are placed in the parameter space, together with the example points of Table~\ref{fig:benchmarkTable1} and the rest of the points in the scan not excluded by constraints. We can see that the benchmark points, complemented by the example ones, are well distributed in the parameter space. We finally list the benchmark points and their main features in Table~\ref{fig:benchmarkTableClusteringMCHM5}, where we can verify many of the features visible in the distribution plots of Figs.~\ref{bigclusterfig.mchm5.p1} and~\ref{bigclusterfig.mchm5.p2}. Points in Region I ($C_4$, $C_5$, $C_6$ and $C_7$) have on average higher couplings $y_L$ and $y_R$ and wider $T^{(1)}$ than those in Region II, although the extreme cases in each region can be similar. The point $C_7$, which contains a narrow $T^{(1)}$ for Region I standards, is quite similar to $C_{10}$, in which $T^{(1)}$  is exceptionally wide for Region II.

Another striking feature of Table~\ref{fig:benchmarkTableClusteringMCHM5} is that for all of the benchmark points (but $C_3$) there is at least a $10\%$ branching ratio in 3-body decays. Similarly striking and linked to the previous observation, we note that for all points but this same $C_3$ point, the 2-body decay into $b W$ is very small, namely between 0\% and 2.4\%.
That is relevant as most top partner searches were done under the assumption that the three 2-body channels ($t h$, $t Z$ and $b W$) comprise the full width. We will explore the phenomenology of these non-standard branching ratios in a future work.

\begin{table}[!!th]
  \centering
  \scriptsize
  \begin{tabular}{|c|c|ccccccccccc|}
    \cline{1-13}
    \multicolumn{2}{|c|}{} & \cellcolor{tableblue} C$_1$ & \cellcolor{tableblue}  C$_2$ & \cellcolor{tableblue}  C$_3$ & \cellcolor{tablered} C$_4$ & \cellcolor{tablered} C$_5$ & \cellcolor{tablered} C$_6$ & \cellcolor{tablered} C$_7$ & \cellcolor{tableblue} C$_8$ & \cellcolor{tableblue} C$_9$ & \cellcolor{tableblue}  C$_{10}$ & \cellcolor{tableblue} C$_{11}$ \\
    \hline
     \multirow{5}{*}{\rotatebox{90}{parameters}}
        & M$_1$(GeV) & -1323 & -1809 & -1483 & 2965 & 2882 & 2999 & 3000 & -1400 & -1618 & -2384 & -2892\\
        & M$_4$(GeV) & 1357 & 1479 & 2235 & 1370 & 1339 & 1479 & 1295 & 1339 & 1309 & 1519 & 1437 \\
        & f(GeV) & 1199 & 1593 & 1071 & 1393 & 1220 & 1168 & 1484 & 1265 & 1229 & 1110 & 1646\\
        & y$_L$ & 0.91 & 2.25 & 1.38 & 2.35 & 1.83 & 2.33 & 1.98 & 1.34 & 1.22 & 0.51 & 1.03\\
        & y$_R$ & 0.88 & 0.58 & 0.72 & 3.38 & 3.57 & 3.28 & 3.25 & 0.66 & 0.74 & 2.30 & 0.85\\
    \hline
    \multicolumn{2}{|c|}{$\mu(\ttH)~\text{(All Energies)}$} & 0.90 & 0.94 & 0.86 & 0.83 & 0.78 & 0.79 & 0.84 & 0.91 & 0.90 & 0.81 & 0.94 \\
    \hline
    \multicolumn{2}{|c|}{$\mu(\ttHH)~\text{(14~TeV)}$} & 2.14 & 1.47 & 0.80 & 1.51 & 1.53 & 1.02 & 2.00 & 2.25 & 2.41 & 1.39 & 1.58 \\
    \multicolumn{2}{|c|}{$\mu(\ttHH)~\text{(100~TeV)}$} & 14.58 & 8.84 & 3.28 & 10.28 & 11.18 & 7.04 & 13.42 & 15.20 & 16.11 & 13.68 & 10.57\\
    \hline
    \multicolumn{2}{|c|}{NR-\ttHH/\ttHH~(14~TeV)} & 0.37 & 0.59 & 0.88 & 0.45 & 0.40 & 0.61 & 0.35 & 0.36 & 0.33 & 0.46 & 0.55 \\
    \multicolumn{2}{|c|}{NR-\ttHH/\ttHH~(100~TeV)} & 0.05 & 0.10 & 0.22 & 0.07 & 0.05 & 0.09 & 0.05 & 0.05 &  0.05 & 0.05 & 0.08 \\
    \hline
    \multicolumn{2}{|c|}{$M_{T^{(1)}}~\text{(TeV)}$} & 1.36 & 1.48 & 1.66 & 1.40 & 1.38 & 1.51 & 1.32 & 1.34 & 1.31 & 1.54 & 1.44\\
    \multicolumn{2}{|c|}{$M_{T^{(2)}}~\text{(TeV)}$} & 1.63 & 2.02 & 2.24 & 3.55 & 2.61 & 3.10 & 3.22 & 1.61 & 1.80 & 1.63 & 2.20\\
    \multicolumn{2}{|c|}{$M_{T^{(3)}}~\text{(TeV)}$} & 1.79 & 3.88 & 2.68 & 5.55 & 5.21 & 4.85 & 5.67 & 2.17 & 2.02 & 3.47 & 3.21 \\
    \multicolumn{2}{|c|}{$M_{B^{(1)}}~\text{(TeV)}$} & 1.74 & 3.87 & 2.68 & 3.55 & 2.60 & 3.10 & 3.22 & 2.16 & 1.99 & 1.62 & 2.22 \\
    \multicolumn{2}{|c|}{$M_{X_{5/3}}~\text{(TeV)}$} & 1.36 & 1.48 & 2.24 & 1.37 & 1.34 & 1.48 & 1.29 & 1.34 & 1.31 & 1.52 & 1.44  \\
    \hline
     \multicolumn{2}{|c|}{$\Gamma_{T^{(1)}}$ (GeV)} & 8.83 & 5.49 & 26.22 & 51.92 & 60.01 & 71.68 & 44.33 & 6.44 & 7.49 & 43.78 & 10.63 \\
    \multicolumn{2}{|c|}{BR(T$^{\text{(1)}}$ $\to$th)} & 0.49 & 0.45 & 0.31 & 0.44 & 0.43 & 0.42 & 0.44 & 0.47 & 0.47 & 0.34 & 0.45 \\
    \multicolumn{2}{|c|}{BR(T$^{\text{(1)}}$ $\to$W$^+$b)} & 0.018 & 0 & 0.47 & 0.004 & 0.004 & 0.003 & 0.006 & 0.024 & 0.016 & 0.005 & 0.010\\
    \multicolumn{2}{|c|}{BR(T$^{\text{(1)}}$ $\to$tZ)} & 0.39 & 0.41 & 0.22 & 0.42 & 0.43 & 0.42 & 0.43 & 0.40 & 0.41 & 0.50 & 0.41 \\
    \multicolumn{2}{|c|}{BR(T$^{\text{(1)}}$ $\to$W$^+$W$^-$t)} & 0.11 & 0.13 & 0 & 0.13 & 0.13 & 0.16 & 0.12 & 0.10 & 0.10 & 0.14 & 0.12 \\
    \hline
    \end{tabular}
    \caption{Benchmark points for the MCHM$_5$ at low scale and their main features. Red and blue column headings indicate points belonging to Region I and II respectively. }
    \label{fig:benchmarkTableClusteringMCHM5}
\end{table}
%

\subsubsection{Clustering of the MCHM$_{14}$}
\label{clusteringLS14}

\begin{figure}[h]
\centering
\includegraphics[width=0.45\textwidth]{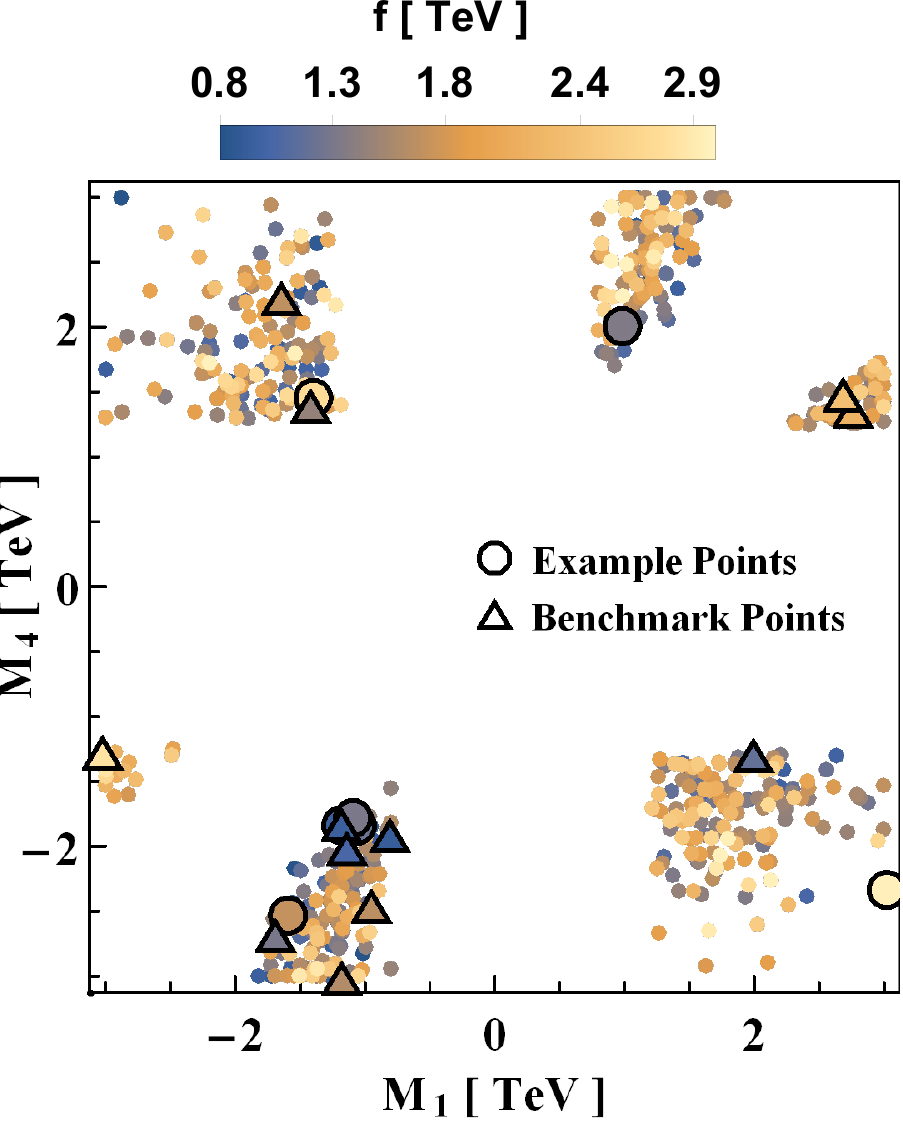}
\caption{Low scale scan  of the \fourt parameter space, including example points of Table~\ref{fig:benchmarkTable2}, benchmark points of Table~\ref{fig:benchmarkTableClusteringMCHM14} and the points consistent with constraints in Eqs.~\ref{const.tth} and~\ref{const.m}. }
\label{fig:mchm14.example.benchmark.points}
\end{figure}

We applied the same clustering method to the MCHM$_{14}$, using again Eqs.~\ref{const.tth} and~\ref{const.m} as constraints. The ideal clustering, following the criteria of homogeneity within clusters while keeping the number of clusters small, was obtained including $M[t,h_1]$, $p_T[t]$ and $\Delta R[h_1,t]$ in the samples and stopping at 12 clusters.
The distribution of the benchmark points can be seen in Figure~\ref{fig:mchm14.example.benchmark.points} and their main properties are listed in Table~\ref{fig:benchmarkTableClusteringMCHM14}.  We omit the plots of all distributions and clusters as the general features are very similar to the MCHM$_5$, with the points without light resonances decaying to $t\overline{t}hh$ being grouped into less homogeneous clusters (specially clusters 4 and 9), but all the plots are available online~\cite{ancillaryMat}.
There is still  a tendency towards narrower resonances when $M_1$ and $M_4$ have opposite signs (Regions II and IV), as expected due to the $1-r_1$ term in Eq.~\ref{mtapprox}, but now there is a $\xi$-proportional correction that is also sensitive to the sign of $M_4$ (as $M_9 > 0$) and makes this tendency weaker (and hard to notice if one looks only to the benchmark points).

\begin{table}[!!th]
  \centering
  \scriptsize
  \begin{tabular}{|c|c|cccccccccccc|}
    \cline{1-14}
    \multicolumn{2}{|c|}{} & \cellcolor{tableorange} D$_1$ & \cellcolor{tableorange} D$_2$ & \cellcolor{tablecyan} D$_3$ & \cellcolor{tableblue} D$_4$ & \cellcolor{tablered} D$_5$ & \cellcolor{tableorange} D$_6$ & \cellcolor{tableorange} D$_7$ & \cellcolor{tableorange} D$_8$ & \cellcolor{tableorange} D$_9$ & \cellcolor{tablered} D$_{10}$ & \cellcolor{tableblue} D$_{11}$ & \cellcolor{tableorange} D$_{12}$ \\
    \hline
     \multirow{5}{*}{\rotatebox{90}{parameters}}
        & M$_1$(GeV) & -1173 & -943 & 1979 & -1631 & 2737 & -2998 & -801 & -1130 & -1677 & 2664 & -1408 & -1169 \\
        & M$_4$(GeV) & -1823 & -2447 & -1297 & 2196 & 1340 & -1272 & -1907 & -2005 & -2670 & 1460 & 1373 & -2997\\
        & M$_9$(GeV) & -1382 & 2000 & 3889 & 3236 & 2836 & 3473 & 1500 & 1467 & 2000 & 3230 & 2965 & 1329\\
        & f(GeV) & 881 & 1275 & 1012 & 1288 & 1550 & 1912 & 863 & 931 & 1071 & 1648 & 1155 & 1244\\
        & y$_L$ & 1.98 & 1.33 & 0.85 & 2.68 & 1.67 & 1.11 & 1.23 & 2.93 & 2.06 & 2.40 & 1.04 & 1.25\\
        & y$_R$ & 3.90 & 1.07 & 0.73 & 0.30 & 1.93 & 1.86 & 1.67 & 1.23 & 2.65 & 2.67 & 0.49 & 1.32\\
    \hline
    \multicolumn{2}{|c|}{$\mu(\ttH)~\text{(All Energies)}$} & 1.39 & 0.90 & 0.48 & 0.71 & 0.68 & 0.68 & 0.93 & 0.98 & 1.22 & 0.71 & 0.68 & 0.93 \\
    \hline
    \multicolumn{2}{|c|}{$\mu(\ttHH)~\text{(14~TeV)}$} & 4.27 & 0.97 & 2.82 & 0.55 & 1.34 & 1.81 & 2.15 & 1.57 & 1.70 & 0.93 & 1.83 & 2.86 \\
    \multicolumn{2}{|c|}{$\mu(\ttHH)~\text{(100~TeV)}$} & 27.70 & 6.32 & 28.44 & 3.05 & 10.87 & 13.69 & 19.94 & 11.20 & 5.21 & 7.35 & 16.04 & 21.53\\
    \hline
    \multicolumn{2}{|c|}{NR-\ttHH/\ttHH~(14~TeV)} & 0.46 & 0.82 & 0.08 & 0.87 & 0.33 & 0.25 & 0.37 & 0.61 & 0.87 & 0.53 & 0.23 & 0.30 \\
    \multicolumn{2}{|c|}{NR-\ttHH/\ttHH~(100~TeV)} & 0.07 & 0.13 & 0.01 & 0.16 & 0.04 & 0.03 & 0.04 & 0.09 & 0.28 & 0.07 & 0.03 & 0.04 \\
    \hline
    \multicolumn{2}{|c|}{$M_{T^{(1)}}~\text{(TeV)}$} & 1.38 & 1.62 & 1.31 & 1.70 & 1.38 & 1.31 & 1.42 & 1.46 & 2.00 & 1.50 & 1.38 & 1.33\\
    \multicolumn{2}{|c|}{$M_{T^{(2)}}~\text{(TeV)}$} & 1.38 & 2.00 & 1.52 & 2.20 & 2.66 & 2.45 & 1.50 & 1.47 & 2.00 & 3.16 & 1.49 & 1.33\\
    \multicolumn{2}{|c|}{$M_{T^{(3)}}~\text{(TeV)}$} & 1.41 & 2.00 & 2.11 & 3.16 & 2.84 & 3.47 & 1.50 & 1.47 & 2.02 & 3.23 & 1.81 & 1.36\\
    \multicolumn{2}{|c|}{$M_{B^{(1)}}~\text{(TeV)}$} & 1.38 & 2.00 & 1.54 & 3.18 & 2.69 & 2.45 & 1.50 & 1.47 & 2.00 & 3.17 & 1.80 & 1.33 \\
    \multicolumn{2}{|c|}{$M_{X^{(1)}_{5/3}}~\text{(TeV)}$} & 1.38 & 2.00 & 1.30 & 2.20 & 1.34 & 1.27 & 1.50 & 1.47 & 2.00 & 1.46 & 1.37 & 1.33  \\
    \hline
     \multicolumn{2}{|c|}{$\Gamma_{T^{(1)}}$ (GeV)} & 12.17 & 91.45 & 22.06 & 157.07 & 67.39 & 53.12 & 64.23 & 79.22 & 17.70 & 82.52 & 17.55 & 3 .63\\
    \multicolumn{2}{|c|}{BR(T$^{\text{(1)}}$ $\to$th)} & 0.40 & 0.25 & 0.44 & 0.26 & 0.42 & 0.42 & 0.28 & 0.11 & 0.26 & 0.42 & 0.46 & 0.17 \\
    \multicolumn{2}{|c|}{BR(T$^{\text{(1)}}$ $\to$W$^+$b)} & 0.35 & 0.51 & 0.05 & 0.48 & 0.01 & 0.01 & 0.45 & 0.61 & 0.36 & 0.005 & 0.11 & 0.43\\
    \multicolumn{2}{|c|}{BR(T$^{\text{(1)}}$ $\to$tZ)} & 0.16 & 0.24 & 0.41 & 0.22 & 0.43 & 0.44 & 0.24 & 0.27 & 0.20 & 0.42 & 0.34 &  0.32 \\
    \multicolumn{2}{|c|}{BR(T$^{\text{(1)}}$ $\to$W$^+$W$^-$t)} & 0.09 & 0 & 0.10 & 0.04 & 0.14 & 0.13 & 0.02 & 0.02 & 0.18 & 0.16 & 0.08 & 0.08 \\
    \hline
    \end{tabular}
    \caption{Benchmark points for the low scale MCHM$_{14}$ scan and their main features. Column headings indicate region, with red and orange meaning respectively Regions I and III (with same sign $M_1$ and $M_4$) and blue and cyan respectively for regions II and IV (with opposite sign $M_1$ and $M_4$). }
    \label{fig:benchmarkTableClusteringMCHM14}
\end{table}

Figure~\ref{fig:mchm14.example.benchmark.points} shows that, between the example points of section~\ref{MCHM14ExamplePts} and the benchmark points obtained here we are covering the parameter space quite well. One can see that Table~\ref{fig:benchmarkTableClusteringMCHM14} contains many points that have a sizeable 3-body decay in all regions and, like in the case of the \five, a fair amount of the scanned points have overlapping resonances (which is the case for the benchmark points of clusters 3, 8 and 11), both of which are relevant concerns for top partner searches and constraints.

We would finally like to stress the presence of the benchmark points $D_1$ and $D_9$, both with a marked increase in $\sigma(t\overline{t}h)$, showing that such a situation is not uncommon in the \fourt and is a really interesting possibility to evidence this realization of the MCHM.

%
\section{MCHM at High Scale}
\label{MCHMHighScale}
We continue the analysis of the two studied MCHM scenarios, extending the dimensionful parameters to higher scales that can be accessible only in the context of the high CM energy $pp$ colliders in project~\cite{Arkani-Hamed:2015vfh,CEPC-SPPCStudyGroup:2015csa,JGao:2020, Golling:2016gvc,Contino:2016spe,Benedikt:2018csr}. A higher CM energy around 100 TeV is, indeed, requested to confront scenarios with very high mass resonances, with a possible deviation of the \ttH cross section from the SM value at the percent level (or even less), with the requested precision to study the \ttHH process (e.g. the various NR components) and to measure the branching ratios of various decays of the produced resonances, even if, for instance, the lightest one is detected at the HL-LHC.

\subsection{Scanning Over Parameter Space}
\label{parameterspaceHS}

The numerical strategy we use here is the same as the one we followed for the Low Scale analysis and described in Subsection~\ref{parameterspace} but with the following extended range of parameters in order to perform a high scale scan.
For the MCHM$_5$ we consider:
\begin{align*}
|M_1|& \in [2, 30]~{\rm TeV}, &  M_4 &\in [2, 30]~{\rm TeV},  \\
f&\in [0.8, 8.0]~{\rm TeV},          &  y_L&\in [0.5, 3.0],
\end{align*}
and for the MCHM$_{14}$, we use:
\begin{align*}
|M_1|& \in [2, 30]~{\rm TeV}, &  |M_4| &\in [2, 30]~{\rm TeV},  &  M_9 &\in [2, 30]~{\rm TeV},  \\
f&\in [0.8, 8.0]~{\rm TeV},          &  y_L&\in [0.5, 3.0].
\end{align*}

\begin{figure}[h]
    \centering
    \includegraphics[width=0.47\textwidth]{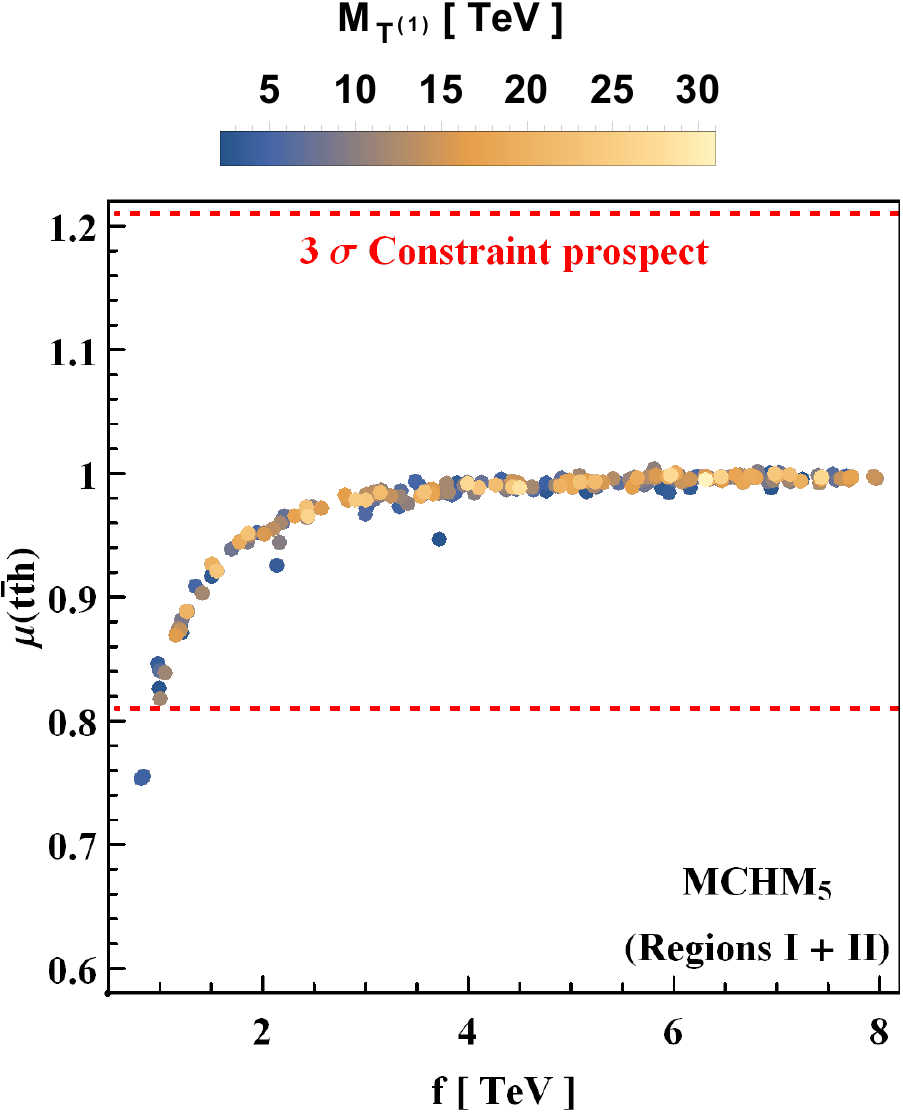}
    \hspace{0.6cm}
    \includegraphics[width=0.47\textwidth]{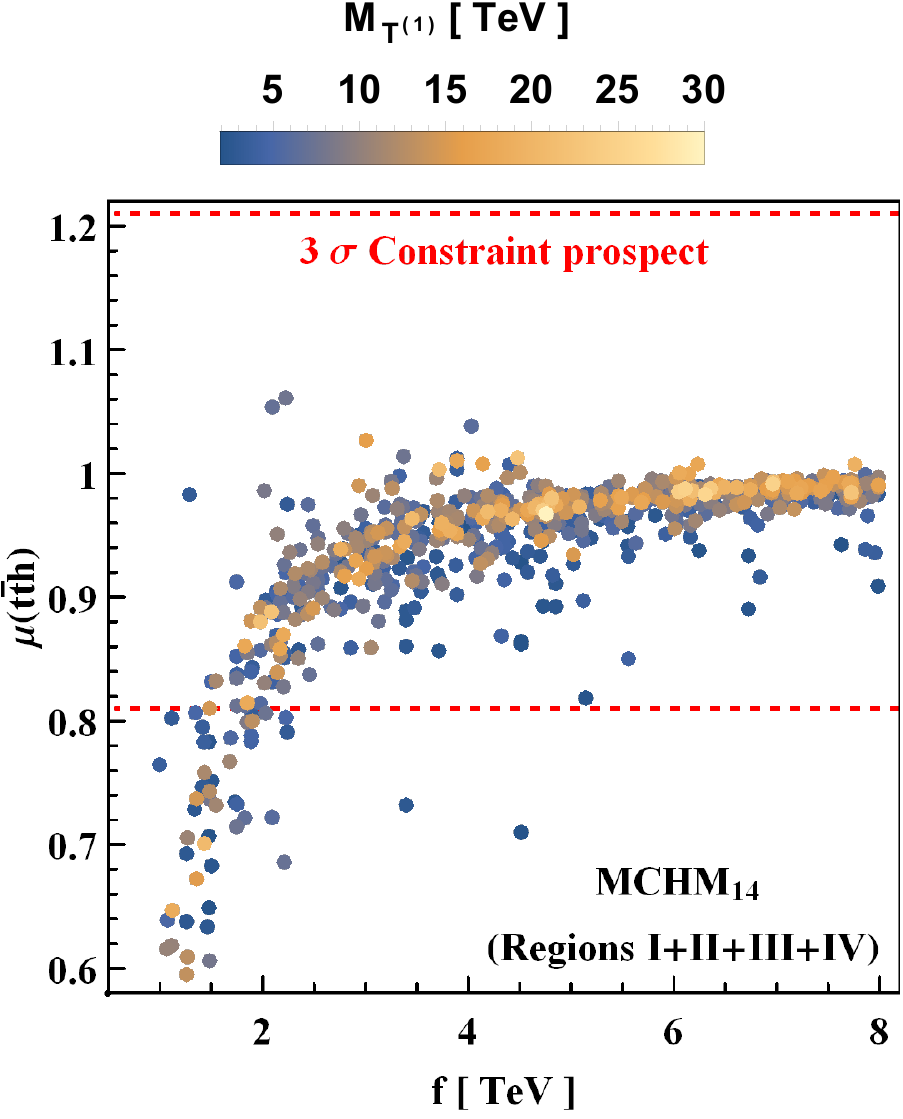}
\caption{Normalized $t\overline{t}h$ cross section as a function of $f$. We also color code the lightest vector-like mass. These points are obtained by joining all the regions in each model. The dashed red lines represent the prospects for the 3$\sigma$ uncertainty on the \ttH signal strength
after the HL-LHC measurements (see Sec.~\ref{Cluster Points HS}).}
    \label{fig:tthvsfHS}
\end{figure}

\begin{figure}[h]
    \centering
    \includegraphics[width=0.47\textwidth]{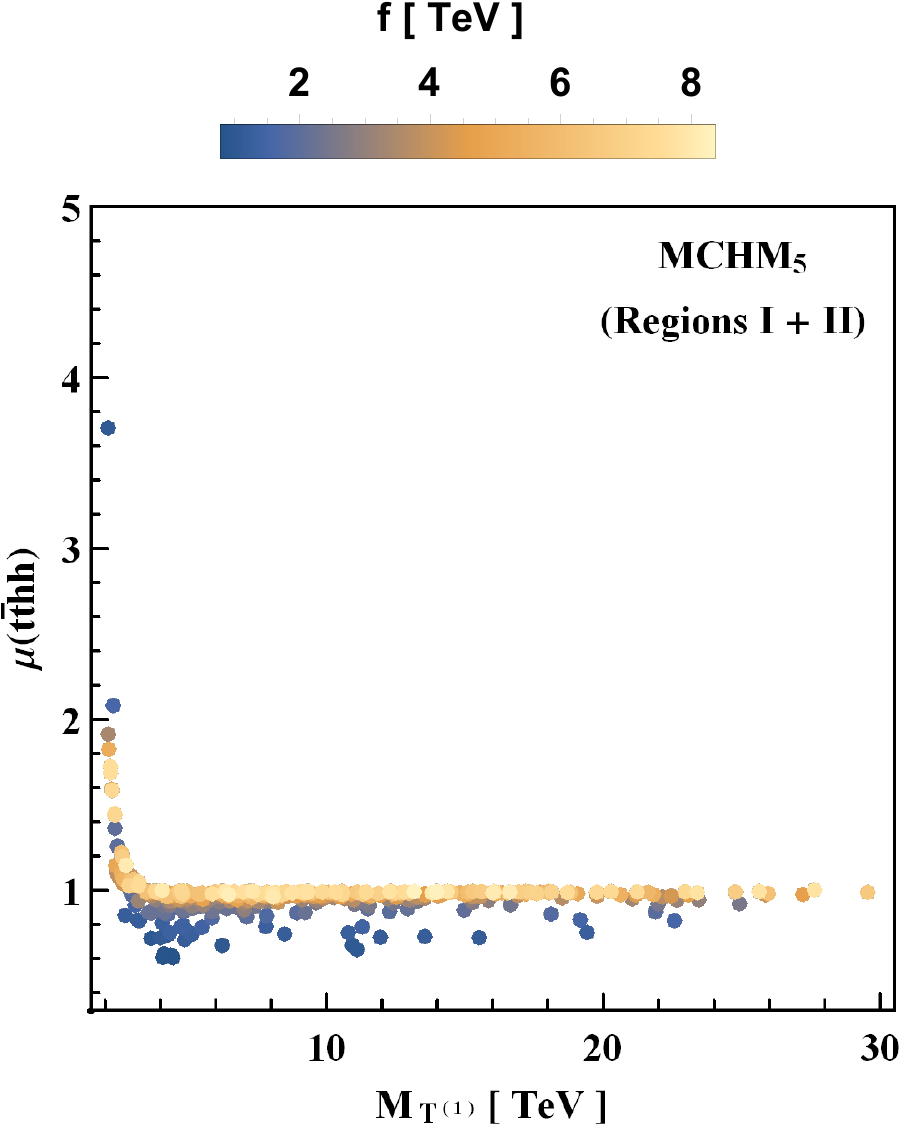}
    \hspace{0.6cm}
    \includegraphics[width=0.47\textwidth]{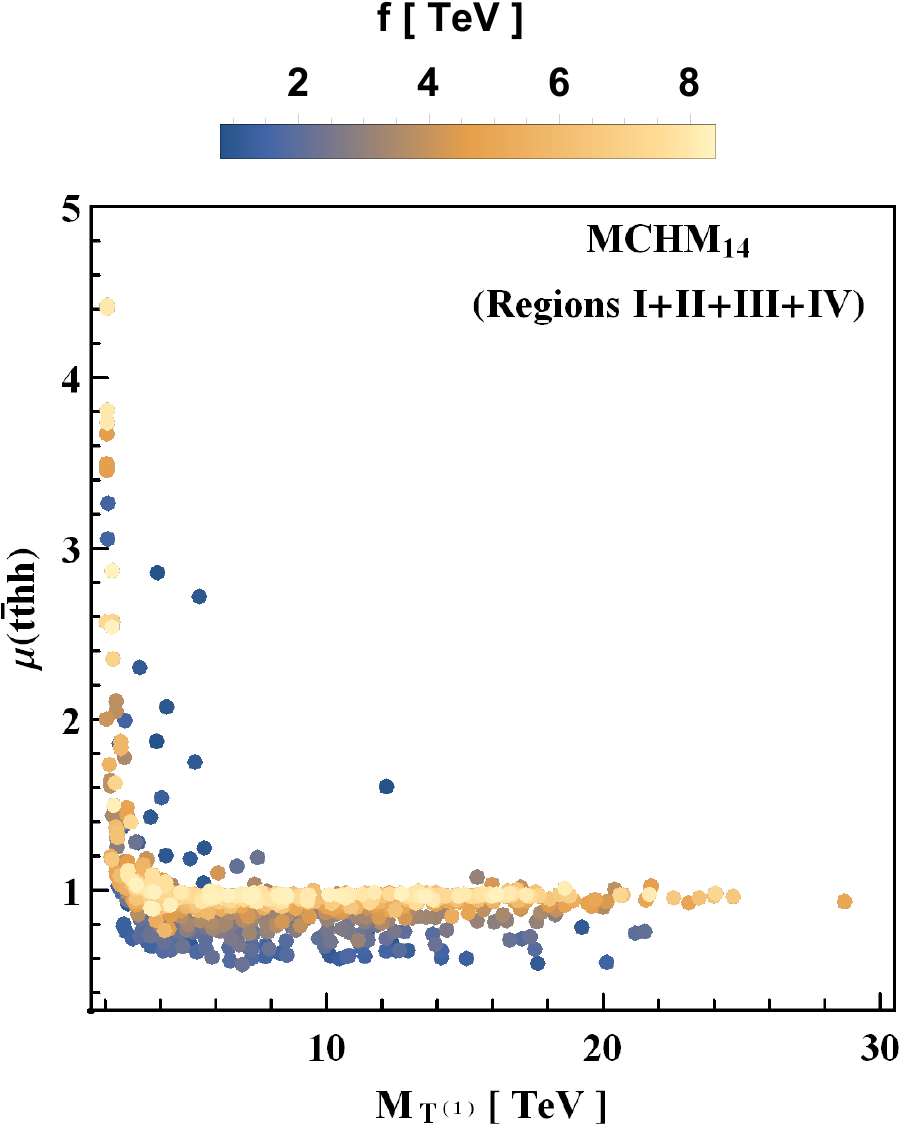}
    \caption{Normalized $t\overline{t}hh$ cross section as a function of the lightest $Q=2/3$ resonance mass at $\sqrt{s}= 100$ TeV. The value of the global scale of symmetry breaking is also color coded. These points are obtained by joining all the regions in each model.}
    \label{fig:tthhvsMT4HS}
\end{figure}

We divide the parameter space of the \five and the \fourt in the same regions we used for the Low Scale analysis (Sec.~\ref{scans}). 
In each region for each model we scanned 200 random points. We present the main results of the scans in Fig.~\ref{fig:tthvsfHS} and Fig.~\ref{fig:tthhvsMT4HS}, where we joined all the possible regions for each model.

Fig.~\ref{fig:tthvsfHS} displays the normalized $t\overline{t} h$ cross section as a function of $f$ for a center of mass energy of $100$ TeV.  As shown already in the low scale study, the $t\overline{t} h$ cross section depends mainly on the scale of global symmetry breaking $f$ and, as expected, for high values of $f$ we get smaller deviations from the SM, either suppressed or enhanced (only in the MCHM$_{14}$).

Fig.~\ref{fig:tthhvsMT4HS} shows the normalized $t\overline{t}hh$ cross section (at $\sqrt{s} = 100$ TeV) as a function of the mass of the lightest top resonance, $M_{T^{(1)}}$, with color coded values of $f$. As explained in Subsection~\ref{tthh}, at low scales the main contribution to this cross section comes from the QCD vector-like pair production. However, the later process undergoes a quick drop for heavy resonances, so it is expected that beyond certain value of $M_{T^{(1)}}$ the QCD resonance pair production becomes a subdominant part of the total $t\overline{t}hh$ cross section.
In the plots of Fig.~\ref{fig:tthhvsMT4HS}, we see that this value is about $M_{T^{(1)}}\approx 4$ TeV as the cross section becomes basically independent from $M_{T^{(1)}}$ beyond that point. In the left plot, corresponding to the MCHM$_5$, above this value there is no point with a $\mu(\ttHH)$ bigger than one, meaning that the main contributor is now the non-resonant part of the process, which, as we saw in Fig.~\ref{fig:nrtthhvstth}, is directly related to $\mu(\ttH)$ and always suppressed w.r.t the SM. Since the later is, in turn, mainly controlled by the compositeness scale $f$ (see Fig.\ref{fig:tthvsfHS}), the points with smaller values of $f$ have a more strongly suppressed cross section and as it increases, the cross section approaches the SM limit.

\begin{figure}[h]
\centering
\includegraphics[width=0.28\textwidth]{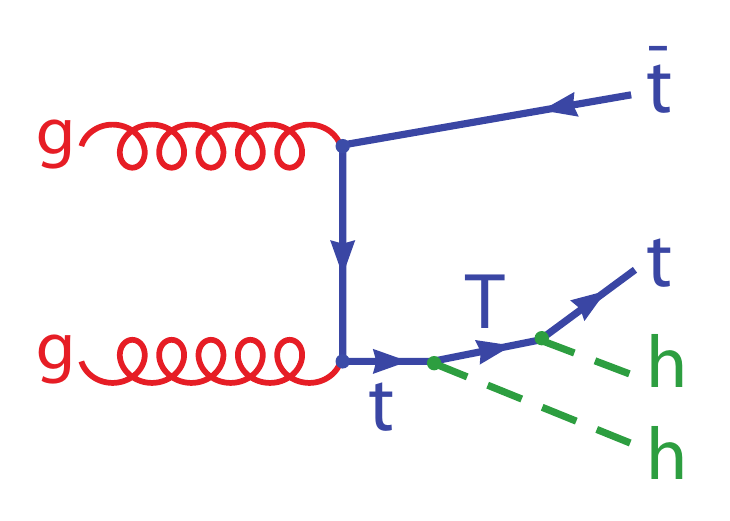}
\caption{Example diagram for the Yukawa mediated single $T$ contribution to $\overline{t}thh$.}
\label{fig:singley}
\end{figure}

We see that, in the right plot of Fig.~\ref{fig:tthhvsMT4HS} (corresponding to the MCHM$_{14}$) there are still points with considerable enhancements with respect to the SM beyond $M_{T^{(1)}}\approx 4$ TeV. There are two kinds of points with this behaviour: some points present enhancements in the $tth$ Yukawa coupling which reflect in enhancements in the NR-$t\overline{t}hh$ and consequently in an increase of order 10\% in \ttHH. The second case is more subtle and is responsible for the biggest cross sections in the region above $M_{T^{(1)}}\approx 4$ TeV. Those are fined tuned points, in which the parameters lead to strong coupling interaction between the top, a vector-like resonance and the Higgs. This vertex enters in diagrams like the one in Fig.~\ref{fig:singley}, in which a single intermediate top partner is produced (not to be confused with weak single production),
increasing the cross section as well. %
It is important to mention that the second type of points usually has a low compositeness scale ($f < 2$ TeV) and can be constrained by the increased precision on the measurement of the top Yukawa attainable by the HL-LHC (see Fig~\ref{fig:tthvsfHS}).

\subsection{Cluster analysis applied to the MCHM at High scale}
\label{Cluster Points HS}
Following the analysis of Secs.~\ref{benchmarks} and~\ref{clusteringLScale} we look now at specific points in the parameter space in order to examine the phenomenological features of the MHCM at high scale. As we show below, the increased mass range in the scan demanded a slightly different strategy in regards to
the clustering technique of Section~\ref{selectedpts}. In what follows, we divided the points of the scan in smaller sets before clustering. This approach, together with the fact the phenomenological behaviour at higher scales is more uniform, provides enough points to showcase all the interesting features of the model. For that reason we have not chosen additional example points for the High Scale scan, and we proceed directly to the results of the clustering algorithm.

\subsubsection{Clustering of the MCHM$_5$}

We start with the 400 points of the scan in the previous section (200 in each region)
and discard all points violating the following constraint on $\kappa_t$, the top Yukawa coupling normalized w.r.t. to the SM:
\begin{equation}
    0.9 < \kappa_t < 1.1
    \label{constraintHS}
\end{equation}
This constraint is based on the projected precision on the top Yukawa measurement in the High Luminosity phase of the LHC. In \cite{Cepeda:2019klc} the 1$\sigma$ uncertainty on $\kappa_t$ is projected to be $3.4\%$ for an accumulated luminosity of 3000 fb$^{-1}$. We constrain our points to the region that is 3 times that uncertainty around the SM value $\kappa_t  = 1$, as a rough estimate of the 3$\sigma$ region. We do this in order to keep points that have a smaller chance of being constrained by measurements at the time when the 100 TeV $pp$ collider starts its operation.
We can directly translate eq.~\ref{constraintHS} into constraints in $\mu(\ttH)$, and we show these limits as red dashed lines in Fig.~\ref{fig:tthvsfHS}, where one can see that this allows for fairly small values of $\mu(\ttH)$ ($\gtrapprox 0.8$). By the end of the HL-LHC phase the points close to that limit will be of course at the edge of the allowed region, and could be even excluded, specially if the central value of $\mu(\ttH)$ turns out to be above $1$ or the precision is better than expected. But on the other hand, it is still a possibility that the opposite happens (central value of $\mu(\ttH) < 1$ for instance) so we understand it is too early to disregard those points.

The remaining points are clustered following the method described in Section~\ref{parameterspaceHS}. Different combinations of kinematic distributions are considered but, in all of them, the clustering procedure was giving too much weight to the position of peaks on
points with lighter resonances, creating many low population clusters for those and leaving
the ones with heavier resonances combined in a unique cluster. This is similar to what happened in the clusterization of the low Scale MCHM$_5$ (see cluster 3 in Fig,~\ref{bigclusterfig.mchm5.p2}), but the problem is exacerbated by the fact that, in this case, we are working with a bigger range of parameters and, consequently, with a wider variety of resonance masses. As we are interested in selecting points distributed in the whole parameter space, we decide to first group the points in slices, guided by the mass of the lightest resonance, and then to apply the clustering algorithm to each slice individually.
The following five slices, are used:

\begin{equation}
    \begin{split}
2~ \text{TeV} < M_{T^{(1)}} < &~3 ~\text{TeV}\\
3~\text{TeV} < M_{T^{(1)}} < &~4 ~\text{TeV}\\
4~\text{TeV} < M_{T^{(1)}} < &~5~\text{TeV}\\
5~\text{TeV} < M_{T^{(1)}} < & ~6~\text{TeV}\\
6~\text{TeV} < M_{T^{(1)}} < &~30~\text{TeV}
\end{split}
\label{slices}
\end{equation}

In this way we ensure we are clustering the points in a more homogeneous way. The last slice groups all the points with resonances heavier than 6 TeV since, as we see in Fig.~\ref{fig:tthhvsMT4HS}, it is expected that all of them possess similar kinematic characteristics because the contribution of the resonances for them  are negligible. The selected clustering is realized by using the $M[t,h_1]$, $p_T[t]$ and $\theta[t]$ distributions and stopping at 2 clusters in each slice, for a total of 10 clusters. The plots showing the clustering in full detail are available online~\cite{ancillaryMat}. The benchmark points corresponding to each cluster are listed in Table~\ref{fig:benchmarkTableClusteringMCHM5HS}.

\begin{table}[h]
  \centering
  \scriptsize
  \begin{tabular}{|c|c|cccccccccc|}
    \cline{1-12}
    \multicolumn{2}{|c|}{} & \cellcolor{tablered}E $_1$ & \cellcolor{tablered} E$_2$ & \cellcolor{tablered} E$_3$ & \cellcolor{tablered} E$_4$ & \cellcolor{tablered} E$_5$ & \cellcolor{tablered} E$_6$ & \cellcolor{tablered} E$_7$ & \cellcolor{tablered} E$_8$ & \cellcolor{tableblue} E$_9$ & \cellcolor{tableblue} E$_{10}$ \\
    \hline
     \multirow{5}{*}{\rotatebox{90}{parameters}}
        & M$_1$(TeV) & 22.7 & 19.2 & 11.1 & 23.0 & 26.5 & 3.6 & 19.3 & 10.5 & -10.7 & -22.9 \\
        & M$_4$(TeV) & 2.4 & 2.1 & 3.2 & 3.2 & 4.0 & 22.5 & 5.1 & 5.1 & 25.6 & 13.5 \\
        & f(GeV) & 1913 & 3273 & 7144 & 1190 & 1300 & 1711 & 1288 & 2812 & 2432 & 1199 \\
        & y$_L$ & 2.45 & 0.87 & 2.85 & 2.43 & 0.99 & 2.00 & 2.35 & 1.84 & 2.57 & 2.46\\
        & y$_R$ & 1.10 & 1.24 & 2.01 & 1.54 & 3.53 & 1.31 & 2.35 & 3.13 & 1.11 & 2.62 \\
    \hline
    \multicolumn{2}{|c|}{$\mu(\ttH)~\text{(All Energies)}$} & 0.95 & 0.97 & 0.99 & 0.88 & 0.83 & 0.94 & 0.88 & 0.97 & 0.96 & 0.87 \\
    \hline
    \multicolumn{2}{|c|}{$\mu(\ttHH)~\text{(100~TeV)}$} & 1.26 & 1.91 & 1.03 & 0.82 & 0.81 & 0.86 & 0.75 & 0.91 & 0.92 & 0.73\\
    \hline
    \multicolumn{2}{|c|}{NR-\ttHH/\ttHH~(100~TeV)} & 0.71 & 0.48 & 0.95 & 0.90 & 0.82 & 1.00 & 1.00 & 1.02 & 1.01 & 1.01\\
    \hline
    \multicolumn{2}{|c|}{$M_{T^{(1)}}~\text{(TeV)}$} & 2.45 & 2.12 & 3.21 & 3.23 & 4.07 & 4.28 & 5.08 & 5.15 & 11.02 & 13.55\\
    \multicolumn{2}{|c|}{$M_{T^{(2)}}~\text{(TeV)}$} & 5.27 & 3.55 & 18.13 & 4.32 & 4.28 & 22.50 & 5.90 & 7.31 & 25.62 & 13.86\\
    \multicolumn{2}{|c|}{$M_{T^{(3)}}~\text{(TeV)}$} & 22.81 & 19.65 & 20.61 & 23.12 & 26.91 & 22.76 & 19.51 & 13.68 & 26.37 & 23.15 \\
    \multicolumn{2}{|c|}{$M_{B^{(1)}}~\text{(TeV)}$} & 5.28 & 3.55 & 20.61 & 4.33 & 4.24 & 22.76 & 5.90 & 7.30 & 26.37 & 13.86\\
    \multicolumn{2}{|c|}{$M_{X_{5/3}}~\text{(TeV)}$} & 2.44 & 2.11 & 3.20 & 3.22 & 4.04 & 22.50 & 5.06 & 5.14 & 25.62 & 13.54\\
    \hline
    \multicolumn{2}{|c|}{$\Gamma_{T^{(1)}}$ (TeV)} & 0.04 & 0.04 & 0.08 & 0.14 & 0.96 & 0.28 & 0.76 & 0.84 & 1.22 & 11.8\\
    \multicolumn{2}{|c|}{$\Gamma_{T^{(1)}}/M_{T^{(1)}}$} & 1.6\%  & 1.9\% & 2.5\% & 4.3\% & 24\% & 6.5\% & 15\% & 16\% & 11\% & 87\% \\
    \multicolumn{2}{|c|}{BR(T$^{\text{(1)}}$ $\to$th)} & 0.35 & 0.38 & 0.29 & 0.29 & 0.15 & 0.26 & 0.18 & 0.17 & 0.25 & 0.04 \\
    \multicolumn{2}{|c|}{BR(T$^{\text{(1)}}$ $\to$W$^+$b)} & 0.003 & 0.004 & 0 & 0.001 & 0 & 0.50 & 0 & 0 & 0.50 & 0 \\
    \multicolumn{2}{|c|}{BR(T$^{\text{(1)}}$ $\to$tZ)} & 0.34 & 0.37 & 0.28 & 0.28 & 0.33 & 0.25 & 0.18 & 0.18 & 0.25 & 0.04 \\
    \multicolumn{2}{|c|}{BR(T$^{\text{(1)}}$ $\to$W$^+$W$^-$t)} & 0.30 & 0.25 & 0.43 & 0.43 & 0.52 & 0 & 0.64 & 0.65 & 0 & 0.92\\
    \hline
    \end{tabular}
    \caption{Benchmark points for the MCHM$_{5}$ at high scale and their main features. Red and blue column headings indicate points belonging to Region I and II respectively.}
    \label{fig:benchmarkTableClusteringMCHM5HS}
\end{table}

The first general feature to be noticed in Table~\ref{fig:benchmarkTableClusteringMCHM5HS} is the importance of the NR-$\ttHH$ contribution to the total cross-section. Only in point $E_2$ it is not the dominant contribution, although it is still about half of the cross section. Even in the points with resonances in the $2\sim 3$ TeV range the NR contribution is high  (0.7 in $E_1$, $\geq 0.9$ in $E_3$ and $E_4$). As the resonances get heavier, the NR component becomes almost all of the cross section. The increasing suppression of the $T^{(1)}$ pair production also causes $\mu(\ttHH)$ to be close to or below 1 for the points with heavier ressonances ($E_3$ to $E_{10}$), as expected for the NR contribution in the MCHM$_5$.

The second striking feature is the number of points with a sizeable or 3-body decay (all points with the exception of $E_6$ and $E_9$). The effect happens when both $M_{T^{(1)}}$ and $M_{X_{5/3}}$ are largely controlled by $M_4$, and are thus almost degenerate (whereas in $E_6$ and $E_9$, $M_{T^{(1)}}$ is closer to $M_1$). In these cases, there is also a marked suppression of BR(T$^{\text{(1)}}$ $\to$W$^+$b).
This was already present in the low scale scan but becomes more prominent here, and one can check in the Table that the 3-body decay becomes more important as $M_{T^{(1)}}$ increases. Considering the points with this property (which are the majority of the benchmark points) we see that the 3-body branching ratio increases from $\sim 10\%$ at $M_{T^{(1)}} \approx 1.3$ TeV (see Table~\ref{fig:benchmarkTableClusteringMCHM5}) to quickly becoming the dominant decay channel around $M_{T^{(1)}} \approx 3$ TeV, being already $25\%$ in point $E_2$ (with $M_{T^{(1)}} = 2.1$ TeV). This makes it extremely important to take three body decays into consideration if one wishes to push the direct search exclusion of top partners beyond the present ballpark of 1.3 TeV. Additionally, the fact that this decay happens through the exotically charged $X_{5/3}$, also highlights the importance of examining more complete models of the fermionic sector, as opposed to simplified models that only include a single electroweak doublet or singlet top partner.

On the other hand, points $E_6$ and $E_9$ exhibit the behaviour typically well captured by simplified models. In these, $M_{T^{(1)}} \approx M_1$ is well separated from the rest of the spectrum, which is dominated by masses close to a much higher $M_4$. The branching ratios follow then the expected pattern: the $W^+ b$ channel branching ratio is twice that of the $th$ and $tZ$ channels.

Other features already present in the low scale scan are also present at higher scale, specially the presence of double peaked structures in many points. Among the benchmark points, we highlight points $E_5$ and $E_{10}$ where we find two top partners close in mass. In fact in these two points, the exotic $X_{5/3}$ and the bottom partner $B^{(1)}$ are also close by, allowing the 3-body decay to go through either of those channels. This causes the width of $T^{(1)}$ to increase a lot: $(\Gamma/M)_{T^{(1)}} = 24\%$ for $E_5$ and even up to $(\Gamma/M)_{T^{(1)}} = 87\%$ for $E_{10}$. Points $E_7$ and $E_8$ also exhibit similar behaviour, but there is a bigger mass gap and smaller $(\Gamma/M)_{T^{(1)}}$.

Figure~\ref{fig:mchm14HS.benchmark.points} shows the scanned points as well as the benchmark points in the $M_1$ - $M_4$ plane with color coded $f$. There, we see that the benchmark points obtained with the clustering technique are covering both regions of the model. Besides verifying the homogeneity of each cluster we also searched for points that have behaviours that deviate significantly from the benchmarks, and found none. This indicates that the behaviour of the model is quite uniform and well represented by the benchmark points, even in the region with a small number of them, as for instance in Region II (negative $M_1$). The same is true for regions with high $M_1$, $M_4$, and specially high $f$, as we are moving towards the decoupling limit of the model and the points become more similar to each other and to the SM (which explains why only one point, $E_3$, was chosen to represent the high $f$ region).  For these reasons, it would be superfluous
to choose and include additional example points.

\begin{figure}[h]
\centering
\includegraphics[width=0.45\textwidth]{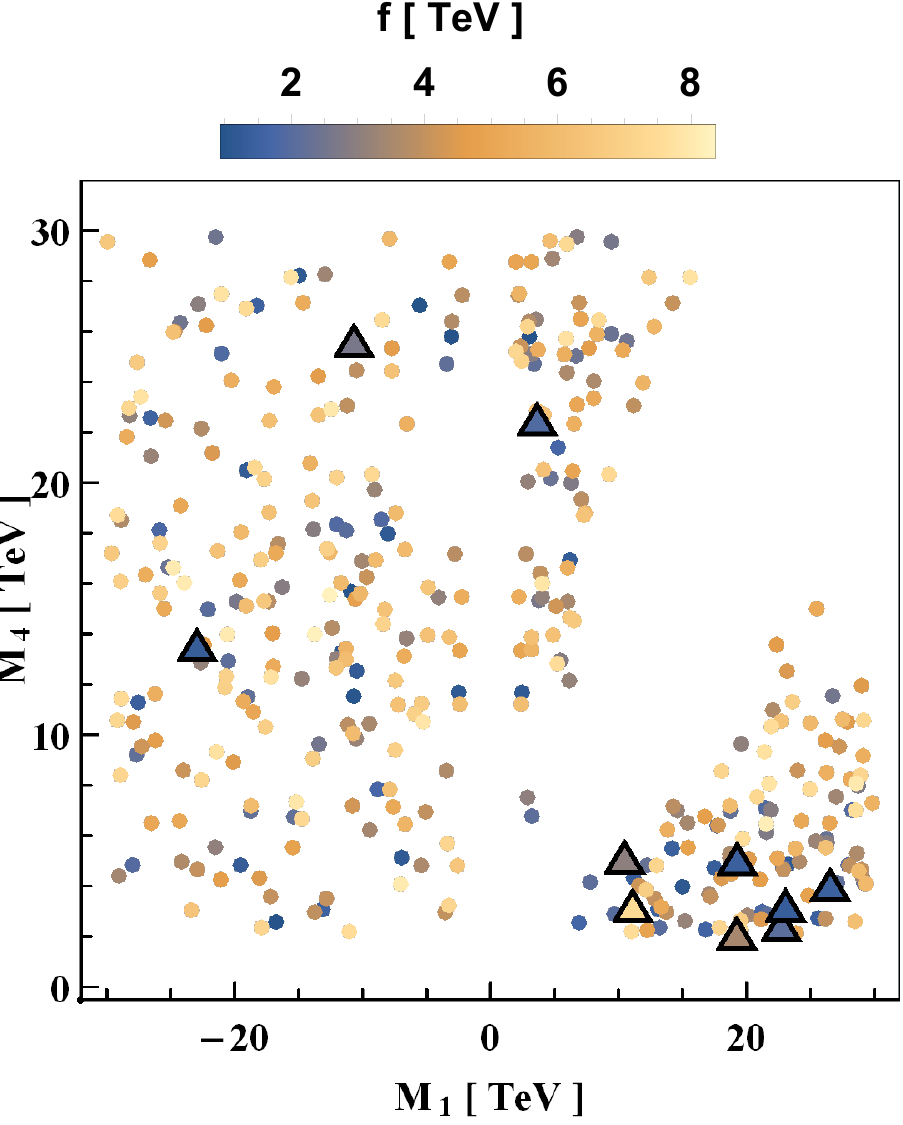}
\caption{High scale scan of the \five parameter space, including the benchmark points of Table~\ref{fig:benchmarkTableClusteringMCHM5HS} represented by triangles and the points satisfying the constraint in Eq.~\ref{constraintHS}. The compositeness scale $f$ is color coded.}
\label{fig:mchm5HS.benchmark.points}
\end{figure}

\subsubsection{Clustering of the MCHM$_{14}$}

In the case of the \fourt at high scale we started with 800 points evenly distributed in the four regions of the $M_1$ - $M_4$ space as described in Section \ref{parameterspaceHS}. Then, we selected all points allowed by the constraint in Eq. \ref{parameterspaceHS} and proceed with the clustering technique. As in the MCHM$_5$, we divide the points in the five slices of Eq.~\ref{slices}. We stopped at 2 or 3 clusters in each slice, depending on how homogeneous were the obtained clusters, and the corresponding benchmark points are listed in Table~\ref{fig:benchmarkTableClusteringMCHM14HS}. The plots showing the clustering in full detail are available online~\cite{ancillaryMat}.

\begin{table}[h]
  \centering
  \scriptsize
  \begin{tabular}{|c|c|ccccccccccc|}
    \cline{1-13}
    \multicolumn{2}{|c|}{} & \cellcolor{tableorange} F$_1$ & \cellcolor{tablecyan} F$_2$ & \cellcolor{tablered} F$_3$ & \cellcolor{tablered} F$_4$ & \cellcolor{tableblue} F$_5$ & \cellcolor{tablered} F$_6$ & \cellcolor{tablered} F$_7$ & \cellcolor{tablecyan} F$_8$ & \cellcolor{tablecyan} F$_9$ & \cellcolor{tableorange} F$_{10}$ & \cellcolor{tableorange} F$_{11}$ \\
    \hline
     \multirow{5}{*}{\rotatebox{90}{parameters}}
        & M$_1$(TeV) & -10.1 & 4.10 & 25.2 & 14.5 & -4.20 & 27.5 & 12.0 & 4.93 & 11.1 & -17.6 & -16.7 \\
        & M$_4$(TeV) & -2.27 & -15.9 & 3.71 & 29.6 & 29.7 & 11.3 & 4.04 & -10.5 & -5.62 & -7.20 & -27.0  \\
        & M$_9$(TeV) & 6.16 & 2.06 & 10.2 & 3.31 & 29.9 & 4.38 & 18.7 & 25.8 & 24.3 & 8.31 & 26.9 \\
        & f(TeV) & 3.41 & 4.52 & 4.89 & 6.84 & 7.89 & 4.35 & 2.13 & 4.81 & 1.98 & 3.13 & 4.48 \\
        & y$_L$ & 1.29 & 1.83 & 0.53 & 2.07 & 2.41 & 2.20 & 2.23 & 2.58 & 1.93 & 1.33 & 2.38 \\
        & y$_R$ & 0.83 & 0.26 & 1.15 & 1.53 & 0.13 & 1.47 & 1.13 & 0.24 & 0.67 & 2.01 & 3.13 \\
    \hline
    \multicolumn{2}{|c|}{$\mu(\ttH)~\text{(All Energies)}$} & 0.93 & 0.86 & 0.96 & 0.92 & 0.97 & 0.98 & 0.87 & 0.97 & 0.86 & 0.88 & 1.01 \\
    \hline
    \multicolumn{2}{|c|}{$\mu(\ttHH)~\text{(100~TeV)}$} & 1.44 & 3.46 & 0.93 & 1.03 & 0.91 & 0.99 & 0.74 & 0.92 & 0.71 & 0.74 & 1.02\\
    \hline
    \multicolumn{2}{|c|}{NR-\ttHH/\ttHH~(100~TeV)} & 0.59 & 0.22 & 0.97 & 0.81 & 1.02 & 0.97 & 0.99 & 1.01 & 1.00 & 1.02 & 0.99 \\
    \hline
    \multicolumn{2}{|c|}{$M_{T^{(1)}}~\text{(TeV)}$} & 2.28 & 2.06 & 3.72 & 3.31 & 4.36 & 4.38 & 4.05 & 5.07 & 5.62 & 7.22 & 21.7 \\
    \multicolumn{2}{|c|}{$M_{T^{(2)}}~\text{(TeV)}$} & 4.90 & 2.06 & 4.53 & 3.31 & 29.7 & 4.38 & 6.18 & 10.45 & 6.74 & 8.23 & 26.9 \\
    \multicolumn{2}{|c|}{$M_{T^{(3)}}~\text{(TeV)}$} & 6.16 & 2.11 & 10.2 & 3.35 & 29.9 & 4.40 & 12.2 & 16.2 & 11.2 & 8.31 & 26.9 \\
    \multicolumn{2}{|c|}{$M_{B^{(1)}}~\text{(TeV)}$} & 4.92 & 2.06 & 4.52 & 3.31 & 29.9 & 4.38 & 6.22 & 16.2 & 6.77 & 8.23 & 27.0  \\
    \multicolumn{2}{|c|}{$M_{X^{(1)}_{5/3}}~\text{(TeV)}$} & 2.27 & 2.06 & 3.71 & 3.31 & 29.7 & 4.38 & 4.04 & 10.5 & 5.62 & 7.20 & 26.9   \\
    \hline
     \multicolumn{2}{|c|}{$\Gamma_{T^{(1)}}$ (TeV)} & 0.05 & 0.06 & 0.24 & 0.18 & 0.83 & 0.25 & 0.28 & 0.63 & 0.23 & 3.00 & 45.8 \\
     \multicolumn{2}{|c|}{$\Gamma_{T^{(1)}}/M_{T^{(1)}}$} & 2.2\%  & 2.9\% & 6.4\% & 5.4\% & 19\% & 5.7\% & 6.9\% & 12\% & 4.1\% & 42\% & 211\% \\
    \multicolumn{2}{|c|}{BR(T$^{\text{(1)}}$ $\to$th)} & 0.36 & 0 & 0.24 & 0 & 0.25 & 0 & 0.23 & 0.25 & 0.16 & 0.10 & 0.04 \\
    \multicolumn{2}{|c|}{BR(T$^{\text{(1)}}$ $\to$W$^+$b)} & 0.006 & 0.27 & 0.001 & 0.21 & 0.50 & 0.16 & 0.005 & 0.50 & 0.007 & 0 & 0.07\\
    \multicolumn{2}{|c|}{BR(T$^{\text{(1)}}$ $\to$tZ)} & 0.35 & 0.53 & 0.25 & 0.41 & 0.25 & 0.32 & 0.23 & 0.25 & 0.16 & 0.11 & 0.03 \\
    \multicolumn{2}{|c|}{BR(T$^{\text{(1)}}$ $\to$W$^+$W$^-$t)} & 0.28 & 0.20 & 0.50 & 0.39 & 0 & 0.52 & 0.54 & 0 & 0.68 & 0.79 & 0.86 \\
    \hline
    \end{tabular}
    \caption{Benchmark points for the high scale MCHM$_{14}$ scan and their main features. Column headings indicate the region at which the point belongs, with red and orange meaning respectively Regions I and III (with same sign $M_1$ and $M_4$) and blue and cyan respectively for regions II and IV (with opposite sign $M_1$ and $M_4$).}
    \label{fig:benchmarkTableClusteringMCHM14HS}
\end{table}

Many features are similar to the MCHM$_5$: the NR-\ttHH once again becomes dominant once $M_{T^{(1)}}$ goes above 3 TeV, and both $\mu(\ttH)$ and $\mu(\ttHH)$ go below 1 when that happens. In principle, the MCHM$_{14}$ could allow for an increased top Yukawa and thus $\mu(\ttH) > 1$ and $\mu(\ttHH) > 1$ even in the non-resonant regime. Only one point with that behaviour was selected by the algorithm ($F_{11}$), but all scales for that point are so high that it is almost SM-like.

In regard to the spectrum and decay of the lightest top partner, many points mimic the behaviour of the MCHM$_5$, but the situation is richer here:

\begin{itemize}
    \item The points $F_1$, $F_3$, $F_7$, $F_9$ and $F_{10}$ reproduce the dominant behaviour in the MCHM$_5$, with the masses of $T^{(1)}$ and $X_{2/3}$ essentially set by $M_4$, large branching ratios in 3-body decays and suppressed $Wb $ decays. All the important remarks made in that case apply.
    \item The points $F_5$ and $F_8$ have the split spectrum similar to $E_6$ and $E_9$. The top partner has a mass set by $M_1$ and the rest of the spectrum is much heavier. Decay channels follow the pattern assumed in simplified models.
    \item The points $F_2$, $F_4$ and $F_6$ are qualitatively new. The mass scale is largely set by $M_9$ and we have three top partners that are degenerate not only among themselves, but also coincide with the masses of the lightest exotically charged fermion and the bottom partner. The width of $T^{(1)}$ does not increase as in the degenerate cases of the MCHM$_5$, and there is a suppression of the $th$ channel. A complete survey of all the decay channels of $T^{(2)}$ and $T^{(3)}$ would be needed to disentangle these states (and determine which are contributing to \ttHH and in what degree), but that is beyond the scope of this work. We intend to explore this point in a future study.

    Moroever, unlike the points $F_4$ and $F_6$, the point $F_2$, if still valid after HL-LHC, will be evidenced at a 100 TeV machine thanks to the high precision measurement of $\mu(tth)$  and, more importantly, to the large deviation from the SM in $\mu(tthh)$.
     \item The point $F_{11}$ is really SM-like from several experimental viewpoints; even with the high precision reachable on $\mu(tth)$ and $\mu(tthh)$ measurements at a 100 TeV collider, it will not be possible to disentangle them from SM. Besides, the very high mass and highly degenerate in $M_4$ and $M_9$ resonance spectrum,  only shows a highly dominant 3 body decay of $T^{(1)}$, as particular features. Let's note that points $F_5$ and $F_8$ have the same highly degenerate in $M_4$ and $M_9$ resonance spectrum but, as mentioned above, with a pattern of decay channels similar to simplified models. The  $F_{11}$ point would deserve some specific phenomenological and experimental study, that goes beyond the scope of this paper.

\end{itemize}

\begin{figure}[h]
\centering
\includegraphics[width=0.45\textwidth]{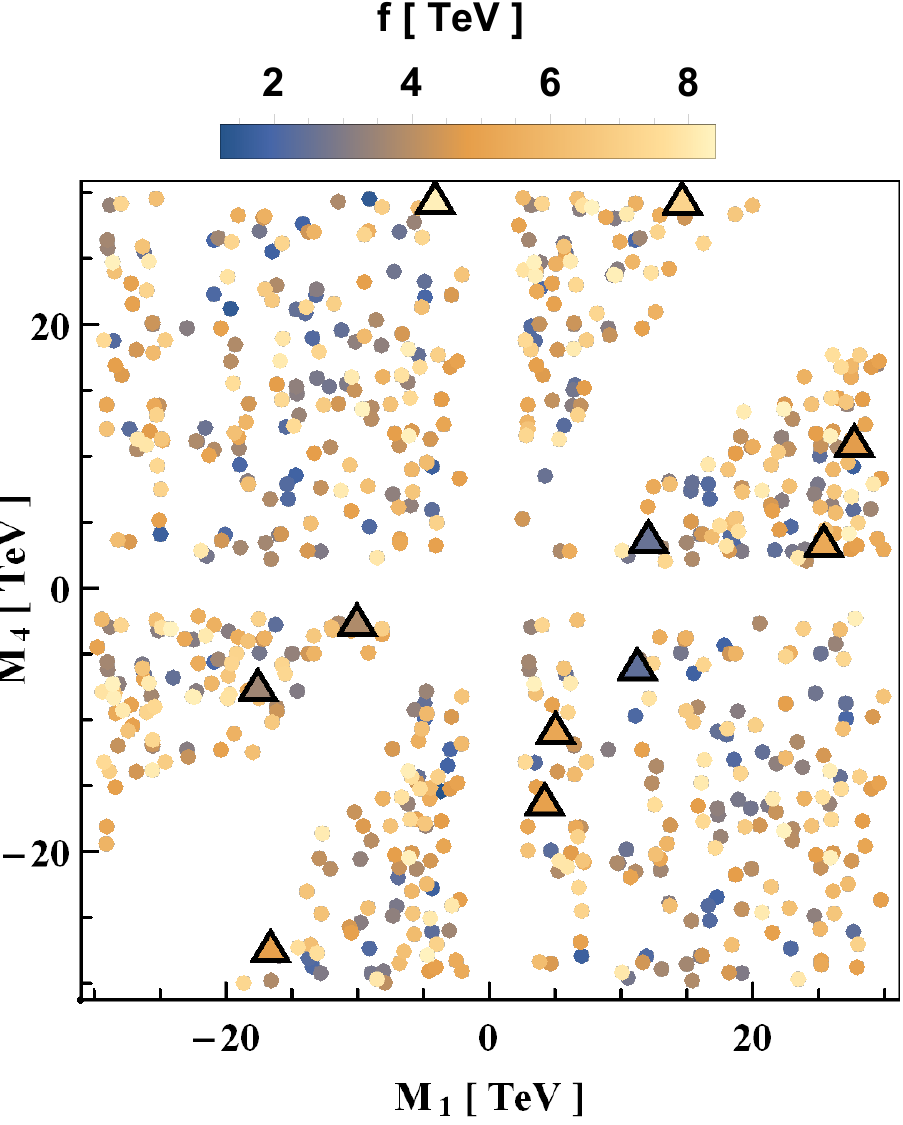}
\caption{High scale scan of the \fourt parameter space, including the benchmark points of Table~\ref{fig:benchmarkTableClusteringMCHM14HS} represented by triangles and the points satisfying the constraint in Eq.~\ref{constraintHS}. The compositeness scale $f$ is color coded.}
\label{fig:mchm14HS.benchmark.points}
\end{figure}

Figure~\ref{fig:mchm14HS.benchmark.points} shows the benchmark points in the $M_1$ - $M_4$ plane with color coded values of $f$. The benchmark points obtained with the clustering cover this space homogeneously and then we decided also not to add example points.

\section{Effective Field Theory Perspective}
\label{EFT}
In the analysis reported in the previous sections, we always consider the effect of the fermionic resonances in the cross sections, however, for the points in parameter space with larger masses, decoupling occurs and the processes are then largely non-resonant. In this case, it can be useful to present expressions for the modifications of the SM couplings, in the context of an effective field theory, in which the heavy degrees of freedom are out of experimental reach and can be safely integrated out\footnote{In the case of the pair production of top partners decaying to \ttHH, this occurs for $M_{T^{(1)}} > 4$ TeV, as can be seen in figure~\ref{fig:tthhvsMT4HS} as the curve bends sharply when we go from resonant to non-resonant production.}, in order to facilitate a comparison with other non-resonant studies in the literature.

The aim of this section is to more generally show the interplay between the considered MCHM scenarios in their overall scanned parameter spaces, and the EFT framework. To do so we  present our results
in terms of modifications to the SM couplings, which are more directly comparable to the increasing number of experimental results on these parameters. It is straightforward to express these results in terms of a non-redundant operator basis such as the strongly-interacting light Higgs (SILH) basis \cite{silh}.
The Higgs effective Lagrangian we consider here, after EWSB and neglecting the light fermion interactions, is given below:
\bea
\begin{split}
\mathcal{L}_h=&\frac{1}{2}\partial_\mu h\partial^\mu h -\frac{1}{2}m_h^2h^2-\kappa_\lambda \lambda_{\text{SM}}vh^3-\frac{m_t}{v}\left(v+\kappa_t h +\frac{c_2}{v}hh\right)\left(\overline{t}_Lt_R + \text{h.c.}\right)\\
&+\frac{1}{4}\frac{\alpha_s}{3\pi v}\left(c_g h -\frac{c_{2g}}{2v}hh\right)G^{\mu\nu}G_{\mu\nu}
\end{split}
\eea
where $\kappa_\lambda$, $\kappa_t$, $c_2$, $c_g$ and $c_{2g}$ are coefficients that encode the modifications to the SM.
We do not present here the expression for $c_{2g}$ as this operator will not contribute to the tth and tthh processes at tree level. The coupling $\kappa_\lambda$ is the same in both the  MCHM$_5$ and the  MCHM$_{14}$:\footnote{Some of the expressions below are reported also in \cite{Liu:2017dsz}, and presented here again for completeness.}
\begin{flalign}
\label{e:klambda}
\hspace{1cm}\kappa_\lambda=\frac{1-2\xi}{\sqrt{1-\xi}}&&
\end{flalign}

For the MCHM$_5$, the remaining couplings are given by:
\begin{flalign}
\hspace{1cm}\begin{split}
        \kappa^5_t&
        =1+\left[-\frac{3}{2}+\frac{1}{2}\left(1-\frac{1}{r_1^2}\right)\text{sin}^2\theta_L+\left(1-r_1^2\right)\text{sin}^2\theta_R\right]\xi+...
\end{split}&&
\end{flalign}
\begin{flalign}
    \hspace{1cm}c^5_2=\left[-2+\frac{3}{4}\left(1-\frac{1}{r_1^2}\right)\text{sin}^2\theta_L+\frac{3}{2}\left(1-r_1^2\right)\text{sin}^2\theta_R\right]\xi+...&&
\end{flalign}
\begin{flalign}
\hspace{1cm}c^5_g=\frac{1-2\xi}{\sqrt{1-\xi}}&&
\end{flalign}
Here we defined $\theta_L$ and $\theta_R$ by $\text{tan}\,\theta_L=y_L\, f/M_4$ and $\text{tan}\,\theta_R=y_R\, f/M_1$.

For the MCHM$_{14}$, the corresponding couplings are given by:
\begin{flalign}
\hspace{1cm}\begin{split}
        \kappa^{14}_t&
         =1+\left[-4-\frac{3}{2}\frac{1-r_9}{1-r_1}+\frac{5}{4}\left(2-\frac{1}{r_1^2}-\frac{1}{r_9^2}\right)\text{sin}^2\theta_L+\frac{5}{2}\left(1-r_1^2\right)\text{sin}^2\theta_R\right]\xi+...
\end{split}&&
\end{flalign}
\begin{flalign}
    \hspace{1cm}c^{14}_2= \left[-6-\frac{9}{4}\frac{1-r_9}{1-r_1}+\frac{15}{8}\left(1-\frac{1}{r_1^2}+1-\frac{1}{r_9^2}\right)\text{sin}^2\theta_L+\frac{15}{4}\left(1-r_1^2\right)\text{sin}^2\theta_R\right]\xi+...&&
\end{flalign}
\begin{flalign}
\hspace{1cm}c^{14}_g=c^{t}_g+c^b_g&&
\end{flalign}
with $c^t_g$ and $c^b_g$ defined in equations \ref{cgt14} and \ref{cgb14}, respectively.

\begin{figure}[h]
\centering
\includegraphics[width=0.32\textwidth]{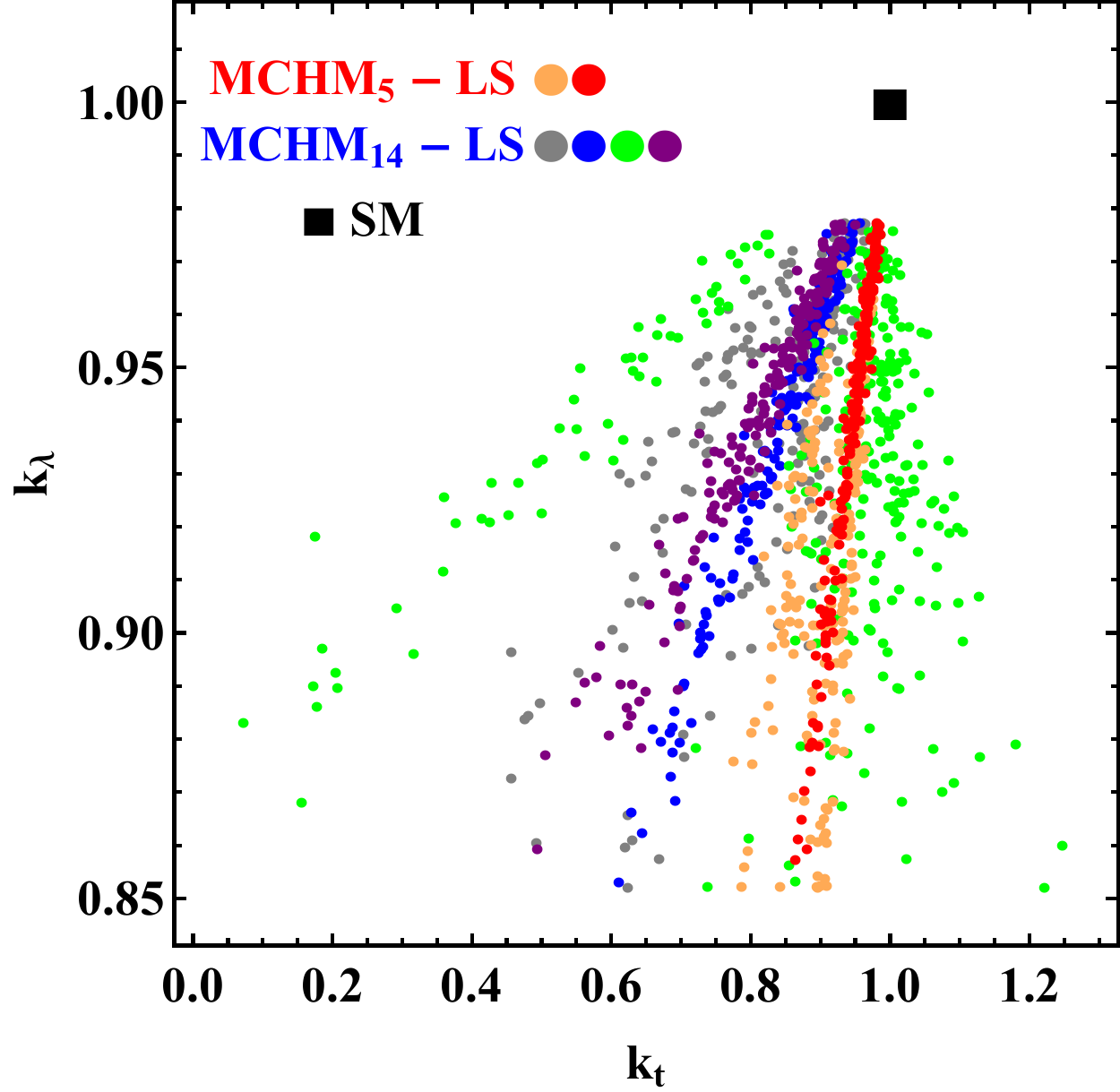}
\includegraphics[width=0.32\textwidth]{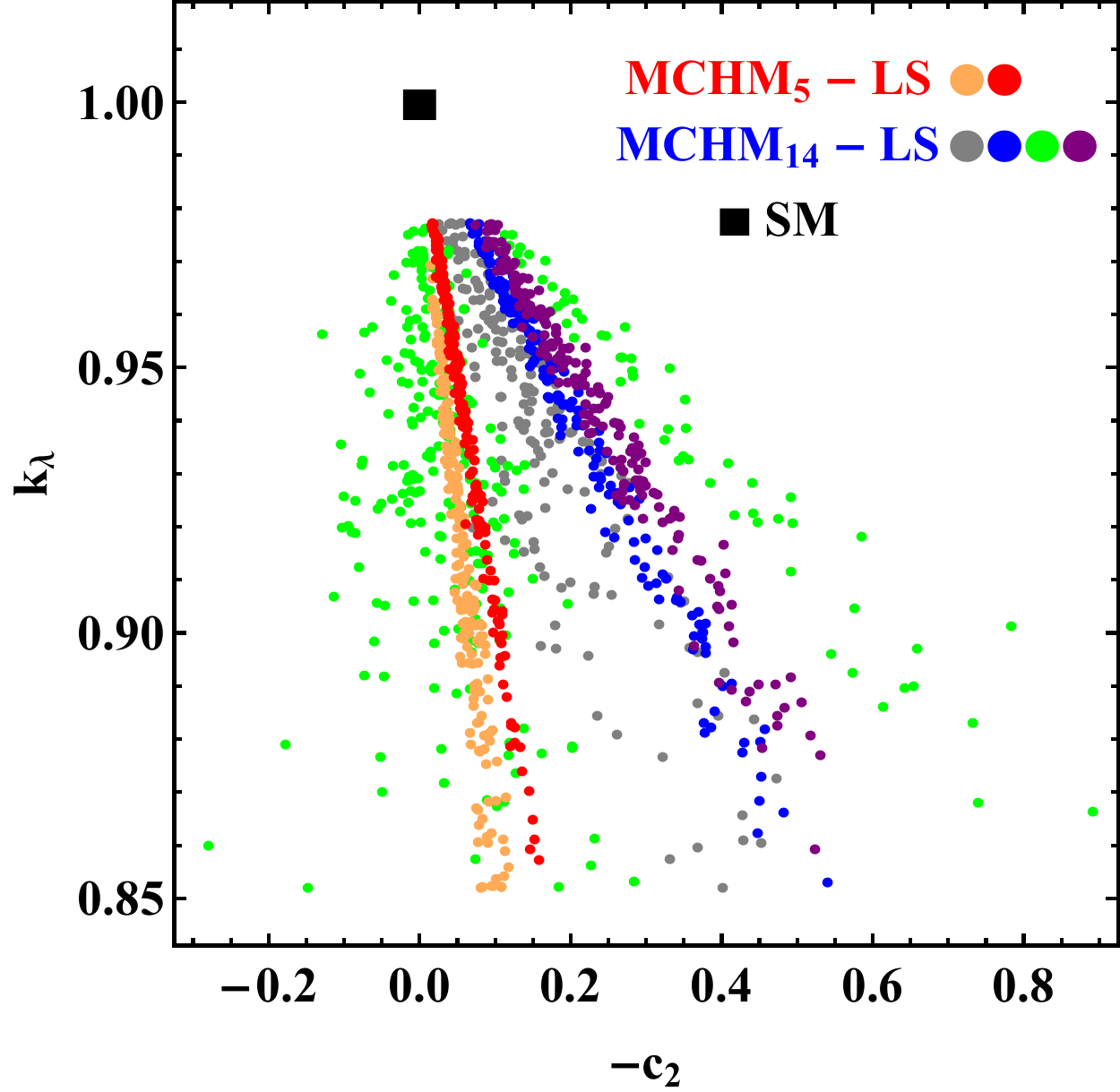}
\includegraphics[width=0.32\textwidth]{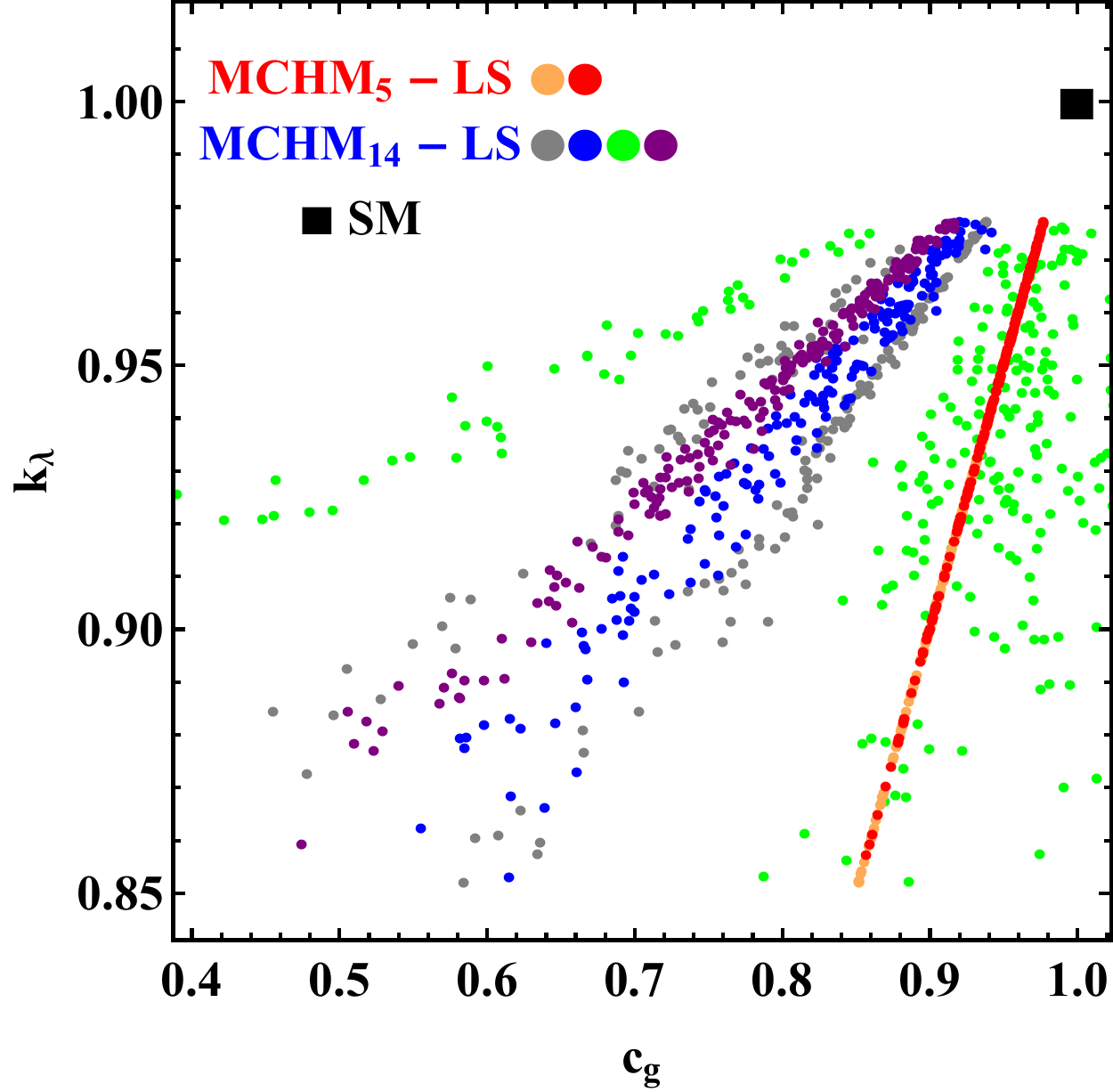}
\caption{Values of some selected EFT parameters in the low scale scan of the MCHM$_5$ and the  MCHM$_{14}$ parameter spaces. The colors indicate the different Regions in each model (I and II for the MCHM$_5$ and I, II, III and IV for the MCHM$_{14}$, in that order). The SM 
is represented by the black square.}
\label{fig:eftplotsLS}
\end{figure}
\begin{figure}[h]
\centering
\includegraphics[width=0.32\textwidth]{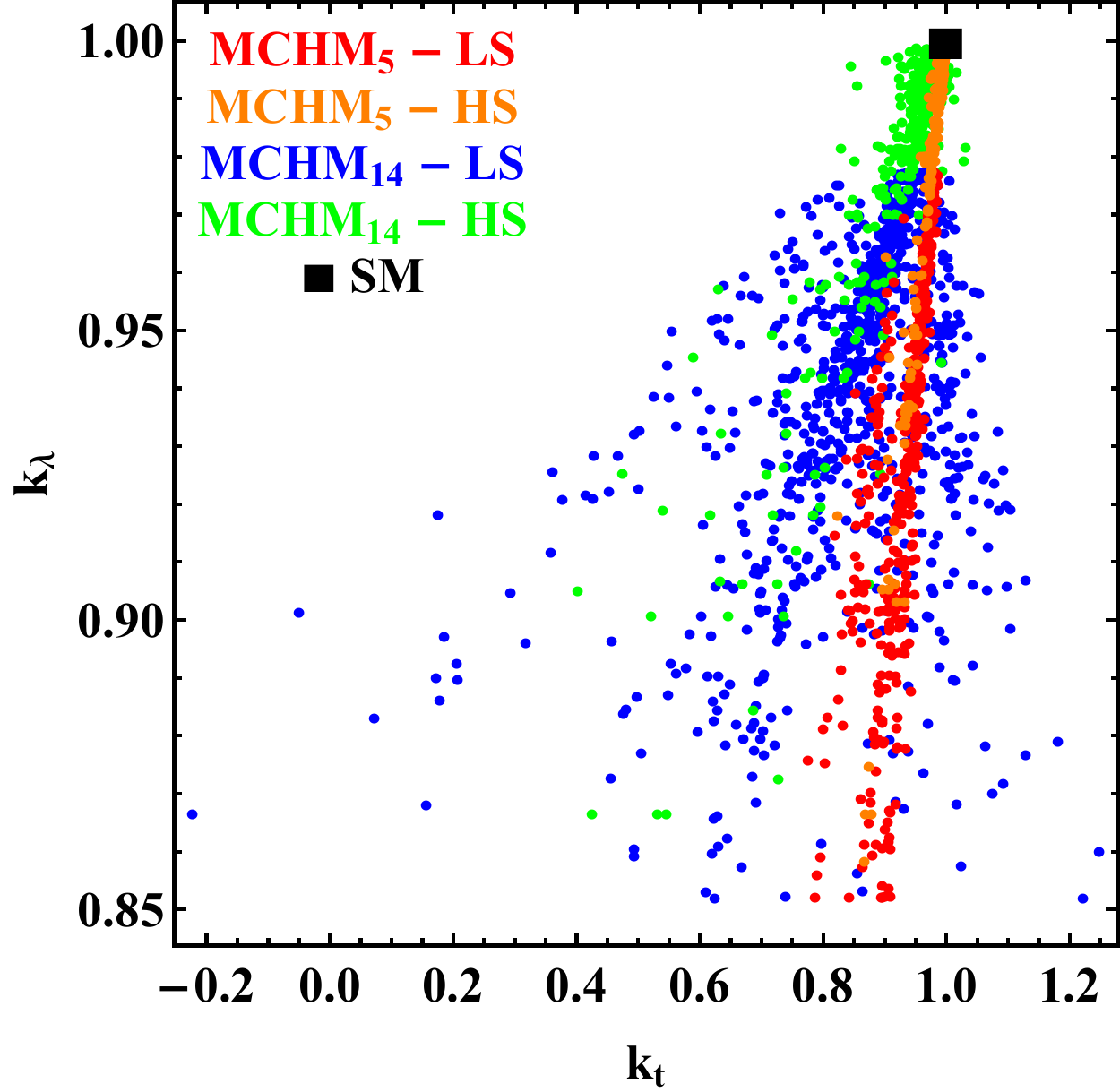}
\includegraphics[width=0.32\textwidth]{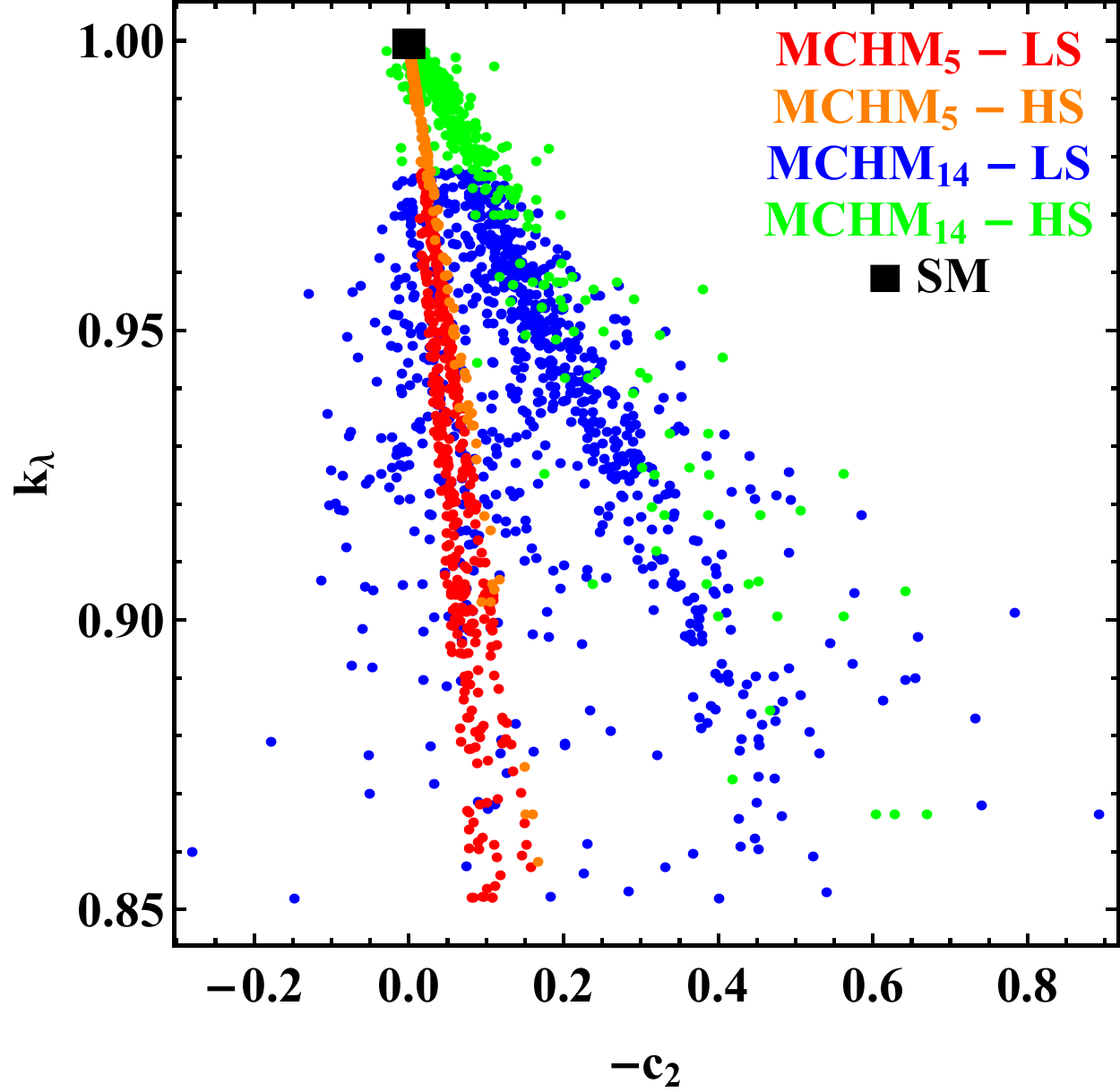}
\includegraphics[width=0.32\textwidth]{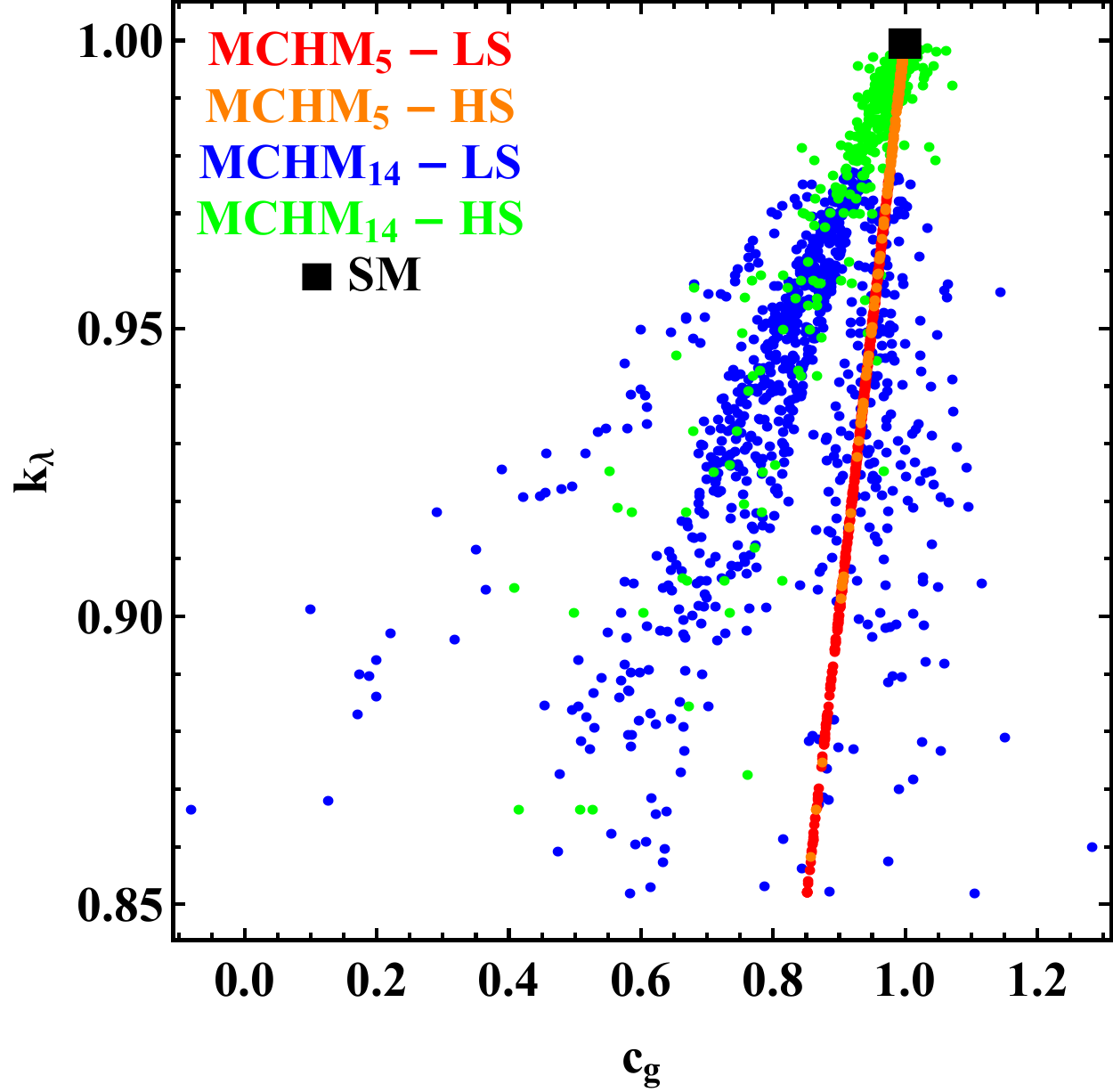}
\caption{Values of some selected EFT parameters in the low and high scale scans of the MCHM$_5$ and the  MCHM$_{14}$ parameter spaces. The SM parameters position is represented by the black square.}
\label{fig:eftplots1LSHS}
\end{figure}

In Figs. \ref{fig:eftplotsLS} and \ref{fig:eftplots1LSHS}, we have plotted selected EFT parameters against each other, for the points scanned in Sections.~\ref{results} and~\ref{MCHMHighScale}.
The colors identify different regions in the parameter space of each model for Fig. \ref{fig:eftplotsLS}, which covers the low scale region of the parameter space only, while in Fig. \ref{fig:eftplots1LSHS} the colors  differentiate the low and high scale scans for the \five and the MCHM$_{14}$. The vertical axis is chosen to be $\kappa_\lambda$, which depends only on the $f$ scale, see Eq. (\ref{e:klambda}). The low energy scan only covers up to $f\leq 2~$TeV, corresponding to $\kappa_\lambda\leq 0.98$, such that points with values closer to the SM will not be present in this case. On the horizontal axis, we show $\kappa_t,~-c_2$ and $c_g$. Clearly the MCHM$_5$ and the MCHM$_{14}$ show distinct behavior, including between the different Regions defined in the $M_1$ - $M_4$ parameter space of each case (especially for MCHM$_{14}$), with considerable more spread for the MCHM$_{14}$, owing to a more involved dependence of the coefficients on the microscopic parameters in this case.
The couplings defined above and their interplay can be used as discriminators in case a deviation from the SM is found; furthermore, by combining the tth and tthh channels with other channels such as the double higgs production, which is sensitive to $c_g$, it should be possible to exclude some combinations of Region and MCHM scenarios or pinpoint which of them is realized in nature. In particular, one can verify that the \five is much easier to discard than the MCHM$_{14}$. For instance, it can be seen that the aforementioned enhancement in $\kappa_t$ is only present for Region III in the $\bf 14$, while the relation $\kappa_\lambda = c_g$ is only satisfied for the MCHM$_5$; therefore, a measurement of $\kappa_t > 1$ or $\kappa_\lambda \ne c_g$ would argue strongly against the MCHM$_5$.
Evidence of non-zero $c_2$ is also particularly interesting since this vertex is absent in the SM.

As expected, for smaller values of the compositeness scale $f$, the deviations from the SM are of course larger, and the separation between the different models is easier.

\section{Conclusions and Outlook}
\label{conclusions}

The Minimal Composite Higgs Models MCHM$_5$ and MCHM$_{14}$ are studied here as an important show-case for the exploration of the beyond standard Model world, in the top-Higgs sector at high energy hadron colliders.  The full generation and simulation framework is developed which allows a phenomenological analysis over the whole parameter space of these two models. Besides, the developed generation and simulation software defines a preliminary experimental framework; they are ready for use by more detailed data analysis at the LHC and HL-LHC experiments and by advanced detector simulations developed for the 100 TeV machines in project.

The focus is on the analysis of the $t{\overline t}h$ and  $t{\overline t}hh$ production processes, covering in the later both the production of heavy fermions and the non-resonant contributions.
In this paper we have gauged the relative contribution of these two cases to  $t{\overline t}hh$  and shown that the non-resonant is sizeable, in fact dominant once the fermions are heavier than 4 TeV (even in the case with 100 TeV of center of mass energy), and gives  access not only to the trilinear Higgs self-coupling, as it does in the SM case, but also to the double Yukawa coupling ($t{\overline t}hh$ vertex) which is introduced by the MCHM. The relative contributions of the top-Yukawa, the trilinear Higgs and the double Yukawa couplings were also analysed, showing that the trilinear Higgs coupling contributes approximately with $15\%$ of the non-resonant cross section (see tables \ref{fig:NRtthhMCHM5Table} and \ref{fig:NRtthhMCHM14Table}).

A systematic exploration of the parameter space of both models was conducted. Using a clustering algorithm, it was possible to find a small number of benchmark points that showcase the phenomenology in a comprehensive way. These points were then complemented by exceptional points where needed, and the most important phenomenological data is summarized in tables \ref{fig:benchmarkTable1}, \ref{fig:benchmarkTableClusteringMCHM5} and \ref{fig:benchmarkTableClusteringMCHM5HS} for the MCHM$_5$, and tables \ref{fig:benchmarkTable2}, \ref{fig:benchmarkTableClusteringMCHM14} and
 \ref{fig:benchmarkTableClusteringMCHM14HS} for the MCHM$_{14}$. This exploration was done in two steps, first covering physics at ``Low Scale'', where we focus on the part of the parameter space that is within reach of the HL-LHC. On a second step we extend the analysis to the ``High Scale'' region, with mass parameters reaching up to 30 TeV, which will be of interest for planned future colliders such as the 100 TeV FCC-hh and the SppC.

The first observations from the phenomenological and preliminary experimental analysis are that a deviation from the SM in the $t{\overline t}h$ production is also an essential measurement for MCHM. An increase will reject the MCHM$_5$ scenario and greatly refine the areas of the parameter space where MCHM$_{14}$ would be valid. A deficit instead, would make MCHM$_5$ and MCHM$_{14}$ both possible. The measurement of this observable is expected to be achieved with 3.4\% accuracy at the HL-LHC and thus with a very high accuracy (at least at the percent level) at the future high energy hadron collider in project at 100 TeV.

Regarding the search for fermionic resonances that are present in the partial compositeness scenario we explored, we find that in most of the benchmark points the 3-body decay channel of the lightest top-partner starts to become increasingly important as its mass grows. For all the benchmark points in this category, the 3-body decay is already 10\% of the branching ratio when the mass is around 1.3 TeV, increasing to around 25\% at 2 TeV and becomes the main decay channel when the top-partner reaches 3 TeV. The same points also have a marked suppression of decays to $W^+ b$.
As expected (see e.g. \cite{De_Simone_2013}), we find that it is not so common to have a ``split spectrum'' with a low lying top-partner separated in mass from other fermionic resonances since a full $SO(4)$ multiplet is controlled by a single mass parameter. This results in complicated interplay between the many states present, and is part of the reason for the increase in three 3-body decay channel. This makes it extremely important to take 3-body decays into consideration in future top-partner searches, and we intend to explore the decay patterns of all the fermionic resonances, and the resulting search strategies, in a upcoming work.

A comparison with the EFT, valid in energy regimes where the masses of the fermionic resonances are not reached, is also provided (see figures \ref{fig:eftplotsLS} and \ref{fig:eftplots1LSHS}). This is specially interesting in the situation we are right now, and most probably will be at least until the end of the HL-LHC phase, in which no new states have been discovered but great improvements on the measurements of the coupling constants are being made. If deviations from the SM in more than one of these couplings are found, the specific combination of deviations can be used to differentiate not only between the MHCM$_5$ and the MCHM$_{14}$, but also between different regions of those models, as these regions generate different correlations between the effective couplings.

For the process $t{\overline t}hh$, both MCHM scenarios can deviate significantly from the SM expectation, either as a deficit or an increase. For this reason also this channel will play an  important role in searches of composite Higgs models.
 The issue is the relatively low SM cross section of about 1 fb at tree level at 14 TeV, whereas $t{\overline t}h$ is roughly a factor 500 higher. Thus, the aim at HL-LHC will be to evidence this process and to get a first indication of a strong deviation from the SM. For really exploring MCHM, higher energy together with higher luminosity as foreseen in the pp colliders in project towards the second half of this century (100 TeV and exceeding 20 ab$^{-1}$ total integrated luminosity) are not only a plus but indeed even a necessity.

\acknowledgments

The authors thank Geum Bong Yu for useful discussions and references. One of us L. A. F. do Prado thanks IRFU-CEA, University Paris Saclay for hospitality. This work was supported by the S\~ao Paulo Research Foundation
(FAPESP) under grants \#2016/01343-7, \#2013/01907-0,
 \#2015/26624-6 and  \#2018/11505-0, by Science Without Borders / CAPES for
UNESP-SPRACE under the Grant No.~88887.116917/2016-00 and in part by the Coordena\c{c}\~{a}o de Aperfei\c{c}oamento de Pessoal de N\'{i}vel Superior -- Brasil (CAPES) -- Finance Code 001.

\appendix

\section{Representations of SO(5)}
\label{app:generators}

We use the following $5\times5$ matrix representation of the
generators $T^B$ of SO(5):
\begin{eqnarray}
T^{1}_L=\left(
\begin{array}{ccccc}
 0 & 0 & 0 & -\frac{i}{2} & 0 \\
 0 & 0 & -\frac{i}{2} & 0 & 0 \\
 0 & \frac{i}{2} & 0 & 0 & 0 \\
 \frac{i}{2} & 0 & 0 & 0 & 0 \\
 0 & 0 & 0 & 0 & 0
\end{array}
\right)\ , \
T^{2}_L=\left(
\begin{array}{ccccc}
 0 & 0 & \frac{i}{2} & 0 & 0 \\
 0 & 0 & 0 & -\frac{i}{2} & 0 \\
 -\frac{i}{2} & 0 & 0 & 0 & 0 \\
 0 & \frac{i}{2} & 0 & 0 & 0 \\
 0 & 0 & 0 & 0 & 0
\end{array}
\right)\ , \
T^{3}_L=\left(
\begin{array}{ccccc}
 0 & -\frac{i}{2} & 0 & 0 & 0 \\
 \frac{i}{2} & 0 & 0 & 0 & 0 \\
 0 & 0 & 0 & -\frac{i}{2} & 0 \\
 0 & 0 & \frac{i}{2} & 0 & 0 \\
 0 & 0 & 0 & 0 & 0
\end{array}
\right)\ , \nonumber
\end{eqnarray}
\begin{eqnarray}
T^{1}_R=\left(
\begin{array}{ccccc}
 0 & 0 & 0 & \frac{i}{2} & 0 \\
 0 & 0 & -\frac{i}{2} & 0 & 0 \\
 0 & \frac{i}{2} & 0 & 0 & 0 \\
 -\frac{i}{2} & 0 & 0 & 0 & 0 \\
 0 & 0 & 0 & 0 & 0
\end{array}
\right)\ , \
T^{2}_R=\left(
\begin{array}{ccccc}
 0 & 0 & \frac{i}{2} & 0 & 0 \\
 0 & 0 & 0 & \frac{i}{2} & 0 \\
 -\frac{i}{2} & 0 & 0 & 0 & 0 \\
 0 & -\frac{i}{2} & 0 & 0 & 0 \\
 0 & 0 & 0 & 0 & 0
\end{array}
\right)\ , \
T^{3}_R=\left(
\begin{array}{ccccc}
 0 & -\frac{i}{2} & 0 & 0 & 0 \\
 \frac{i}{2} & 0 & 0 & 0 & 0 \\
 0 & 0 & 0 & \frac{i}{2} & 0 \\
 0 & 0 & -\frac{i}{2} & 0 & 0 \\
 0 & 0 & 0 & 0 & 0
\end{array}
\right)\ , \nonumber
\end{eqnarray}
\begin{eqnarray}
T^{\hat 1}=\left(
\begin{array}{ccccc}
 0 & 0 & 0 & 0 & -\frac{i}{\sqrt{2}} \\
 0 & 0 & 0 & 0 & 0 \\
 0 & 0 & 0 & 0 & 0 \\
 0 & 0 & 0 & 0 & 0 \\
 \frac{i}{\sqrt{2}} & 0 & 0 & 0 & 0
\end{array}
\right)\ , \
T^{\hat 2}=\left(
\begin{array}{ccccc}
 0 & 0 & 0 & 0 & 0 \\
 0 & 0 & 0 & 0 & -\frac{i}{\sqrt{2}} \\
 0 & 0 & 0 & 0 & 0 \\
 0 & 0 & 0 & 0 & 0 \\
 0 & \frac{i}{\sqrt{2}} & 0 & 0 & 0
\end{array}
\right)\ ,\nonumber
\end{eqnarray}
\begin{eqnarray}
T^{\hat 3}=\left(
\begin{array}{ccccc}
 0 & 0 & 0 & 0 & 0 \\
 0 & 0 & 0 & 0 & 0 \\
 0 & 0 & 0 & 0 & -\frac{i}{\sqrt{2}} \\
 0 & 0 & 0 & 0 & 0 \\
 0 & 0 & \frac{i}{\sqrt{2}} & 0 & 0
\end{array}
\right)\ , \
T^{\hat 4}=\left(
\begin{array}{ccccc}
 0 & 0 & 0 & 0 & 0 \\
 0 & 0 & 0 & 0 & 0 \\
 0 & 0 & 0 & 0 & 0 \\
 0 & 0 & 0 & 0 & -\frac{i}{\sqrt{2}} \\
 0 & 0 & 0 & \frac{i}{\sqrt{2}} & 0
\end{array}
\right)\ ,
\label{T5}
\end{eqnarray}
which act on the fundamental representation {\bf 5} of SO(5) as
$\delta \Psi_{\bf 5} = T^B \Psi_{\bf 5}$.  We write the {\bf 5}
representation (using a notation appropriate for states with $U(1)_X$
charge $X = 2/3$) as
\bea
\Psi^{(2/3)}_{\bf 5} &=&
X_{5/3} \, v_{{\frac{1}{2},\frac{1}{2}}} +
X_{2/3} \, v_{-\frac{1}{2},\frac{1}{2}} +
T \, v_{{\frac{1}{2},-\frac{1}{2}}} +
B \, v_{-\frac{1}{2},-\frac{1}{2}} +
\tilde{T} \, v_{0}~,
\eea
where we used the normalized basis
{ \small
\bea
\begin{array}{cclcclcclccl}
v_{\frac{1}{2},\frac{1}{2}} &=& \frac{1}{\sqrt{2}}\left(\begin{array}{c}i\\-1\\0\\0\\0\end{array}\right),
&
v_{-\frac{1}{2},\frac{1}{2}} &=& \frac{1}{\sqrt{2}}\left(\begin{array}{c}0\\0\\-i\\-1\\0\end{array}\right),
&
v_{\frac{1}{2},-\frac{1}{2}} &=& \frac{1}{\sqrt{2}}\left(\begin{array}{c}0\\0\\-i\\1\\0\end{array}\right),
&
v_{-\frac{1}{2},-\frac{1}{2}} &=& \frac{1}{\sqrt{2}}\left(\begin{array}{c}-i\\-1\\0\\0\\0\end{array}\right),
\nonumber
\end{array}
\eea
}
and
\bea
v^T_{0} = \left( 0, 0, 0, 0, 1 \right)~.
\nonumber
\eea
The notation is such that the subindices $a,b$ denote the
$T^{3}_{L,R}$ eigenvalues: $T^{3}_{L} \, v_{a,b} = a \, v_{a,b}$,
$T^{3}_{R} \, v_{a,b} = b \, v_{a,b}$, while $v_{0}$ denotes the
complete singlet, $T^{i}_{L} \, v_{0} = T^{i}_{R} \, v_{0} = 0$.
Furthermore, one has $T^+_L \, v_{\frac{1}{2}, \pm \frac{1}{2}} = 0$,
$T^+_L \, v_{-\frac{1}{2}, \pm \frac{1}{2}} = v_{\frac{1}{2}, \pm
\frac{1}{2}}$, $T^-_L \, v_{\frac{1}{2}, \pm \frac{1}{2}} =
v_{-\frac{1}{2}, \pm \frac{1}{2}}$, $T^-_L \, v_{-\frac{1}{2}, \pm
\frac{1}{2}} = 0$, and similarly $T^+_R \, v_{\pm
\frac{1}{2},\frac{1}{2}} = 0$, $T^+_R \, v_{\pm \frac{1}{2},
-\frac{1}{2}} = v_{\pm \frac{1}{2},\frac{1}{2}}$, $T^-_R \, v_{\pm
\frac{1}{2},\frac{1}{2}} = v_{\pm \frac{1}{2}, - \frac{1}{2}}$, $T^-_R
\, v_{\pm \frac{1}{2}, -\frac{1}{2}} = 0$, where $T^{\pm}_{L,R} \equiv
T^1_{L,R} \pm i T^2_{L,R}$ are the standard $SU(2)$ raising and
lowering operators, satisfying
\bea
\begin{array}{cclcccl}
[T^3_{L} , T^\pm_{L}] &=& \pm T^\pm_{L}~,  &\hspace{0.7em} & [T^3_{R} , T^\pm_{R}] &=& \pm T^\pm_{R}~,
\nonumber \\ [0.5em]
[T^{+}_{L} , T^{-}_{L}] &=& 2 \, T^3_{L}~,  &\hspace{0.7em} &  [T^{+}_{R} , T^{-}_{R}] &=& 2 \, T^3_{R}~,
\end{array}
\eea
and $[T^i_L, T^j_R] = 0$.  The states satisfy the standard
normalization
\bea
T^\pm_L |s_L, m_L ; s_R, m_R \rangle &=& \sqrt{s_L (s_L + 1) - m_L (m_L \mp 1)} \, |s_L, m_L \pm 1; s_R, m_R \rangle~,
\nonumber
\\ [0.5em]
T^\pm_R |s_L, m_L ; s_R, m_R \rangle &=& \sqrt{s_R (s_R + 1) - m_R (m_R \mp 1)} \, |s_L, m_L ; s_R, m_R \pm 1 \rangle~.
\nonumber
\eea
Thus, $\Psi_{\bf 4} \sim (X_{5/3}, X_{2/3}, T, B)$ is a bi-doublet
$({\bf 2}, {\bf 2})$ of $SU(2)_L \times SU(2)_R$ with $(s_L, s_R) =
(1/2, 1/2)$, while $\Psi_{\bf 1} \sim \tilde{T}$ is a singlet with
$(s_L, s_R) = (0,0)$.

The ${\bf 14}$ representation of SO(5) can be written in terms of a
$5\times5$ symmetric and traceless matrix, such that the $SO(5)$
generators (\ref{T5}) act as $\delta \Psi_{\bf 14}=[T^B,\Psi_{\bf
14}]$.  Using again the notation for states with $X = 2/3$, we write
\bea
\Psi^{(2/3)}_{\bf 14} &=&
\tilde{T} \, S_0 +
(X_{\frac{5}{3}} \, S_{\frac{1}{2},\frac{1}{2}} +
X_{\frac{2}{3}} \, S_{-\frac{1}{2},\frac{1}{2}} +
T \, S_{\frac{1}{2},-\frac{1}{2}} +
B \, S_{-\frac{1}{2},-\frac{1}{2}}
)
\nonumber
\\ [0.5em]
&& \mbox{} +
[U_{\frac{8}{3}} S_{1,1} +
U_{\frac{5}{3}} S_{0,1} +
U_{\frac{2}{3}} S_{-1,1}
\nonumber
\\ [0.5em]
&& \mbox{} +
~V_{\frac{5}{3}} \, S_{1,0} +
V_{\frac{2}{3}} \, S_{0,0} +
V_{-\frac{1}{3}} \, S_{-1,0}
\\ [0.5em]
&& \mbox{} +
~F_{\frac{2}{3}} \, S_{1,-1} +
F_{-\frac{1}{3}} \, S_{0,-1} +
F_{-\frac{4}{3}} \, S_{-1,-1}
]
\nonumber
\eea
which exhibits the decomposition {$\bf 14\sim (3, 3)+(2, 2)+(1,1)$}
under ${\rm SU(2)}_L \times {\rm SU(2)}_R$.

The notation for the basis $S_{a,b}$ is the same as explained above,
e.g. $[T^{3}_{L} , S_{a,b}] = a \, S_{a,b}$, $[T^{3}_{R} , S_{a,b} ] =
b \, S_{a,b}$, while $S_{0}$ denotes the complete singlet, $[T^{i}_{L}
, S_{0}] = [T^{i}_{R} , S_{0}] = 0$.  The states $\Psi_{\bf 9} \sim
(U_{8/3}, U_{5/3}, U_{2/3}, V_{5/3}, V_{2/3}, V_{-1/3}, F_{2/3},
F_{-1/3}, F_{-4/3})$ transform as a bi-triplet of $SU(2)_L \times
SU(2)_R$ with $(s_L, s_R) = (1, 1)$, and the bi-doublet $\Psi_{\bf 4}$
and singlet $\Psi_{\bf 1}$ follow the same notation used for the ${\bf
5}$ of $SO(5)$ above.  The subindices on the fields denote the
electric charge, given by
\bea
Q = T^3_L + T^3_R + X~.
\eea
Explicitly, the $S_{a,b}$ and $S_0$ matrices are given by:
{ \small
\bea
S_{1,1} = \frac{1}{2}
\left(
\begin{array}{ccccc}
 1 & i & 0 & 0 & 0 \\
 i & -1 & 0 & 0 & 0 \\
 0 & 0 & 0 & 0 & 0 \\
 0 & 0 & 0 & 0 & 0 \\
 0 & 0 & 0 & 0 & 0 \\
\end{array}
\right),
 &
S_{1,0} = \frac{1}{2\sqrt{2}}
\left(
\begin{array}{ccccc}
 0 & 0 & -1 & -i & 0 \\
 0 & 0 & -i & 1 & 0 \\
 -1 & -i & 0 & 0 & 0 \\
 -i & 1 & 0 & 0 & 0 \\
 0 & 0 & 0 & 0 & 0 \\
\end{array}
\right),
 &
S_{1,-1} = \frac{1}{2}
\left(
\begin{array}{ccccc}
 0 & 0 & 0 & 0 & 0 \\
 0 & 0 & 0 & 0 & 0 \\
 0 & 0 & 1 & i & 0 \\
 0 & 0 & i & -1 & 0 \\
 0 & 0 & 0 & 0 & 0 \\
\end{array}
\right),
\nonumber
\eea
}
{ \small
\bea
S_{0,1} = \frac{1}{2\sqrt{2}}
\left(
\begin{array}{ccccc}
 0 & 0 & -1 & i & 0 \\
 0 & 0 & -i & -1 & 0 \\
 -1 & -i & 0 & 0 & 0 \\
 i & -1 & 0 & 0 & 0 \\
 0 & 0 & 0 & 0 & 0 \\
\end{array}
\right),
 &
S_{0,0} = \frac{1}{2}
\left(
\begin{array}{ccccc}
 -1 & 0 & 0 & 0 & 0 \\
 0 & -1 & 0 & 0 & 0 \\
 0 & 0 & 1 & 0 & 0 \\
 0 & 0 & 0 & 1 & 0 \\
 0 & 0 & 0 & 0 & 0 \\
\end{array}
\right),
 &
S_{0,-1} = \frac{1}{2\sqrt{2}}
\left(
\begin{array}{ccccc}
 0 & 0 & 1 & i & 0 \\
 0 & 0 & -i & 1 & 0 \\
 1 & -i & 0 & 0 & 0 \\
 i & 1 & 0 & 0 & 0 \\
 0 & 0 & 0 & 0 & 0 \\
\end{array}
\right),
\nonumber
\eea
}
{ \small
\bea
S_{-1,1} = \frac{1}{2}
\left(
\begin{array}{ccccc}
 0 & 0 & 0 & 0 & 0 \\
 0 & 0 & 0 & 0 & 0 \\
 0 & 0 & 1 & -i & 0 \\
 0 & 0 & -i & -1 & 0 \\
 0 & 0 & 0 & 0 & 0 \\
\end{array}
\right),
 &
S_{-1,0} = \frac{1}{2\sqrt{2}}
\left(
\begin{array}{ccccc}
 0 & 0 & 1 & -i & 0 \\
 0 & 0 & -i & -1 & 0 \\
 1 & -i & 0 & 0 & 0 \\
 -i & -1 & 0 & 0 & 0 \\
 0 & 0 & 0 & 0 & 0 \\
\end{array}
\right),
 &
S_{-1,-1} = \frac{1}{2}
\left(
\begin{array}{ccccc}
 1 & -i & 0 & 0 & 0 \\
 -i & -1 & 0 & 0 & 0 \\
 0 & 0 & 0 & 0 & 0 \\
 0 & 0 & 0 & 0 & 0 \\
 0 & 0 & 0 & 0 & 0 \\
\end{array}
\right),
\nonumber
\eea
}
for the ${\bf 9}$ of $SO(4) = SU(2)_L \times SU(2)_R$,
{ \normalsize
\bea
\begin{array}{cclccl}
S_{\frac{1}{2},\frac{1}{2}} &=& \frac{1}{2}
\left(
\begin{array}{ccccc}
 0 & 0 & 0 & 0 & i \\
 0 & 0 & 0 & 0 & -1 \\
 0 & 0 & 0 & 0 & 0 \\
 0 & 0 & 0 & 0 & 0 \\
 i & -1 & 0 & 0 & 0 \\
\end{array}
\right)~,
 &
S_{\frac{1}{2},-\frac{1}{2}} &=& \frac{1}{2}
\left(
\begin{array}{ccccc}
 0 & 0 & 0 & 0 & 0 \\
 0 & 0 & 0 & 0 & 0 \\
 0 & 0 & 0 & 0 & -i \\
 0 & 0 & 0 & 0 & 1 \\
 0 & 0 & -i & 1 & 0 \\
\end{array}
\right)~,
\nonumber
 \\ [3.5em]
S_{-\frac{1}{2},\frac{1}{2}} &=& \frac{1}{2}
\left(
\begin{array}{ccccc}
 0 & 0 & 0 & 0 & 0 \\
 0 & 0 & 0 & 0 & 0 \\
 0 & 0 & 0 & 0 & -i \\
 0 & 0 & 0 & 0 & -1 \\
 0 & 0 & -i & -1 & 0 \\
\end{array}
\right)~,
 &
S_{-\frac{1}{2},-\frac{1}{2}} &=& \frac{1}{2}
\left(
\begin{array}{ccccc}
 0 & 0 & 0 & 0 & -i \\
 0 & 0 & 0 & 0 & -1 \\
 0 & 0 & 0 & 0 & 0 \\
 0 & 0 & 0 & 0 & 0 \\
 -i & -1 & 0 & 0 & 0 \\
\end{array}
\right)~,
 \\
\end{array}
\nonumber
\eea
}
for the ${\bf 4}$ of $SO(4)$, and
\bea
S_{0} = \frac{1}{2 \sqrt{5}}
\left(
\begin{array}{ccccc}
 1 & 0 & 0 & 0 & 0 \\
 0 & 1 & 0 & 0 & 0 \\
 0 & 0 & 1 & 0 & 0 \\
 0 & 0 & 0 & 1 & 0 \\
 0 & 0 & 0 & 0 & -4 \\
\end{array}
\right)~.
\nonumber
\eea
for the $SO(4)$ singlet.  These matrices form an orthonormal basis:
${\rm Tr}(S^*_A S_B) = \delta_{AB}$.

\section{Embeddings of $SO(4)$ into $SO(5)$}
\label{app:embeddings}

The four NGBs resulting from the spontaneous breaking of $SO(5) \to
SO(4) = SU(2)_L \times SU(2)_R$ can be parametrized as
\bea
U &=& e^{i \frac{\sqrt{2}}{f} h^{\hat a} T^{\hat{a}}}~,
\eea
where $T^{\hat{a}}$ are the (four) broken generators in $SO(5)/SO(4)$
given in Eq~(\ref{T5}) and $f$ is the scale of spontaneous symmetry
breaking.  The $h^{\hat a}$ transform as a 4-plet of $SO(4)$, and can
be arranged in a doublet of $SU(2)_L$ as:
\bea
H &=& \frac{1}{\sqrt{2}}
\left(\begin{array}{c}
        h_2 + i h_1 \\
        h_4 - i h_3
\end{array}\right).
\label{HDoublet}
\eea
One can assume that EWSB proceeds through a non-vanishing vev $h_0 =
\langle h_4 \rangle$, with the vev's of the other components
vanishing.  In unitary gauge, $h_1 = h_2 = h_3 = 0$ and $h_4 = h_0 +
h$, where $h$ is the physical Higgs boson.  Using the explicit form of
the broken generators results in the matrix given in
Eq.~(\ref{Umatrix}).

It is also easy to embed the various fermion $SO(4)$ multiplets used
in the main text into (incomplete) representations of $SO(5)$ that
simplify the writing of the Lagrangian.  For example,
\begin{itemize}

\item For the MCHM$_5$:

The elementary fermions are written as
\bea
Q_L^{\mathbf{5}} &=&
t_L \, v_{{\frac{1}{2},-\frac{1}{2}}} +
b_L \, v_{-\frac{1}{2},-\frac{1}{2}}~,
\nonumber \\
T_R^{\mathbf{5}} &=& t_R \, v_{0}~,
\nonumber
\eea
and the composite fermions are written as
\bea
\Psi_{\bf 4} &=&
X_{5/3} \, v_{{\frac{1}{2},\frac{1}{2}}} +
X_{2/3} \, v_{-\frac{1}{2},\frac{1}{2}} +
T \, v_{{\frac{1}{2},-\frac{1}{2}}} +
B \, v_{-\frac{1}{2},-\frac{1}{2}}~,
\nonumber \\
\Psi_{\bf 1} &=& \tilde{T} \, v_{0}~.
\nonumber
\eea
These result in Eqs.~(\ref{e:embed5}) and (\ref{e:res5}).

\item For the MCHM$_{14}$:

The elementary fermions are written as
\bea
Q_L^{\mathbf{14}} &=&
t_L \, S_{{\frac{1}{2},-\frac{1}{2}}} +
b_L \, S_{-\frac{1}{2},-\frac{1}{2}}~,
\nonumber \\
T_R^{\mathbf{14}} &=& t_R \, S_{0}~,
\nonumber
\eea
and the composite fermions are written as
\bea
\Psi_{\bf 9} &=&
[U_{\frac{8}{3}} S_{1,1} +
U_{\frac{5}{3}} S_{0,1} +
U_{\frac{2}{3}} S_{-1,1}
\nonumber
\\ [0.5em]
&& \mbox{} +
~V_{\frac{5}{3}} \, S_{1,0} +
V_{\frac{2}{3}} \, S_{0,0} +
V_{-\frac{1}{3}} \, S_{-1,0}
\nonumber
\\ [0.5em]
&& \mbox{} +
~F_{\frac{2}{3}} \, S_{1,-1} +
F_{-\frac{1}{3}} \, S_{0,-1} +
F_{-\frac{4}{3}} \, S_{-1,-1}
]
\nonumber \\ [0.5em]
\Psi_{\bf 4} &=&
X_{5/3} \, S_{{\frac{1}{2},\frac{1}{2}}} +
X_{2/3} \, S_{-\frac{1}{2},\frac{1}{2}} +
T \, S_{{\frac{1}{2},-\frac{1}{2}}} +
B \, S_{-\frac{1}{2},-\frac{1}{2}}~,
\nonumber \\ [0.5em]
\Psi_{\bf 1} &=& \tilde{T} \, S_{0}~.
\nonumber
\eea
All of these are traceless, symmetric $5 \times 5$ matrices.  It is
then straightforward to form a complete ${\bf 14}$ of $SO(5)$ as
\bea
\Psi_{\bf 14} &=& \Psi_{\bf 9} + \Psi_{\bf 4} + \Psi_{\bf 1}~.
\nonumber
\eea

\end{itemize}
%
\section{Explicit form of gauge and Higgs interactions}
\label{app:vertices}

In unitary gauge, the gauged $d_\mu$ and $e_\mu$ symbols are given by

\bea
d_\mu &=&  \left\{ \frac{g}{\sqrt{2}} W^1_\mu s_h,~\frac{g}{\sqrt{2}} W^2_\mu s_h,~\frac{g W^3_\mu-g^\prime B_\mu}{\sqrt{2}}  s_h, \frac{\partial_\mu h}{f}\right\}\nonumber\\
e_\mu &=& \frac{i}{2}\left(
          \begin{array}{cccc}
            0 & -g^\prime B_\mu- g W^3_\mu & g W^2_\mu & -g W^1_\mu c_h \\
            g^\prime B_\mu+ g W^3_\mu & 0 & -g W^1_\mu & -g W^2_\mu c_h \\
            -g W^2_\mu & g W^1_\mu & 0 & (g^\prime B_\mu - g W^3_\mu) c_h \\
            g W^1_\mu c_h & g W^2_\mu c_h & -(g^\prime B_\mu - g W^3_\mu) c_h & 0 \\
          \end{array}
        \right),
\eea
where $c_h = \cos\frac{h_0+h}{f}$ and $s_h = \sin\frac{h_0+h}{f}$. Under $SO(4)$, $d_\mu^i$ transforms in the fundamental and $e_\mu^a$ in the adjoint.
In terms of these, as well as the fermion embeddings from the previous appendices, we may write explicitly the vertices from the Lagrangians $\mathcal{L}_{\bf int}^{\rm 5, 14}$ in the interaction basis as

\begin{equation}
\begin{split}
&\mathcal{L}^{\mathbf{5}}_{\rm int}= ~ c_R \left\{ g \sqrt{\xi} \left[\frac{1}{\sqrt{2}}\left(\overline{X}_{5/3}\right)_R \slashed{W}^+ \tilde{T}_R - \frac{1}{\sqrt{2}}\overline{B}_R\, \slashed{W}^- \tilde{T}_R -\frac{1}{2 c_w}\overline{T}_R \, \slashed{Z} \, \tilde{T}_R\right.\right.\\
 &\left.\left.-\frac{1}{2 c_w}\left(\overline{X}_{2/3}\right)_R \slashed{Z}\,  \tilde{T}_R \right]+\,i\left[ \left(\overline{X}_{2/3}\right)_R -\overline{T}_R\right]\frac{\slashed{\partial}h}{f}\tilde{T}_R\right\}+\left(R\rightarrow L\right)+\text{h.c.}
\end{split}
\end{equation}

\beann
\mathcal{L}^{\mathbf{14}}_{\rm int}&=&
        c_4 \left[ g \sqrt{\xi} \left(\frac{1}{\sqrt{2}}\overline{X}_{5/3} \slashed{W}^+ \tilde{T} - \frac{1}{\sqrt{2}}\overline{B} \slashed{W}^- \tilde{T} -\frac{1}{2 c_w}\overline{T} \slashed{Z} \tilde{T}\right.\right.
 \left.\left.-\frac{1}{2 c_w}\overline{X}_{2/3} \slashed{Z} \tilde{T} \right)\right.\\
        & &\hspace{1cm}\left.+\,i\left( \overline{X}_{2/3} -\overline{T}\right)\frac{\slashed{\partial}h}{f}\tilde{T}\right]\\
        &\\
        &-&\frac{1}{4}\,c_9\, \frac{g_2}{c_w}\sqrt{\xi}\left\{\sqrt{2}\left[\overline{V}_{-1/3}+\overline{F}_{-1/3}\right]\slashed{Z} B + \left[\overline{V}_{2/3} + 2 \overline{U}_{2/3}\right]\slashed{Z} X_{2/3}\right.\\
        &+&\left.\left[\overline{V}_{2/3} + 2 \overline{F}_{2/3}\right]\slashed{Z} \,T+\sqrt{2}\left[\overline{U}_{5/3}+\overline{V}_{5/3}\right]\slashed{Z} X_{5/3}\right\}\\
        &&\\
        &-&\frac{1}{4}c_9 g_2\sqrt{\xi}\left\{ \sqrt{2}\overline{V}_{2/3}\slashed{W}^-\overline{X}_{5/3}+2\sqrt{2}\overline{F}_{-4/3}\slashed{W}^- B
        + 2 \overline{F}_{-1/3}\slashed{W}^{-}T+2\overline{V}_{-1/3}\slashed{W}^{-}X_{2/3} \right.\\
        &-&\left.\sqrt{2}\overline{V}_{2/3}\slashed{W}^+B+2\sqrt{2}\overline{U}_{8/3}\slashed{W}^+ X_{5/3}
        +2 \overline{V}_{5/3}\slashed{W}^+T+2\overline{U}_{5/3}\slashed{W}^+X_{2/3}\right\} \\
        &&\\
        &+& \frac{i}{2 f}c_9 \left\{\sqrt{2}\left[\overline{V}_{-1/3}-\overline{F}_{-1/3}\right] \slashed{\partial}h B+\left[\overline{V}_{2/3}-2\overline{F}_{2/3}\right] \slashed{\partial}h\,T+\right.\\
        &&\left.\left[2\overline{U}_{2/3}-\overline{V}_{2/3}\right] \slashed{\partial}h\, X_{2/3}+\sqrt{2}\left[\overline{U}_{5/3}-\overline{V}_{5/3}\right] \slashed{\partial}h\, X_{5/3}\right\}\\
        &&\\
    &+&\frac{i c_{T9}}{4 \pi f^3}\partial_\mu h\,\partial^\mu h\left[\overline{U}_{2/3}-\overline{V}_{2/3}+\overline{F}_{2/3}\right]\tilde{T}\\
    &&\\
    &+&\frac{c_{T9} g_2 \sqrt{\xi}}{4\pi f^2 c_w}Z_\mu \left[ \partial^\mu h\left(\overline{F}_{2/3}-\overline{U}_{2/3}\right)+W_\mu^+\left(\overline{U}_{5/3}-\overline{V}_{5/3}\right)
    +W_\mu^-\left(\overline{F}_{-1/3}-\overline{V}_{-1/3}\right)\right]\tilde{T} \\
    &&\\
    &+&\frac{i c_{T9} g_2^2 \xi}{8 \pi f}\left[-\frac{Z_\mu Z^\mu}{2c_w}\left(\overline{U}_{2/3}+\overline{V}_{2/3}+\overline{F}_{2/3}\right)
    -\,W^+_\mu W^{+\mu}\overline{U}_{8/3}-\,\,W^-_\mu W^{-\mu}\overline{F}_{-4/3}\right.\\
    &+&\,\,W^+_\mu W^{-\mu}\overline{V}_{2/3}-\left.\frac{Z_\mu W^{-\mu}}{c_w}\left(\overline{V}_{-1/3}+\overline{F}_{-1/3}\right)
    +\frac{Z_\mu W^{+\mu}}{c_w}\left(\overline{U}_{5/3}+\overline{V}_{5/3}\right)\right]\tilde{T} +\text{h. c.},
\eeann
where $c_w$ is the cosine of the Weinberg angle.

Furthermore, the $e_\mu$ symbol in the covariant derivatives lead to compositeness corrected electroweak interactions for the resonances in the fourplet and nonet, given by

\bea
\mathcal{L}^{\mathbf{5}}_{\rm gauge} &=& \overline{\Psi}_{\mathbf{4}} \left(\frac{2}{3} g^\prime \slashed{B}-\slashed{e}\right) \Psi_{\mathbf{4}} \nonumber \\ &=& \frac{g}{c_w} \left[\left(-\frac{1}{2}+ \frac{1}{3} s_w^2\right) \overline{B} \slashed {Z} B
+  \left(\frac{1}{2} -\frac{5}{3} s_w^2\right) \overline{X}_{5/3} \slashed {Z} X_{5/3}\right. \nonumber \\
&+& \left. \left(\frac{\sqrt{1-\xi}}{2}  -\frac{2}{3} s_w^2\right) \overline{T} \slashed {Z} T + \left(-\frac{\sqrt{1-\xi}}{2} -\frac{2}{3} s_w^2\right) \overline{X}_{2/3} \slashed {Z} X_{2/3} \right]\nonumber \\
&+& \frac{g}{\sqrt{2}} \frac{1+\sqrt{1-\xi}}{2} \left[ \overline{B} \slashed {W}^{-} T +\overline{X}_{5/3} \slashed{W}^{+} X_{2/3}+  \text{h. c.} \right]\nonumber \\
&+& \frac{g}{\sqrt{2}} \frac{1-\sqrt{1-\xi}}{2} \left[ \overline{X}_{5/3} \slashed{W}^{+} T +\overline{B} \slashed{W}^{-} X_{2/3}  + \text{h. c.}\right],
\eea
plus standard photon couplings, for the fourplet (these are the same as in reference \cite{De_Simone_2013}, reported here again for completeness), and
\bea
\mathcal{L}^{\mathbf{14}}_{\rm gauge} &=& \overline{\Psi}_{\mathbf{9}} \left(\frac{2}{3} g^\prime \slashed{B} \Psi_{\mathbf{9}}-[\slashed{e},\Psi_{\mathbf{9}}]\right) \nonumber \\
&=& \frac{g}{c_w} \left[\left(-\sqrt{1-\xi}+ -\frac{2}{3} s_w^2\right) \overline{U}_{2/3} \slashed {Z} U_{2/3}
+  \left(\frac{1-\sqrt{1-\xi}}{2} -\frac{5}{3} s_w^2\right) \overline{U}_{5/3} \slashed {Z} U_{5/3} \right. \nonumber \\
&+& \left. \left(\frac{1}{2}  -\frac{8}{3} s_w^2\right) \overline{U}_{8/3} \slashed {Z} U_{8/3} +  \left(-\frac{1+\sqrt{1-\xi}}{2} +\frac{1}{3} s_w^2\right) \overline{V}_{-1/3} \slashed {Z} V_{-1/3} \right.\nonumber \\
&+& \left.   -\frac{2}{3} s_w^2 \overline{V}_{2/3} \slashed {Z} V_{2/3} + \left(\frac{1+\sqrt{1-\xi}}{2} +\frac{1}{3} s_w^2\right) \overline{V}_{5/3} \slashed {Z} V_{5/3} \right.\nonumber \\
&+& \left.  \left(-1 +\frac{4}{3} s_w^2\right) \overline{F}_{-4/3} \slashed {Z} F_{-4/3} + \left(-\frac{1-\sqrt{1-\xi}}{2} +\frac{1}{3} s_w^2\right) \overline{F}_{-1/3} \slashed {Z} F_{-1/3} \right.\nonumber \\
&+& \left.  \left(\sqrt{1-\xi} -\frac{2}{3} s_w^2\right) \overline{F}_{2/3} \slashed {Z} F_{2/3} \right]\nonumber \\
&+& \frac{g}{2} (1+\sqrt{1-\xi}) \left[\overline{V}_{5/3} \slashed{W}^{+} V_{2/3} +\overline{V}_{2/3} \slashed {W}^{+} V_{-1/3} \right. \nonumber \\
&+&\left.\overline{F}_{-1/3} \slashed{W}^{+} F_{-4/3} +\overline{F}_{2/3} \slashed{W}^{+} F_{-1/3} \right.\nonumber \\
&+&\left.\overline{U}_{8/3} \slashed{W}^{+} U_{5/3} +\overline{U}_{5/3} \slashed{W}^{+} U_{2/3} +\text{h. c.} \right]\nonumber \\
&+&\frac{g}{2} (1-\sqrt{1-\xi}) \left[\overline{V}_{-1/3} \slashed{W}^{+} F_{-4/3} +\overline{V}_{2/3} \slashed {W}^{+} F_{-1/3} \right. \nonumber \\
&+&\left.\overline{V}_{5/3} \slashed{W}^{+} F_{2/3} +\overline{U}_{5/3} \slashed{W}^{+} V_{2/3} \right.\nonumber \\
&+&\left.\overline{U}_{8/3} \slashed{W}^{+} V_{5/3} +\overline{U}_{2/3} \slashed{W}^{+} V_{-1/3} +\text{h. c.} \right],
\eea
 plus standard photon couplings, for the nonet.

\bibliographystyle{JHEP}
\bibliography{ref}
\end{document}